\documentclass[epj,nopacs,final,english]{svjour}

\usepackage{amsmath,cite,babel}
\usepackage{graphicx,axodraw}
\usepackage{array}

\newcommand{\newc}{\newcommand}
\newc{\gev}{\,GeV}
\newc{\mev}{\,MeV}
\newc{\ra}{\rightarrow}
\newc{\rpv}{$\mathrm{\not\!R_p}$}
\newc{\rp}{$\mathrm{R_p}$}
\newc{\real}{\mathcal{R}e}
\newc{\alsm}{{\displaystyle \sum_{\alpha=1,2}}}
\newc{\besm}{{\displaystyle \sum_{\beta=1,2}}}
\newc{\al}{\alpha}
\newc{\sgn}{\mr{sgn}\,}
\newc{\be}{\beta}
\newc{\ga}{\gamma}
\newc{\de}{\delta}
\newc{\sla}{\!\!\!\!\!\not\:\:\!}
\newc{\slab}{\!\!\!\!\!\not\,\,\,}
\newc{\slac}{\!\!\!\!\!\!\!\not\,\,\,\,}
\newc{\met}{$\not\!\!E_T$}
\newc{\cw}{\cos\theta_W}
\newc{\sw}{\sin\theta_W}
\newc{\ssw}{\sin^2\theta_W}
\newc{\ccw}{\cos^2\theta_W}
\newc{\cbe}{\cos\beta}
\newc{\sbe}{\sin\beta}
\newc{\ort}{\frac1{\sqrt{2}}}
\newc{\sh}{\hat{s}}
\newc{\uh}{\hat{u}}
\newc{\tha}{\hat{t}}
\newc{\sa}{\sin\al}

\newc{\ca}{\cos\al}
\newc{\mz}{M_{\mr{Z}}}
\newc{\mw}{M_{\mr{W}}}
\newc{\bv}{$\mathrm{\not\!B}$}
\newc{\lv}{$\mathrm{\not\!L}$}
\newc{\beq}{\begin{equation}}
\newc{\eeq}{\end{equation}}
\newc{\ie}{{\it i.e.\/}\ }
\newc{\lam}{\lambda}
\newc{\cht}{\tilde{\chi}}
\newc{\glt}{\tilde{g}}
\newc{\upt}{\tilde{u}}
\newc{\qkt}{\tilde{q}}
\newc{\elt}{\tilde{\ell}}
\newc{\hgt}{\tilde{H}}
\newc{\nut}{\tilde{\nu}}
\newc{\dnt}{\tilde{d}}
\newc{\ftl}{\mr{\tilde{f}}}
\newc{\psb}{\bar{\psi}}
\newc{\rtt}{2^{1/2}}
\newc{\mut}{\tilde{\mu}}
\newc{\mr}{\mathrm}
\newc{\bath}{\bar{\theta}}
\newc{\tht}{\theta}
\newc{\JC}{{\bf J}}
\newc{\lra}{\longrightarrow}
\newc{\eg}{{\it e.g.\  }}
\newc{\barr}{\begin{eqnarray}}
\newc{\earr}{\end{eqnarray}}
\newc{\me}{\mathcal{M}}
\newc{\mesqbar}{\bar{\me}}
\newc{\dbm}{\partial_\mu}
\newc{\dbmu}{\stackrel{\leftrightarrow\  }{\partial^\mu}}
\newc{\sgm}{\sigma_\mu}
\newc{\captionB}[2]{\caption[{#1}]{{\small {#2}}}}
\newc{\ahref}[2]{#2}
\newc{\bpc}{\begin{picture}}
\newc{\epc}{\end{picture}}
\newc{\mesq}{|\me|^2}
\newc{\hel}[1]{\lambda_{#1}}
\newc{\diff}[1]{\frac{d^{3}p_{#1}}{(2\pi)^3 2E_{#1}}}
\newc{\mep}[4]{\me^{#1 #2\ra #3 #4}_{\hel{#3}\hel{#4}}}
\newc{\mepc}[4]{\me^{* #1 #2\ra #3 #4}_{\hel{#3}\hel{#4}}}
\newc{\Hw}{\textsf{Herwig++ }}

\title{\textbf{Simulation of Beyond Standard Model Physics in Herwig++}}
\author{Martyn Gigg \and Peter Richardson}
\institute{Institute of Particle Physics Phenomenology, Department of Physics \\ 
  University of Durham,  Durham DH1 3LE, UK. \\
  \email{m.a.gigg@durham.ac.uk, peter.richardson@durham.ac.uk}  }
\abstract{
  We present a new approach for the simulation of Beyond Standard Model (BSM) 
  physics
  within the \Hw event generator. Our approach is more generic than previous
  methods with the aim of  minimising the effort of implementing 
  further new physics models. Spin correlations, which are important for
  BSM models due to new heavy fermions and bosons, are discussed and
  their effects demonstrated for the Minimal Supersymmetric Standard Model
  (MSSM) and Randall-Sundrum Model using our new framework.
}

\begin{document}
\maketitle

  \section{Introduction}
  In the arena of modern particle physics Monte Carlo event 
  generators have become essential tools for analysing experimental data. They 
  are necessary in order to compare the behaviour of theoretical predictions under
  the conditions present within a collider experiment by giving a realisitic
  description of the final-state particles which interact with the detector 
  including any experimental cuts. It is essential that these generators 
  reproduce Standard Model physics as accurately as possible since these processes
  will provide a background to any new physics signals that might be present
  at future colliders as well as being of interest in their own right.  New physics 
  models will also need to be incorporated
  into a Monte Carlo simulation in order for their implications to be fully 
  understood.

  There are a wide variety of new physics models
  and while one could implement each model independently in its own event 
  generator it is more efficient to have a general purpose event 
  generator that can 
  handle a variety of these models but can also offer the full event simulation
  framework, \ie hard process, decay, parton shower and hadronization. This is the 
  approach that will be described in this paper using the 
  \Hw\!\!~\cite{Gieseke:2003hm,Gieseke:2006ga} event generator. 

  \Hw is a new event generator, written in C++, based on the well tested 
  HERWIG \cite{Corcella:2002jc,Corcella:2000bw,Moretti:2002eu} program.
  It is not simply a translation of the old FORTRAN code in to C++, it includes
  significant improvements to both the physics models and simulation framework.
  The object oriented
  aspect of the C++ language will allow future additions and modifications to
  be incorporated more easily. One area where improvements are needed
  is in the simulation of Beyond Standard Model (BSM) physics. In the past 
  each model was hard coded in the generator making the 
  addition of new models a time consuming process. We wish to minimise the 
  effort required in order to add new physics models to \Hw\!\!.
  Our approach is to factorize the problem in to smaller pieces and reuse
  as much information that has already been calculated as possible.

  For example in gluino production our method would first calculate the 
  $2\ra 2$ production matrix element then choose a decay mode for the 
  gluino based on the branching ratios and generate the decay products. 
  In addition, since the gluino is a coloured object, there
  will be QCD radiation which is simulated more easily in our factorized
  approach as it is simply another step between the production and decay of
  the particle. While factorizing the problem in this manner makes many 
  things easier it does 
  introduce complications when considering spin correlations. Additional
  information must be passed between each step to ensure that the final decay 
  products are correctly distributed including correlations between the production
  and various decays.
  
  Other packages, such as MadGraph~\cite{Maltoni:2002qb}, 
  CompHEP~\cite{Pukhov:1999gg},
  Sherpa~\cite{Gleisberg:2003xi} and Omega~\cite{Moretti:2001zz} 
  with WHiZard~\cite{Kilian:2001qz}
  exist which are capable of producing a 
  wide class of BSM physics processes\footnote{A general list of programs for BSM
    physics can be found at http://www.ippp.dur.ac.uk/montecarlo/BSM/} but they have 
  limitations. The main problem is the efficiency with which the  
  variety of possible processes can be generated. The  above programs all
  treat the processes as $2\ra n$ scattering which requires the exact 
  production and decay chain to be specified from the beginning. While this
  does mean that effects such as spin correlations are included automatically, 
  it limits the number of processes that can be generated in a reasonable
  amount of time. For example in order to study two different decay modes 
  of the gluino, with the above generators, one would have to calculate the
  production step twice whereas our method would simply be able to pick
  another decay mode without recalculating the hard subprocess.

  To minimise the amount of work needed for every new model, our approach will 
  require only a set of Feynman rules, the 
  specification of any new non-Standard Model particles and their properties. 
  In addition there is
  a mechanism to read in parameters from a Les Houches~\cite{Skands:2003cj}\footnote{Currently only SLHA1 is supported.} file 
  for a supersymmetric (SUSY) model. The new physics models currently implemented 
  are the Minimal Supersymmetric Standard Model (MSSM) with CP, R-parity and
  flavour conservation and a Randall-Sundrum~\cite{Randall:1999ee} type 
  model with the lowest Kaluza-Klein excitation coupling to Standard Model matter.

  Section~\ref{sec:spin} introduces how spin correlations are dealt with since
  these will be important when dealing with heavy fermions and vector bosons.
  Some details on the technical structure of the code will be given in 
  section~\ref{sec:tech} along with some physics of the models implemented thus far.
  A comparison with the FORTRAN code for some physical distributions will
  be presented in section~\ref{sec:dist} to demonstrate the consistency of
  our approach.
  
  \section{Spin Correlations}\label{sec:spin}
  Many new physics models predict the existence of new particles that are 
  as yet undetected by experiment. Heavy spin-$\frac{1}{2}$, spin-1 
  and spin-2 particles will be produced which will decay to lighter
  states. Their non-zero spin gives correlations between the production
  and decay steps which must be taken into account in order for the final-state
  angular distributions to be correct. An algorithm for dealing with these 
  correlations 
  is demonstrated in~\cite{Richardson:2001df,Collins:1987cp,Knowles:1988hu}. 
  It will briefly be described below for the process 
  $e^+ e^- \ra t \bar{t}$ where the top quark subsequently decays, via a W-boson,
  to a b quark and a pair of light fermions.
    
  Initially the outgoing momenta are generated according to the usual cross-section
  integral,
  \beq\label{eqn:sigma}
  \frac{(2\pi)^4}{2s}\int \diff{t} \diff{\bar{t}} \mep{e^+}{e^-}{t}{\bar{t}} 
  \mepc{e^+}{e^-}{t}{\bar{t}}
  \eeq
  where $\mep{e^{+}}{e^{-}}{t}{\bar{t}}$ is the matrix element for the initial 
  hard process and $\lambda_{t,\bar{t}}$ are the helicity of the $t$ and $\bar{t}$
  respectively. One of the outgoing particles is then picked at random, 
  say the top and a spin density matrix calculated
  \beq\label{rho}
  \rho^{t}_{\hel{t}\hel{t}^{'}}=\frac{1}{N} 
  \me^{e^+ e^- \ra t \bar{t}}_{\hel{t}\hel{\bar{t}}}
  \me^{* e^+ e^- \ra t \bar{t}}_{\hel{t}^{'} \hel{\bar{t}}},
  \eeq
  with $N$ defined such that $\rm{Tr}\,\rho = 1$.
  
  The top is decayed and the momenta of the decay products distributed
  according to  
  \beq
  \frac{(2\pi)^4}{2m_t}\int \diff{b} \diff{W^+} \rho^{t}_{\hel{t}\hel{t}^{'}} 
  \me^{t\ra bW^+}_{\hel{t}\hel{W^+}} \me^{*t\ra bW^+}_{\hel{t}^{'}\hel{W^+}}.
  \eeq
  where the inclusion of the spin density matrix ensures the correct correlation
  between the top decay products and the beam.

  A spin density matrix for the $W^+$ is calculated because the $b$ is stable
  \beq
  \rho^{W^+}_{\hel{W^+}\hel{W^+}^{'}} = \frac{1}{N} 
  \rho^{t}_{\hel{t}\hel{t}^{'}} 
  \me^{t \ra bW^+}_{\hel{t}\hel{W^+}} 
  \me^{*t \ra bW^+}_{\hel{t}^{'}\hel{W^+}^{'}},
  \eeq
  and the $W$ decayed in the same manner as the top with the inclusion of the
  spin density matrix here ensuring the correct correlations between the $W$ decay
  products, the beam and the bottom quark.

  The decay products of the $W$ are stable fermions so the decay chain terminates
  here and a decay matrix for the $W$ 
  \beq
  D^{W^+}_{\hel{W^+}\hel{W^+}^{'}} = \frac{1}{N}
  \me^{t\ra bW^+}_{\hel{t}\hel{W^+}} 
  \me^{*t\ra bW^+}_{\hel{t}\hel{W^+}^{'}},
  \eeq
  is calculated. Moving back up the chain a decay matrix for the top quark is 
  calculated using the decay matrix of the $W$,
  \beq
  D^{t}_{\hel{t}\hel{t}^{'}} = \frac{1}{N} 
  \me^{t\ra bW^+}_{\hel{t}\hel{W^+}}
  \me^{*t\ra bW^+}_{\hel{t}^{'}\hel{W^+}^{'}}
  D^{W^+}_{\hel{W^+}\hel{W^+}^{'}}.
  \eeq
  Since the top came from the hard subprocess we must now deal with the 
  $\bar{t}$ in a similar manner but instead of using $\delta_{\lambda_i \lambda_i}$
  when calculating the initial spin density matrix, the decay matrix of the top
   is used and the $\bar{t}$ decay is generated accordingly. The
  density matrices pass information from one decay chain to the
  associated chain thereby preserving the correct correlations.

 \begin{figure*}
    \begin{center}
    \includegraphics[angle=90,width=0.25\textwidth]{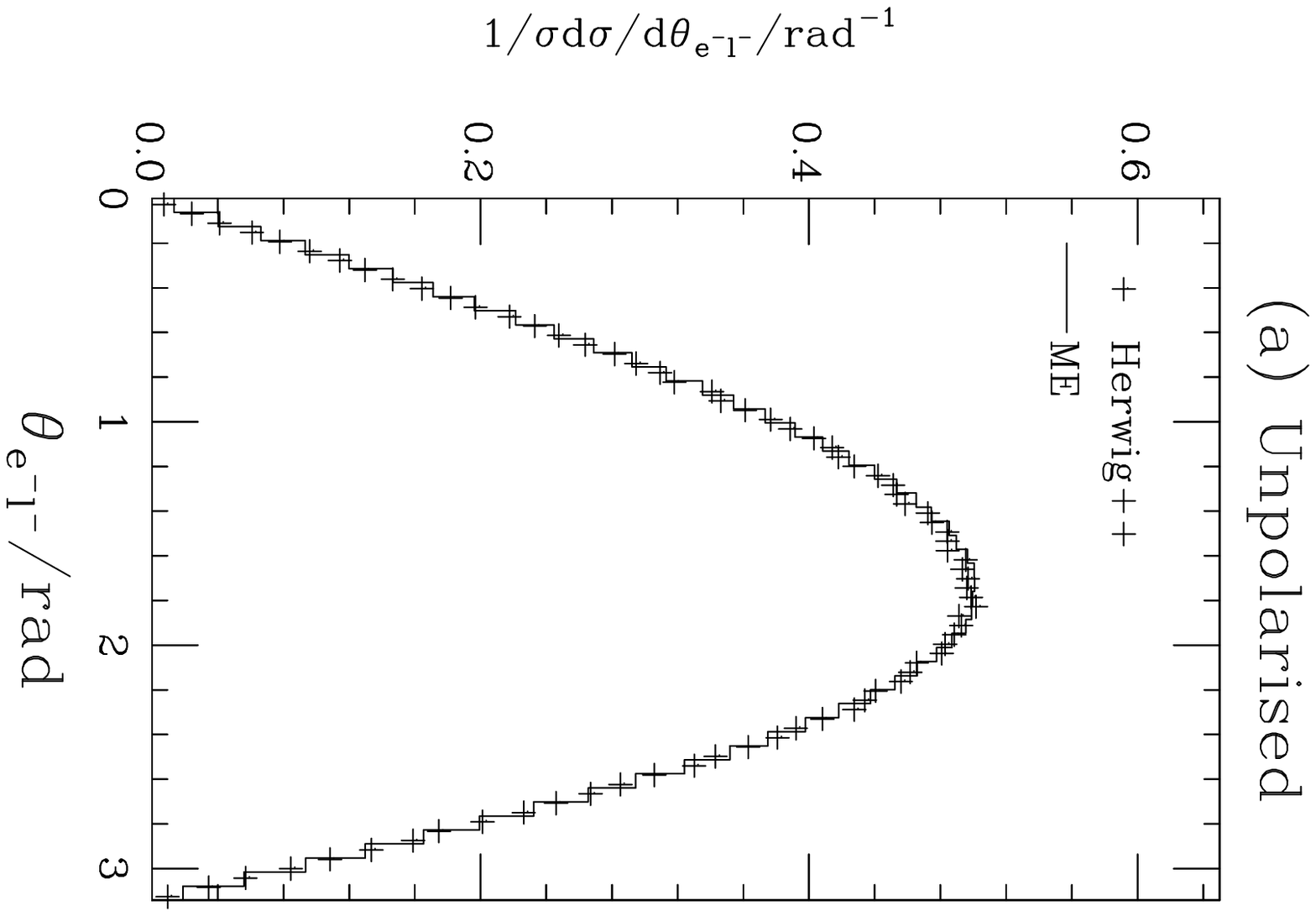}
    \includegraphics[angle=90,width=0.25\textwidth]{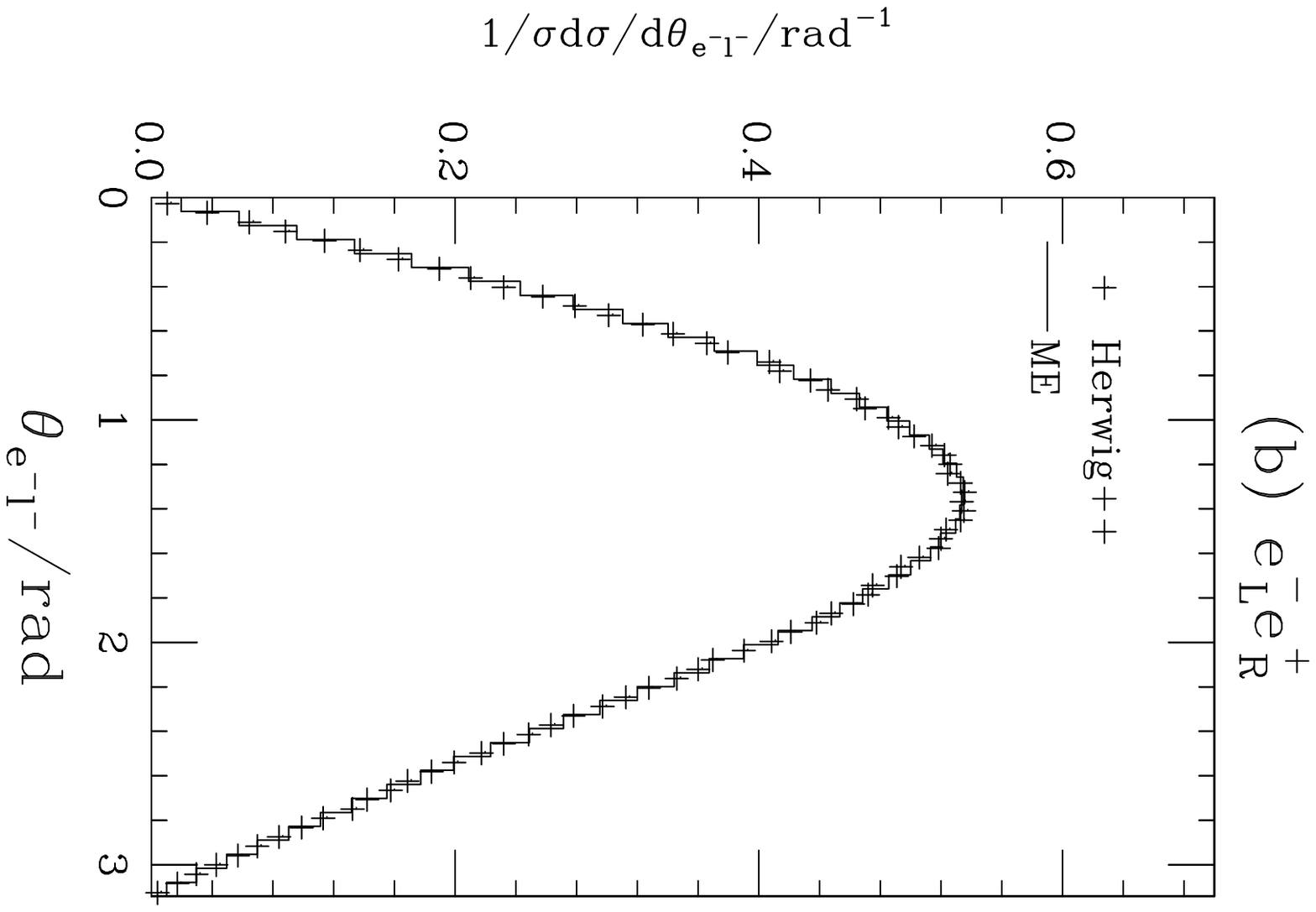}
    \includegraphics[angle=90,width=0.25\textwidth]{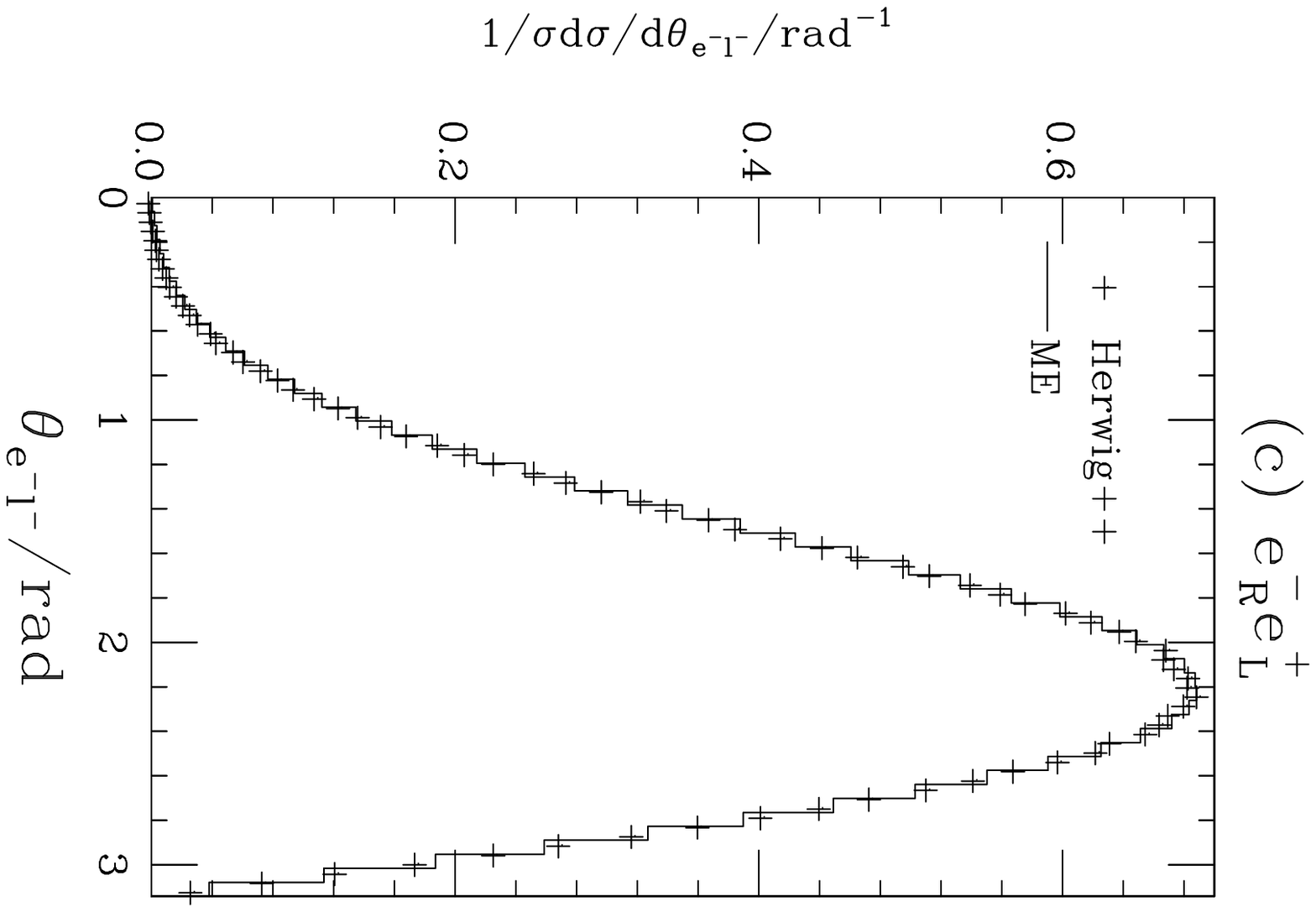}
    \caption{Angle between the beam and the outgoing lepton in $e^+e^-\ra 
      t\bar{t}\ra b\bar{b} l^{+}\nu_l l^{-}\bar{\nu_l} $ in the lab frame for
    a centre-of-mass energy of 500 GeV with (a) unpolarised incoming beams,
    (b) negatively polarised electrons and positively polarised positrons and
    (c) positively polarised electrons and negatively polarised electrons.}
    \label{fig:e-beam}
    \end{center}
  \end{figure*}

\begin{figure*}
   \begin{center}
    \includegraphics[angle=90,width=0.25\textwidth]{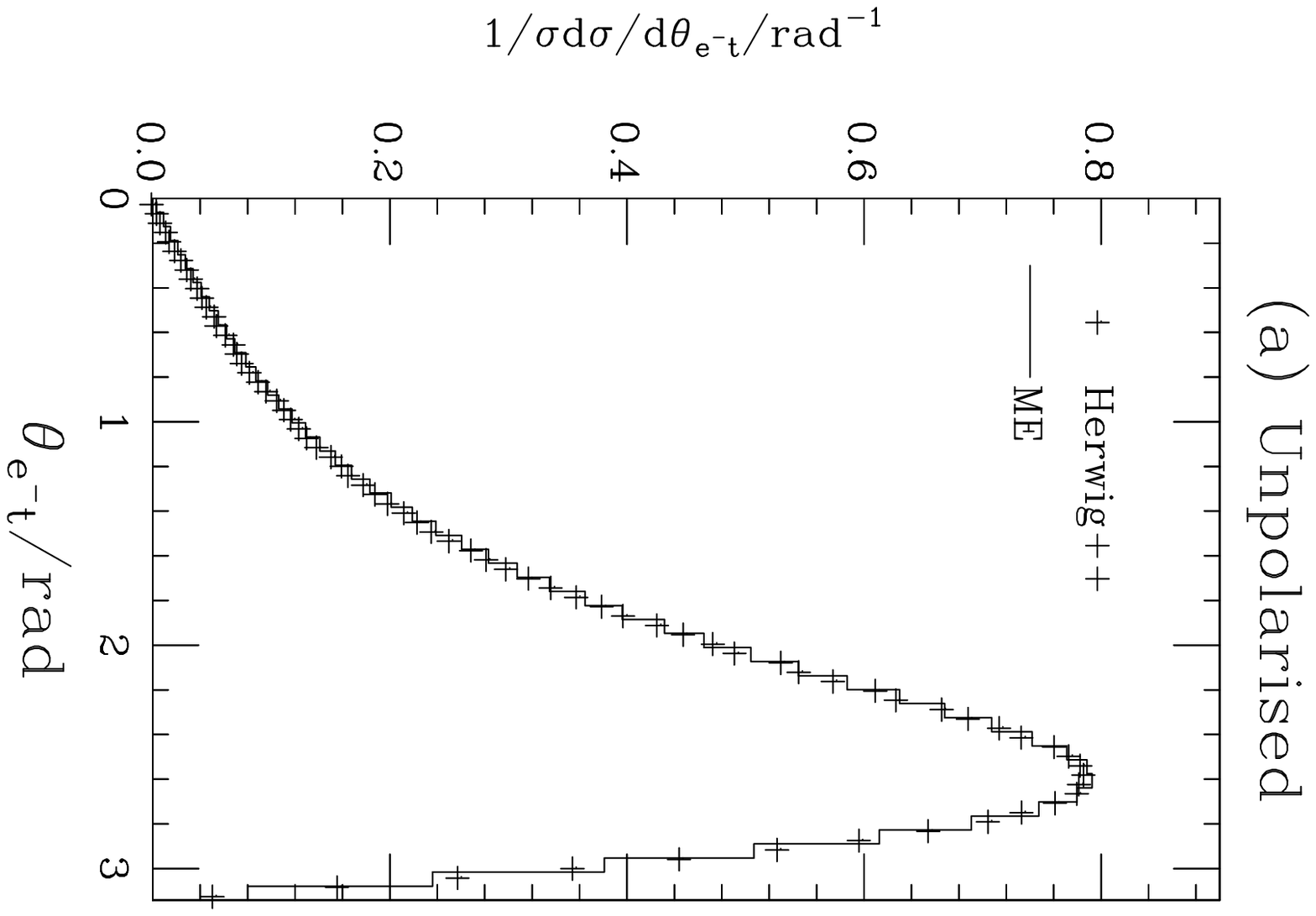}
    \includegraphics[angle=90,width=0.25\textwidth]{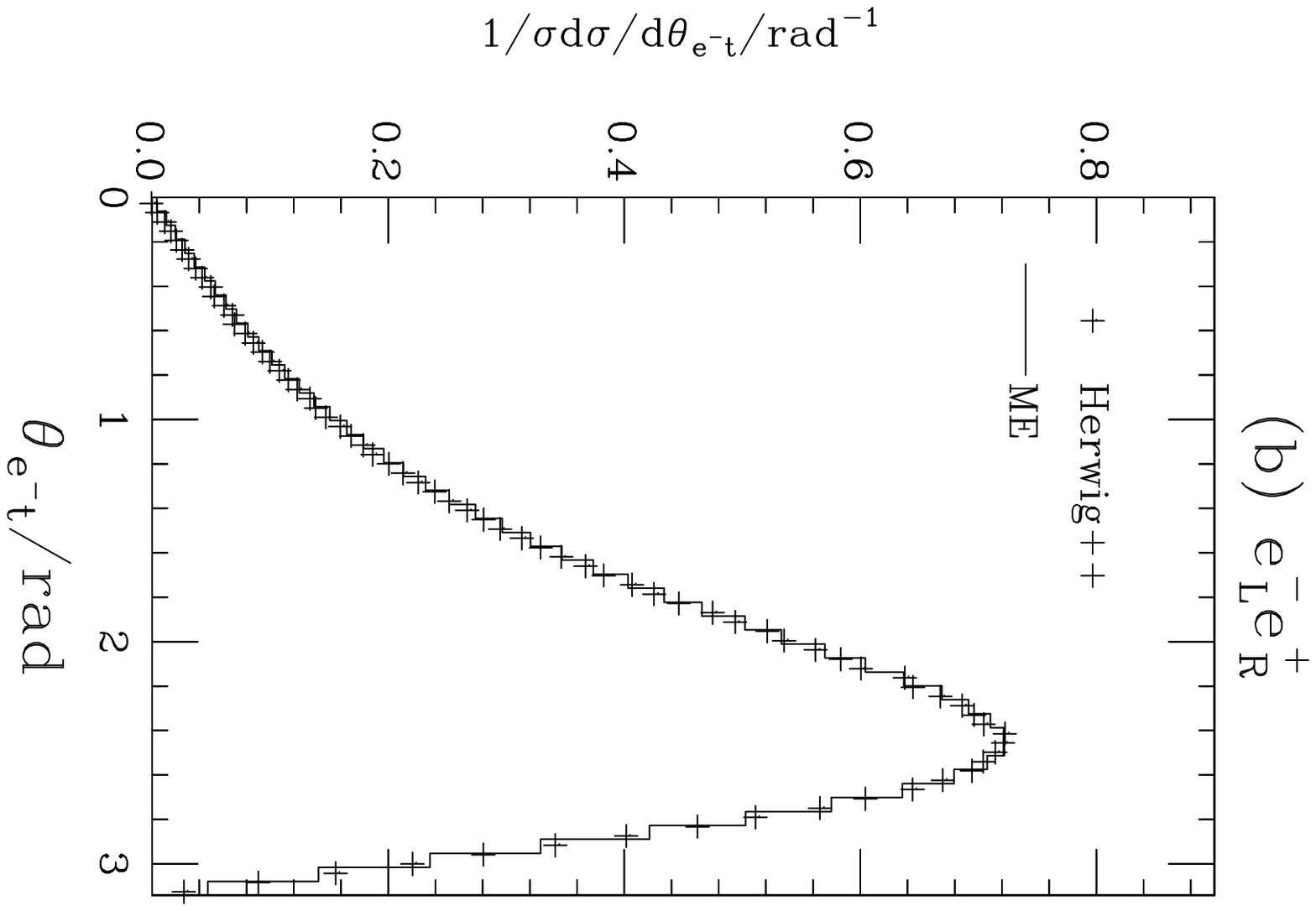}
    \includegraphics[angle=90,width=0.25\textwidth]{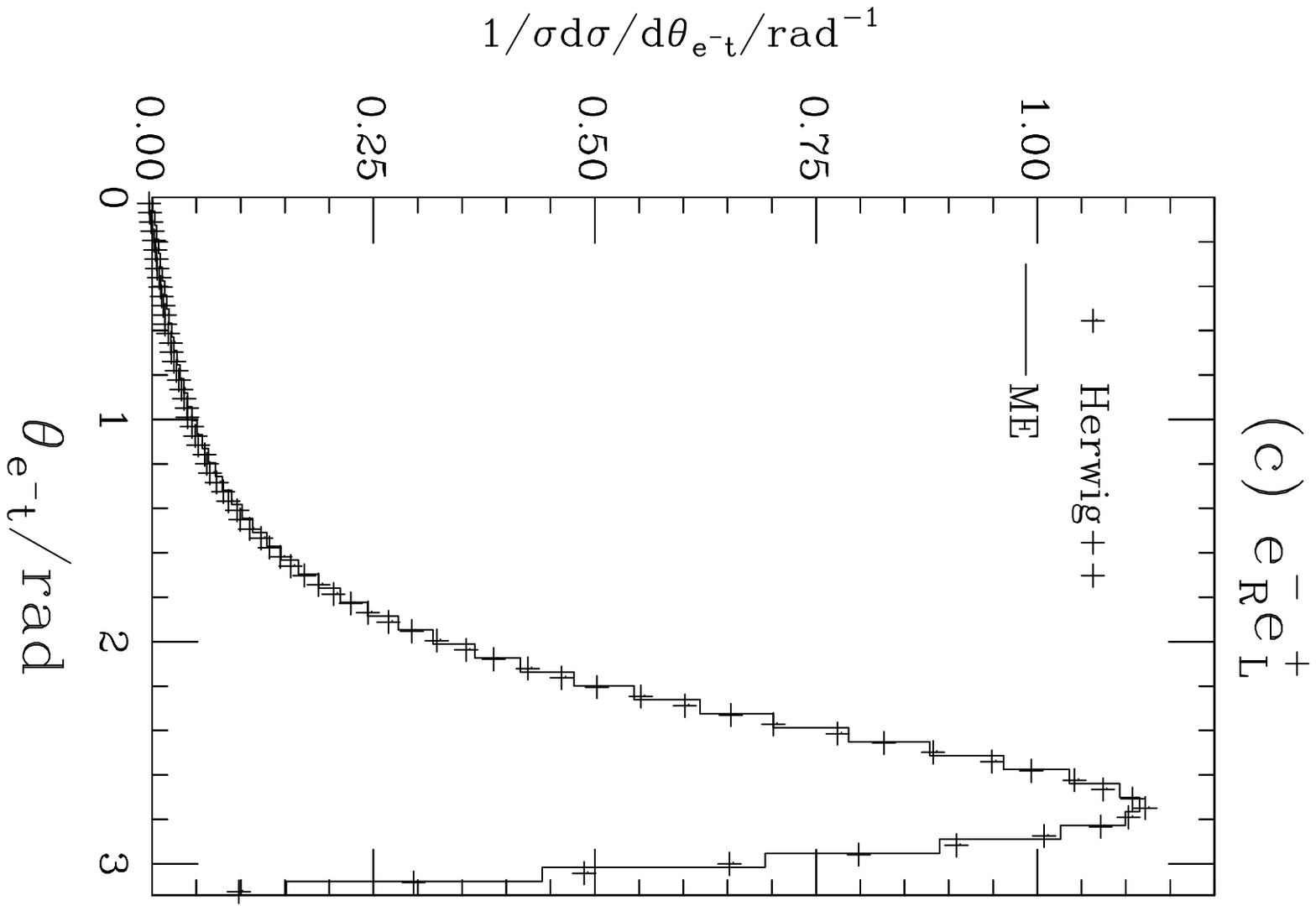}
    \caption{Angle between the lepton and the top quark in $e^+e^-\ra 
      t\bar{t}\ra b\bar{b} l^{+}\nu_l l^{-}\bar{\nu_l} $ in the lab frame for
    a centre-of-mass energy of 500 GeV with (a) unpolarised incoming beams,
    (b) negatively polarised electrons and positively polarised positrons and
    (c) positively polarised electrons and negatively polarised electrons.}
    \label{fig:e-top}
    \end{center}
\end{figure*}

\begin{figure*}
   \begin{center}
    \includegraphics[angle=90,width=0.25\textwidth]{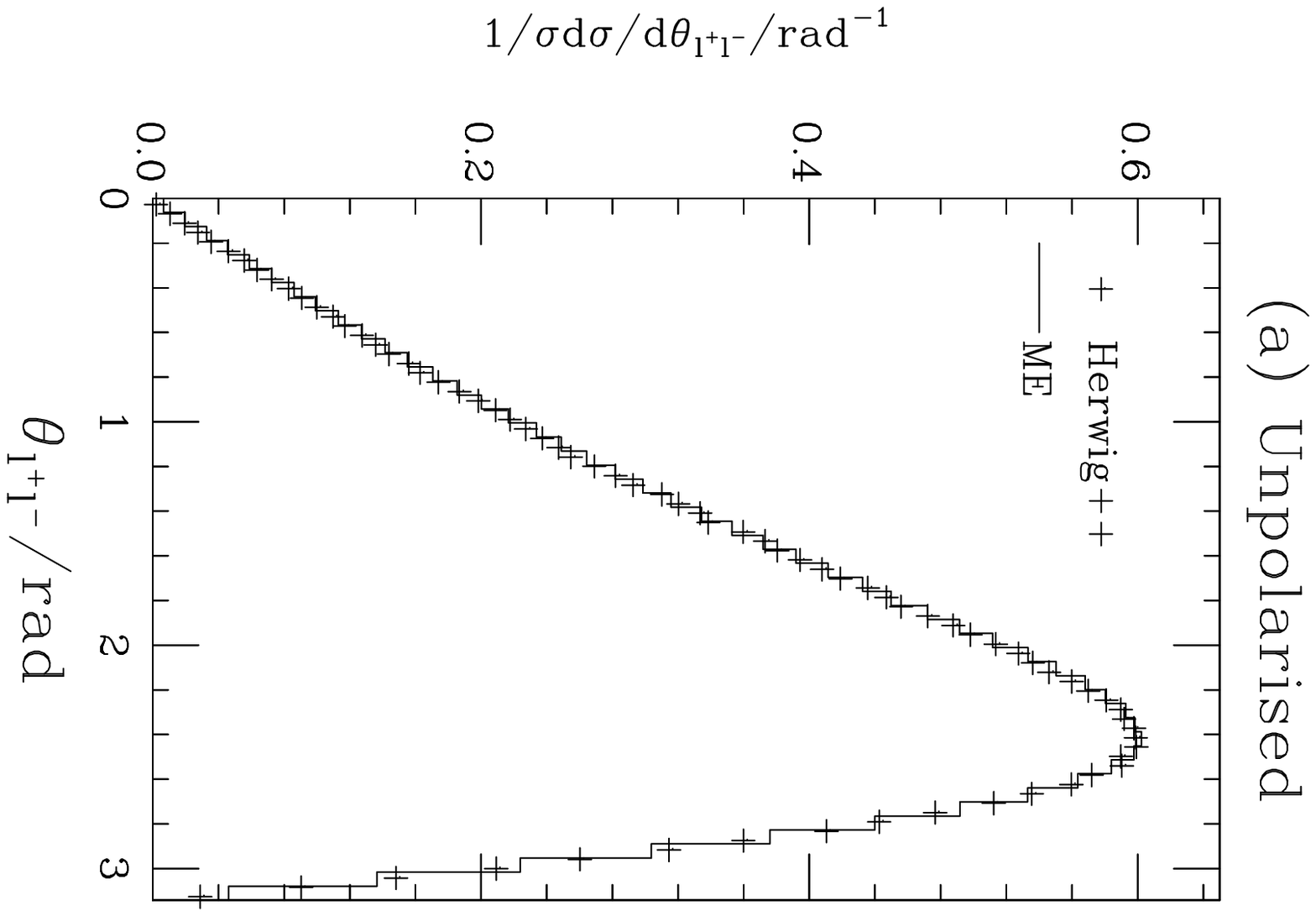}
    \includegraphics[angle=90,width=0.25\textwidth]{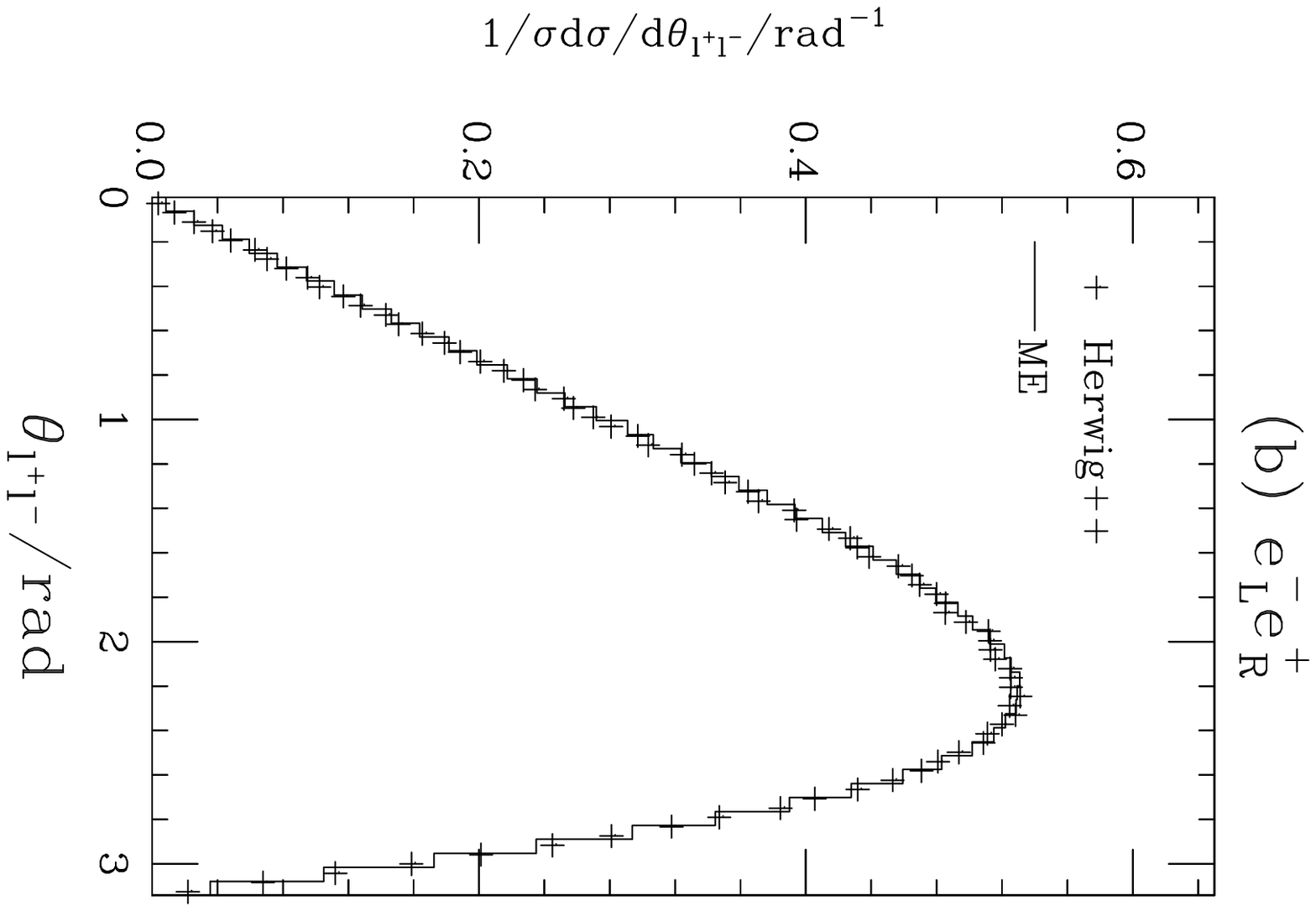}
    \includegraphics[angle=90,width=0.25\textwidth]{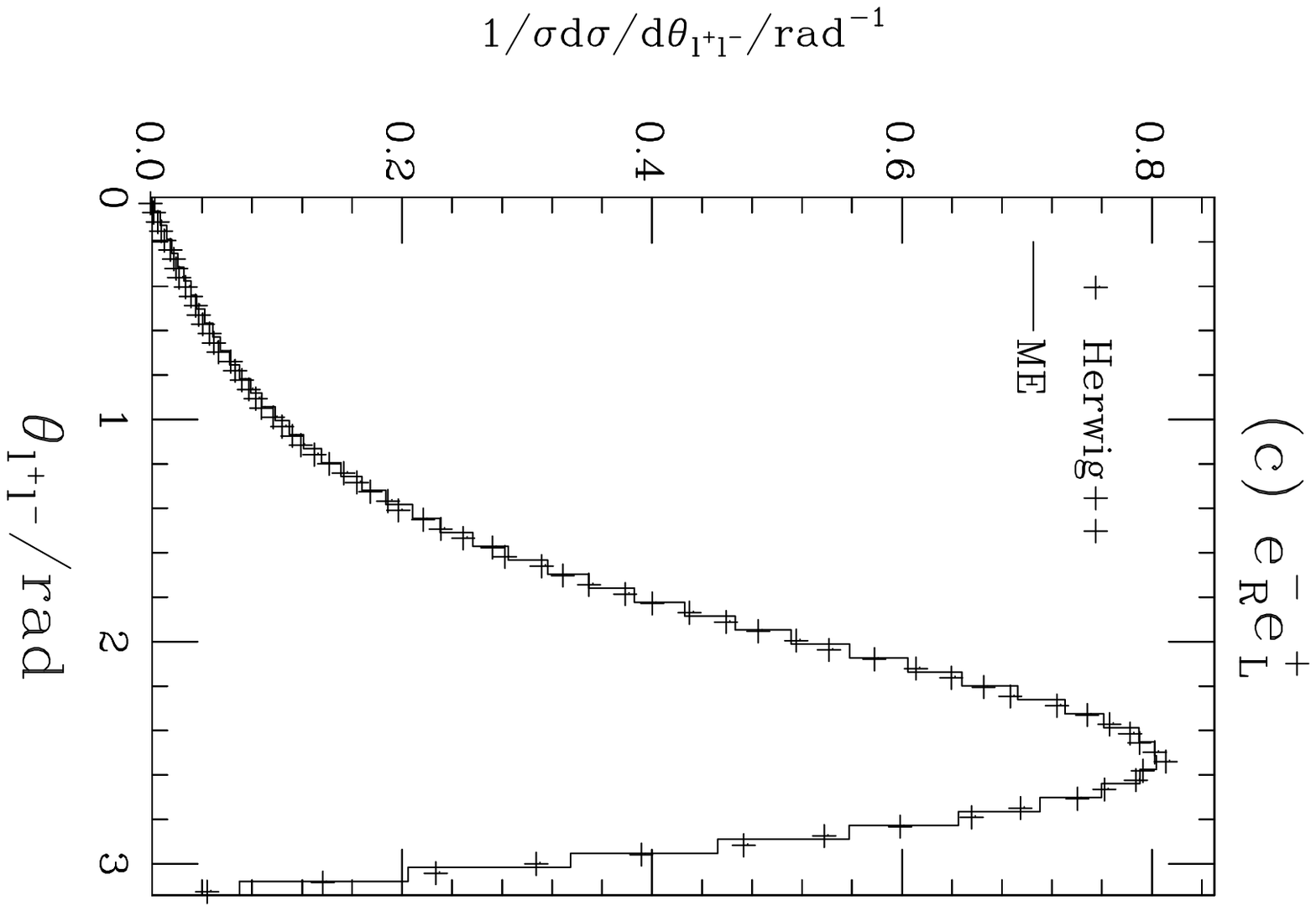}
    \caption{Angle between the outgoing lepton and anti-lepton in $e^+e^-\ra 
      t\bar{t}\ra b\bar{b} l^{+}\nu_l l^{-}\bar{\nu_l} $ in the lab frame for
    a centre-of-mass energy of 500 GeV with (a) unpolarised incoming beams,
    (b) negatively polarised electrons and positively polarised positrons and
    (c) positively polarised electrons and negatively polarised electrons.}
    \label{fig:e+e-}
    \end{center}
  \end{figure*}

  The production and decay of the top, using the spin correlation algorithm,
  is demonstrated in figures~\ref{fig:e-beam},~\ref{fig:e-top} and~\ref{fig:e+e-}.
  The hard subprocess and subsequent decays were generated using our new method.
  The results from the full matrix element
  calculation are also included  to show that the algorithm has
  been correctly implemented. The separate plots illustrate the different 
  stages of the algorithm at work. Figure~\ref{fig:e-beam} gives the angle 
  between the beam and the outgoing lepton.  The results from the simulation agree 
  well with the full matrix element calculation which demonstrates the
  consistency of the algorithm for the decay of the $\bar{t}$.

  Figure~\ref{fig:e-top} gives the angle between the top quark and the produced
  lepton. This shows the same agreement as the previous figure and demonstrates
  the correct implementation of the spin density matrix for the $\bar{t}$ decay.
  Finally figure~\ref{fig:e+e-} gives the results for the angle between the 
  final-state lepton/anti-lepton pair showing the correct implementation of the
  decay matrix that encodes the information about the $\bar{t}$ decay.
  Again there is good agreement between our
  numerical results and the full matrix element calculation giving us
  confidence about the implementation of the spin correlation algorithm.  
  
  The above procedure is well suited for implementation in an event generator, 
  as demonstrated, where one would like additional processes to occur
  between the hard production and decay such as showering of a coloured 
  particle. The algorithm as presented here is implemented in \Hw and will
  be used extensively during the simulation of many BSM physics~models.
  
  \section{Technical Details}\label{sec:tech}
  Instead of following the paradigm of implementing a specific model we have
  chosen a more generic approach to the problem,
  which is intended to be as model independent as possible.
  This will allow a wider variety of models to be implemented 
  within the event generator framework. The structure of the code relies heavily
  on the inheritance facilities available in the C++ language which
  allow independent structures to have a common heritage.

  Due to the existing structure it was sufficient only to consider generation
  of the hard subprocess and decay of the subsequent
  unstable particles since the showering and hadronizing are
  handled (almost) independently of the model details. Both the hard process
  and decay require a knowledge of the Feynman rules, couplings and masses within
  the model and these are currently implemented for the MSSM and Randall-Sundrum 
  model.

  In \Hw the Feynman rules are encoded in \textsf{Vertex} classes.
  They form part of the structure that enables the calculation of matrix
  elements using the \textsf{HELAS} formalism~\cite{Murayama:1992gi}. These classes
  provide the couplings to the particles defined within the model and so they
  must be provided with every new model which is 
  implemented. The way in which the vertices have been set up minimises this
  effort and will be described in the next section.

  \subsection{Vertices}
  For a given combination of spins interacting at a vertex, if we assume the
  perturbative form of the interaction, there is a specific Lorentz structure 
  and a limit on the number of possible couplings for any given interaction.
  This is carried into the implementation of the vertices 
  by defining a base class that holds all common functionality and 
  inheriting from this class to define a specific spin structure. The
  spin structure must then be further specialised into the exact vertex
  required by specifying the coupling and the particles that are able to
  interact at it.

\begin{figure}
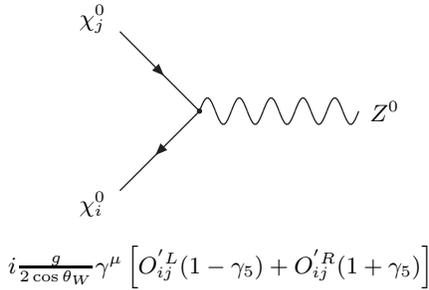

    \begin{center}
     \bpc(300,100)(25,-25)
	\ArrowLine(100,75)(130,45)
	\ArrowLine(130,45)(100,15)
	\Photon(130,45)(190,45){5}{5}
	\Vertex(130,45){1}
	\Text(85,80)[l]{$\chi_j^0$}
	\Text(85,10)[l]{$\chi_i^0$}
	\Text(195,45)[l]{$Z^0$}
	\Text(55,-14)[l]{ $i\frac{g}{2\cos\theta_W} \gamma^{\mu}
	  \left[O^{'L}_{ij}(1-\gamma_5) + O^{'R}_{ij}(1+\gamma_5)\right]$}
		
     \epc
     \caption{Feynman rule for the $\chi_i^0 \chi_j^0 Z^0$ vertex where
	the $O^{'}_{ij}$ are a combination of neutralino mass matrices.}
     \label{fig:nnzrule}
     \end{center}
  \end{figure}

  As an example, consider the $\chi_i^0 \chi_j^0 Z^0$ vertex in the MSSM as 
  shown in figure~\ref{fig:nnzrule}.

  A more general rule than that given in the figure is:
  \beq
  ic \gamma^{\mu}\left[\frac{a^L}{2}(1-\gamma_5) + \frac{a^R}{2}(1+\gamma_5)
    \right],
  \eeq
  where $c$ is the overall normalisation and $a^{\{L,R\}}$ are the left and right
  couplings respectively.
  We would choose $c=g/\cw$, $a^L=O^{'L}_{ij}$ and $a^R=O^{'R}_{ij}$.
  The overall coupling $c$ is stored in the base class and 
  the inherited class for the specific spin stores the left and right 
  couplings, since they may not always be required. Finally the 
  actual vertex class  implements a function to calculate the value of the 
  couplings.

  In addition to storing the couplings, the spin specific vertices have
  functions that can be used to either evaluate the vertex as a complex
  number or return an appropriate off-shell wave function. The ability 
  to calculate not just the entire vertex but off-shell components 
  underlies the \textsf{HELAS}~\cite{Murayama:1992gi} formalism for the 
  calculation of
  matrix elements. As an example consider decay of the top as in 
  section~\ref{sec:spin}.
  The \textsf{HELAS} approach factorizes the problem into two parts. 
  First a vector wave function for an off-shell 
  $W^+$ is calculated for a specified helicity of the top and bottom quarks. This
  is used as an input, along with the spinors for the light fermions, at the
  second vertex to calculate the final matrix element for that helicity 
  combination. To obtain the spin-summed
  matrix element the procedure is repeated for all possible helicities of the
  external particles.
  This dramatically reduces the amount of code required for numerical
  evaluation of the vertices. It is also has the additional benefit of providing
  basis states\footnote{In this case the spinors for the quarks 
    and light fermions.} for the particles that can be stored and
  passed between the production and decay to ensure the spin correlations are 
  consistently implemented. Our implementation and the formulation are described
  in more detail in appendix~\ref{app:HELAS}.
  
  \subsection{Decayers}
  The decay of the top described in section~\ref{sec:spin} 
  is handled by a class that is solely responsible for this decay.
  When just considering the Standard
  Model this is reasonable since the top is the only heavy fermion\footnote{Heavy in this context means that it decays before it hadronizes.} present.
  However, in the case of new physics models
  there will be a wealth of heavy particles which decay
  and creating a class for each decay would be inefficient. Instead we have
  implemented a set of classes for each set of possible external spin states  
  and  each of these classes, called 
  \textsf{Decayers}, is responsible for a specific external spin configuration
  in a decay. At the present time only two body decays have been considered\footnote{Any three body decays read from a decay table are handled with a 
  phase-space decayer and therefore will not include spin correlation 
  informatation.}.
  Appendix~\ref{app:decayers} contains a list of the currently implemented
  \textsf{Decayer} classes.

  The standard way in which \Hw handles particle decay modes is with a
  text file listing each mode, along with branching ratios and the object that 
  will handle the decay. We have instead taken
  the approach of constructing the decay modes and \textsf{Decayer} objects 
  automatically. All that is left to the user is to specify the particle(s)
  which will be decayed. The steps for creating the decay modes and
  \textsf{Decayer} objects are:
  \begin{enumerate}
    \item specify the particles for which decay modes are required;
    \item analyse each vertex to find whether the particle can interact, if it
      cannot skip to the next vertex in the list until one is found that
      is able to;
    \item find the decay products and test whether the decay would 
      be kinematically possible, if not skip to the next possible mode;
    \item if an object already exists that can handle the decay then assign
      it to handle the mode else create a new \textsf{Decayer} and assign this 
      to handle the decay.
  \end{enumerate}
  The created \textsf{Decayer} object contains the appropriate code for 
  calculating the matrix element for all possible helicity combinations which
  can be used in the spin correlation algorithm from section~\ref{sec:spin}.

  There is an additional point to consider when dealing with SUSY models.
  These models contain additional parameters, such as mixing matrices, that 
  are necessary for vertex calculations. A mechanism has been implemented 
  to read this information from a SUSY Les Houches Accord 
  file~\cite{Skands:2003cj}.
  In principle the file can contain decay modes along with branching ratios. 
  If this is the case then the decay mode is not created automatically it is 
  just assigned an appropriate \textsf{Decayer}.

  In order to be able to decay the particles we must first produce them along
  with their associated momenta.
  The next section will describe how this procedure is accomplished within
  the new framework.

  \subsection{Hard Processes}\label{sec:hardproc}
  In the hard process the initial momenta of the outgoing particles from
  a hard collision are calculated via a leading-order matrix element as in 
  equation~\ref{eqn:sigma}. There will be additional PDFs involved if the incoming
  particles are composite. In \Hw the mechanism for this is again ``factorized''
  into pieces concerned with the phase-space evaluation and pieces concerned with 
  the calculation of $\mesq$. Here we only concern ourselves with the
  $2\ra2$ cross sections and as a result the existing structure only requires
  us to implement calculations of $\mesq$ for the new processes.

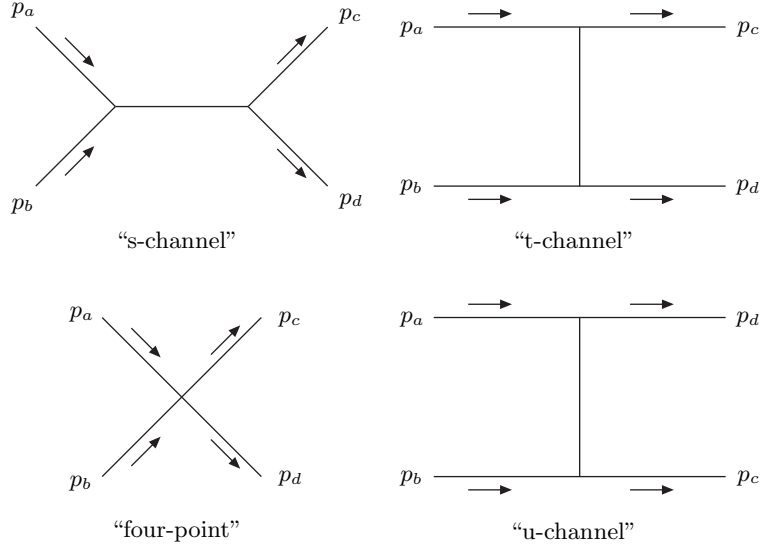
\begin{figure*}
    \begin{center}
      \begin{picture}(300,195)
	\Line(15,185)(45,155) \LongArrow(26,181)(36,171)
	\Text(6,192)[l]{$p_a$}
	\Line(45,155)(15,125) \LongArrow(26,129)(36, 139)
	\Text(6,118)[l]{$p_b$}
	\Line(45,155)(95,155) 
	\Line(95,155)(125,185) \LongArrow(106,171)(116,181)
	\Text(130,190)[l]{$p_c$}
	\Line(95,155)(125,125) \LongArrow(106,139)(116,129)
	\Text(130,120)[l]{$p_d$}
	\Text(70,105)[]{``s-channel'' }
	\SetOffset(25,0)
	\Line(140,185)(250,185) 
	\LongArrow(153,190)(168,190) \LongArrow(214,190)(229,190)
	\Text(128,185)[l]{$p_a$} \Text(264,185)[r]{$p_c$}
	\Line(195,185)(195,125) 
	\Line(140,125)(250,125) \Text(128,125)[l]{$p_b$} \Text(264,125)[r]{$p_d$} 
	\LongArrow(153,120)(168,120) \LongArrow(214,120)(229,120)
	\Text(195,105)[]{``t-channel'' }
	\SetOffset(25,-110)
	\Line(140,185)(250,185) 
	\LongArrow(153,190)(168,190) \LongArrow(214,190)(229,190)
	\Text(128,185)[l]{$p_a$} \Text(264,185)[r]{$p_d$}
	\Line(195,185)(195,125) 
	\Line(140,125)(250,125) \Text(128,125)[l]{$p_b$} \Text(264,125)[r]{$p_c$} 
	\LongArrow(153,120)(168,120) \LongArrow(214,120)(229,120)
	\Text(195,105)[]{``u-channel'' }
	\SetOffset(0,-110)
	\Line(40,185)(70,155) \LongArrow(51,181)(61,171)
	\Text(28,187)[l]{$p_a$}
	\Line(70,155)(40,125) \LongArrow(51,129)(61,139)
	\Text(28,123)[l]{$p_b$}
	\Line(70,155)(100,185) \LongArrow(81,171)(91,181)
	\Text(107,185)[l]{$p_c$}
	\Line(70,155)(100,125) \LongArrow(81,139)(91,129)
	\Text(107,125)[l]{$p_d$} 
	\Text(70,105)[]{``four-point'' }
      \end{picture}
    \end{center}
    \caption{Tree-level topologies for a $2\ra2$ process. The arrows denote 
    the flow of momenta.}
    \label{fig:topol}
\end{figure*}

  Figure~\ref{fig:topol} shows the possible topologies for a $2\ra2$ process
  at tree level. Obviously some processes will not involve all topologies.
  As in the case for the \textsf{Decayers} our approach is not to create 
  a class for each possible process but instead create classes based on the 
  external spins of the particles involved. The user simply specifies the 
  incoming states and an outgoing particle, all of the possible diagrams 
  with this outgoing particle are then created along with the appropriate 
  \textsf{MatrixElement} object. The 
  object is responsible for calculating the spin-averaged matrix 
  element. In the \textsf{HELAS} approach the matrix element is first calculated by 
  computing the complex amplitude for each diagram of a given helicity combination.
  The diagram 
  contributions are then summed and the modulus squared taken. This is
  done for each helicity and the sum of each helicity gives the spin-summed 
  $\mesq$. In the case of strong processes it is easier to separate the colour
  structure from the evaluation of the matrix element. In a given process the 
  diagrams can be split into ``colour flows'', which are a combination of
  diagrams with the same colour structure, reducing the amount of computation
  required.

\begin{figure}
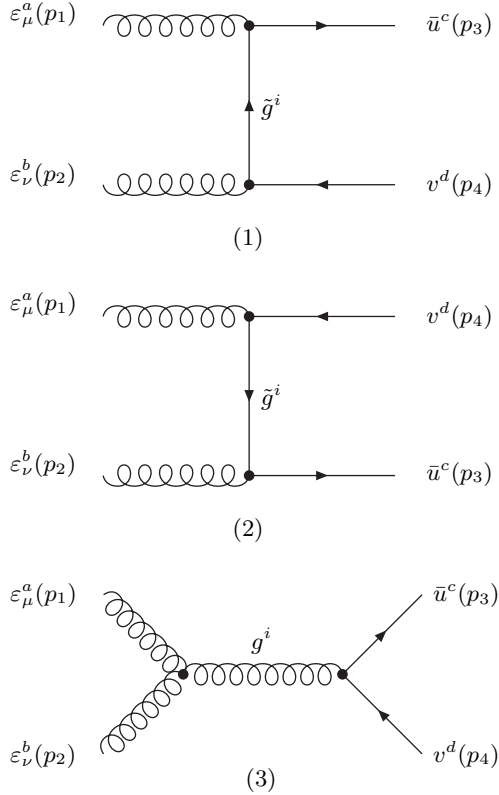

    \begin{center}
      \bpc(200,300)
      \SetOffset(0,110)
	\Gluon(35,185)(90,185){4}{6} 
	\Text(0,190)[l]{$\varepsilon_{\mu}^a (p_1)$} 
	\ArrowLine(90,185)(145,185) \Text(170,186)[]{$\bar{u}^c(p_3)$}
	\Gluon(35,125)(90,125){-4}{6} 
	\Text(0,130)[l]{$\varepsilon_{\nu}^b (p_2)$} 
	\ArrowLine(90,125)(90,185) \Text(95,155)[l]{$\tilde{g}^i$}
	\ArrowLine(145,125)(90,125) \Text(170,126)[]{$v^d(p_4)$}
	\Vertex(90,185){2} \Vertex(90,125){2}
	\Text(90,105)[]{(1)}
	\SetOffset(0,0)
	\Gluon(35,185)(90,185){4}{6} 
	\Text(0,190)[l]{$\varepsilon_{\mu}^a (p_1)$} 
	\ArrowLine(145,185)(90,185) \Text(170,186)[]{$v^d(p_4)$}
	\Gluon(35,125)(90,125){-4}{6} 
	\Text(0,130)[l]{$\varepsilon_{\nu}^b (p_2)$} 
	\ArrowLine(90,185)(90,125) \Text(95,155)[l]{$\tilde{g}^i$}
	\ArrowLine(90,125)(145,125) \Text(170,126)[]{$\bar{u}^c(p_3)$}
	\Vertex(90,185){2} \Vertex(90,125){2}
	\Text(90,105)[]{(2)}
	\SetOffset(-55,-10)
	\Gluon(90,90)(120,60){4}{6} 
	\Text(55,90)[l]{$\varepsilon_{\mu}^a (p_1)$}
	\Gluon(90,30)(120,60){4}{6}
	\Text(55,30)[l]{$\varepsilon_{\nu}^b (p_2)$}
	\Gluon(120,60)(180,60){4}{8} \Text(150,73)[]{$g^i$}
	\ArrowLine(180,60)(210,90) \Text(215,90)[l]{$\bar{u}^c(p_3)$}
	\ArrowLine(210,30)(180,60) \Text(215,30)[l]{$v^d(p_4)$}
	\Vertex(120,60){2} \Vertex(180,60){2}
	\Text(150,20)[]{(3)}
      \epc
    \end{center}
    \caption{Diagrams contributing to the process $gg\ra\tilde{g}\tilde{g}$ 
    where lowered Greek letters denote space-time indices and raised
    Roman letters denote colour indices in the adjoint representation. $u,v$
    are the spinors for the gluinos and $\varepsilon_{\mu,\nu}$ are the
    polarisation vectors of the gluons. The momenta $(p_1, p_2)$ are incoming 
    and $(p_3,p_4)$ are outgoing.}
    \label{fig:cflows}
  \end{figure}

  To demonstrate this procedure consider the process $gg\ra\tilde{g}\tilde{g}$.
  The diagrams that contribute are shown in figure~\ref{fig:cflows}. If the 
  amplitude, stripped of the colour information, for the $i$th diagram is 
  denoted by $\me_i$ the full amplitude for the 3 diagrams is given by,
 \begin{subequations}
  \begin{eqnarray}
    g_1 & = &\phantom{-} if^{aci} if^{bid} \me_1, \label{eqn:ggf1}\\
    g_2 & = &\phantom{-} if^{adi} if^{bic} \me_2, \label{eqn:ggf2}\\
    g_3 & = & -if^{aib} if^{icd} \me_3, \label{eqn:ggf3}
  \end{eqnarray}
  \end{subequations}
  where $g_i$ denotes the full amplitude and $f^{abc}$ denotes the 
  anti-symmetric structure constants. 

  The factors of $i$ associated with 
  the structure constants are present because of the way the vertex
  rules are defined within the code. The vertices in \Hw are stripped of their 
  colour information and therefore require the extra factors of $i$ to be 
  included, where appropriate, with the colour matrices to give the  correct sign.

  The combination of structure constants in equation~\ref{eqn:ggf3} can be 
  rewritten using the Jacobi identity to give
  \beq\label{eqn:ggf3j}
  g_3 = \left(f^{aci}f^{bid} - f^{bci}f^{aid}\right)\me_3,
  \eeq
  making it apparent that the colour structure of the $s$-channel gluon
  exchange diagram is simply a combination of the other two colour
  structures. The full colour amplitude can therefore be written as
  \beq\label{eqn:fullcolamp}
  \me_T = -\left[c_1(\me_1 - \me_3) + c_2(\me_2 + \me_3)\right],
  \eeq
  where $c_i$ denotes the combination of structure constants from above and 
  the combination of diagram amplitudes will be known as ``colour flows'' 
  denoted by $f_i$.
  Note that the overall minus sign can be dropped 
  since it simply corresponds to a phase that will not contribute to the final
  answer. In order to calculate $\mesq$ the constants $c_i$ need to be squared. 
  For the process being considered 
  $|c_1|^2=|c_2|^2=N_c^2(N_c^2-1)$ and $c_1c^*_2=c_2c^*_1=N_c^2(N_c^2-1)/2$ where
  $N_c$ is the number of colours.
  The spin-summed matrix element, averaged over initial colours and 
  polarisations, is then
  \beq\label{eqn:fullme2}
  |\mesqbar|^2=\frac{1}{2}\frac{1}{4}\frac{1}{(N_c^2-1)^2}\sum_{\lambda}
  C_{ij}f_{i}^{\lambda}f_{j}^{* \lambda}
  \eeq
  where $C_{ij}$ is a matrix containing the squared colour factors 
  and $f^{\lambda}_i$ is the $i$th colour flow for the set 
  of helicities $\lambda$. 

  As well as calculating $\mesq$ for a given process each \textsf{MatrixElement}
  object is also responsible for setting up the colour structure of the hard 
  process which is required to generate the subsequent QCD radiation and 
  hadronization. Depending on the colours of the internal and external 
  states involved there may be more than one possible colour structure for 
  each diagram. For the example in figure~\ref{fig:cflows}
  each diagram has 4 possible colour topologies since the both the gluon and
  gluino carry a colour and an anti-colour line in the large $N_c$ limit.
  When an event is generated, if this case presents itself, a colour structure 
  is picked at random from the $N_t$ possiblities. While this does rely on
  using the large $N_c$ limit we believe it to be an good approximation for
  the colour structures that we are dealing with.

  The possible Majorana nature of external states also gives rise to
  complications when calculating the matrix element. If the incoming states are
  a spinor and a barred-spinor then in the case where a $u$-channel diagram 
  is required two additional spinors must be calculated. The reason
  for this is that, using the notation of figure~\ref{fig:topol}, 
  when $c$ and $d$ are crossed their fermion flow can no longer be reversed 
  since the initial fermions
  set the direction in which these arrows point. The two additional spinors
  required are a spinor for the original outgoing barred state and a 
  barred spinor for the original outgoing spinor state with care being taken
  to associate the new spinors with the correct helicity.
  
  Appendix~\ref{app:matrix} contains a list of the currently implemented
  \textsf{MatrixElement} classes\footnote{This includes all of the spin combinations needed in the MSSM and RS model. Additional cases will be implemented when required
    for a new type of model.}. 
  The example given above demonstrates a 
  possible SUSY gaugino 
  production process that is taken into account with our new mechanism.
  We have also implemented another mechanism for the simulation of resonances.
  This will be described now using an example from the Randall-Sundrum model.

  \subsection{Resonant Processes}
\begin{figure}
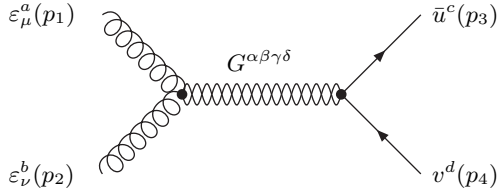

    \begin{center}
  \bpc(100,70)(90,25)
  \Gluon(90,90)(120,60){4}{6} 
  \Text(55,90)[l]{$\varepsilon_{\mu}^a (p_1)$}
  \Gluon(90,30)(120,60){4}{6}
  \Text(55,30)[l]{$\varepsilon_{\nu}^b (p_2)$}
  \Photon(120,60)(180,60){4}{8} 
  \Photon(120,60)(180,60){-4}{8} 
  \Text(150,73)[]{$G^{\alpha\beta\gamma\delta}$}
  \ArrowLine(180,60)(210,90) \Text(215,90)[l]{$\bar{u}^c(p_3)$}
  \ArrowLine(210,30)(180,60) \Text(215,30)[l]{$v^d(p_4)$}
  \Vertex(120,60){2} \Vertex(180,60){2}
  \epc
  \end{center}
  \caption{Resonant graviton exchange from gluon fusion to produce 2 
    fermions.}
    \label{fig:gravEx}
  \end{figure}

  Often we are interested in the study of $s$-channel resonances which decay to 
  Standard Model particles rather than the production of a new particle in a
  $2\ra2$ process. We therefore include a mechanism to study this type of 
  proccess. We will take as an example here the virtual 
  exchange of a graviton, the lowest lying state of a Kaluza-Klein tower.
  The graviton is predicted by various models with extra dimensions where gravity 
  is allowed to propagate in the bulk, an example process is shown in 
  figure~\ref{fig:gravEx}.
  The same matrix element classes which are used to calculate the hard 
  processes are used to calculate the resonant processes. Now, however, there is
  less computation since each \textsf{MatrixElement} will contain only a 
  single $s$-channel diagram and hence a single colour flow.

  The next section will detail some physical distributions for the two models
  discussed previously.

  \section{Results}\label{sec:dist}
  The following sections will show some distributions produced using the new
  BSM code in \Hw\!\!. As there are currently only two models implemented these
  will form the basis of the distributions considered.

  \subsection{Graviton Resonances}
\begin{figure}
  \begin{center}
    \includegraphics[angle=90,clip,width=0.25\textwidth]{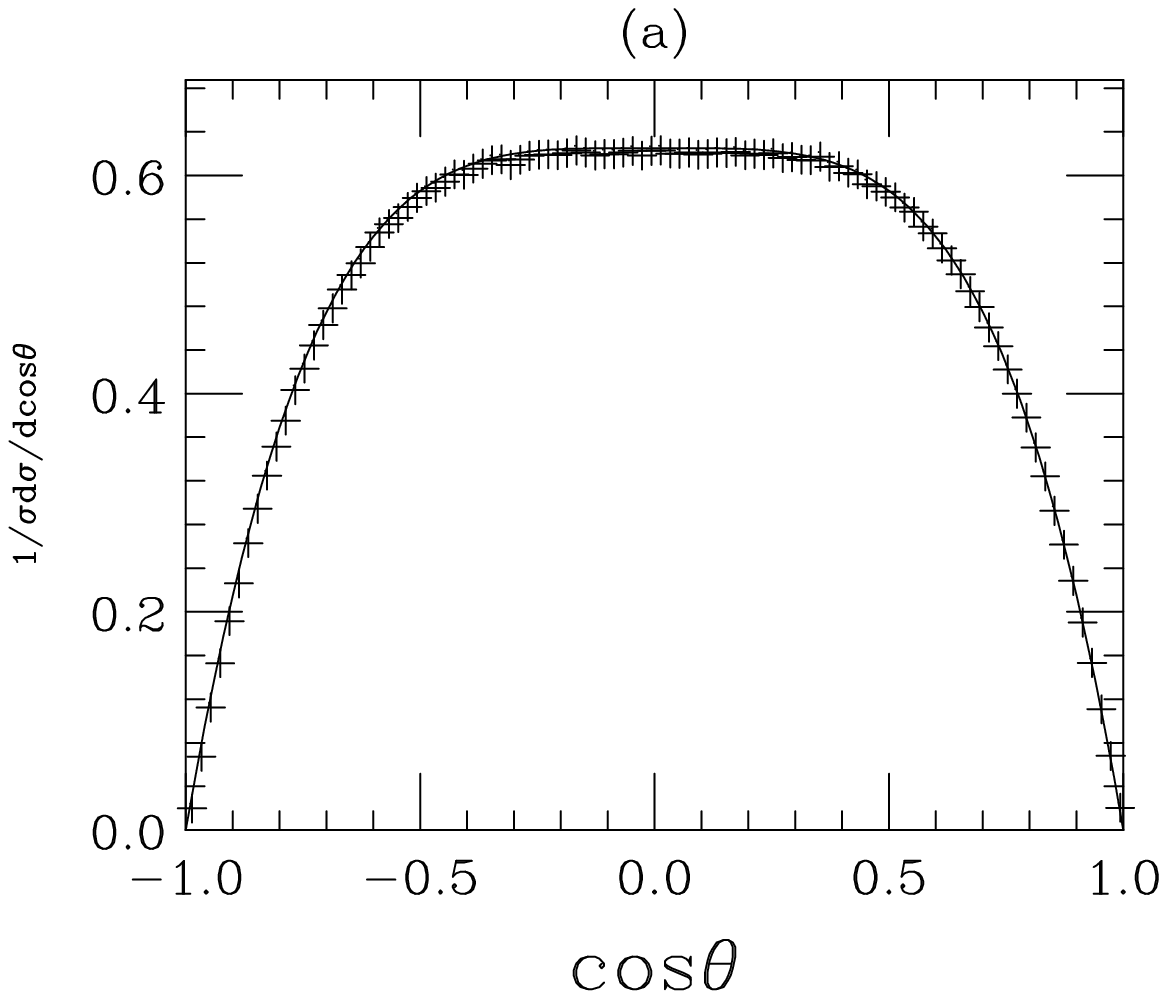}
    \includegraphics[angle=90,clip,width=0.25\textwidth]{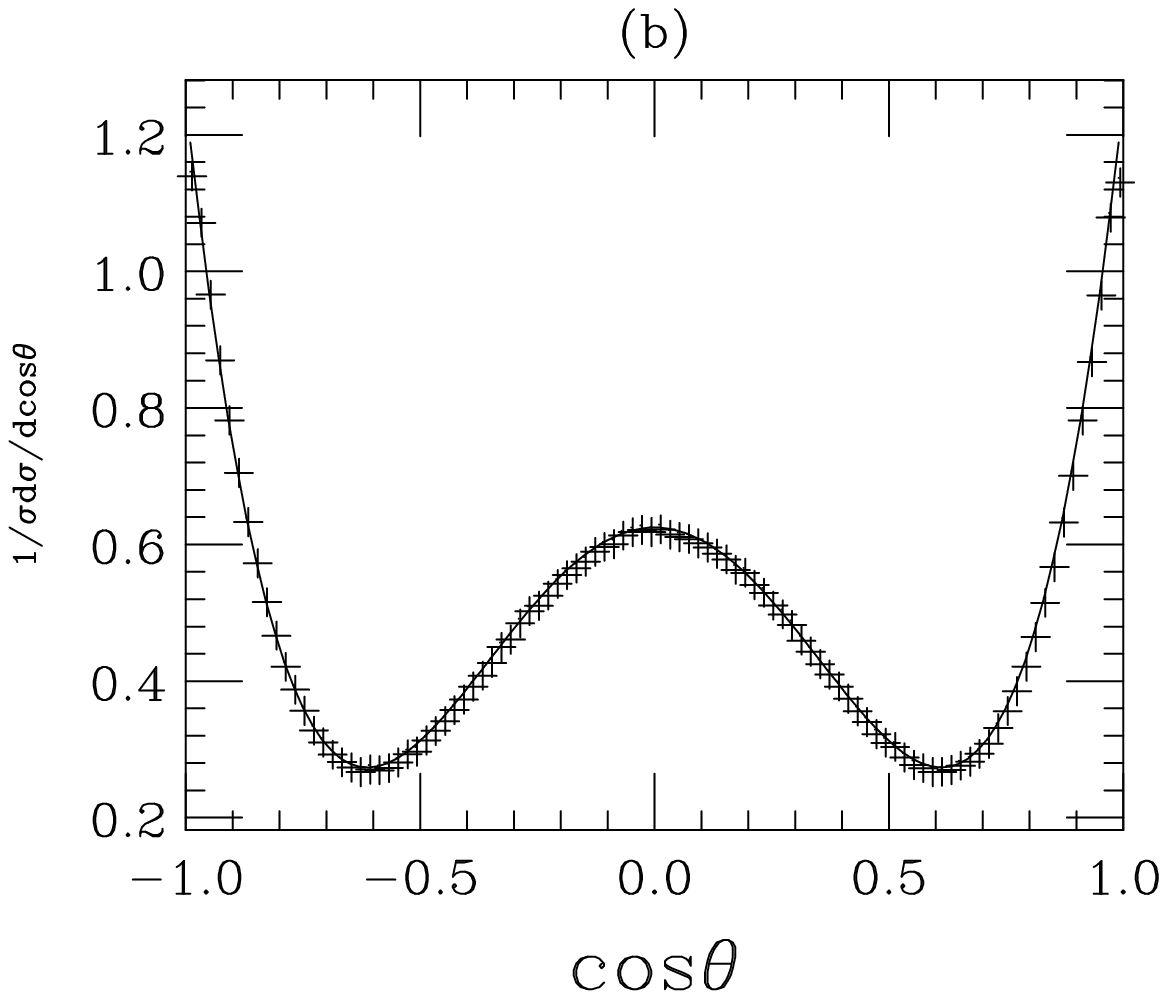}
    \includegraphics[angle=90,clip,width=0.25\textwidth]{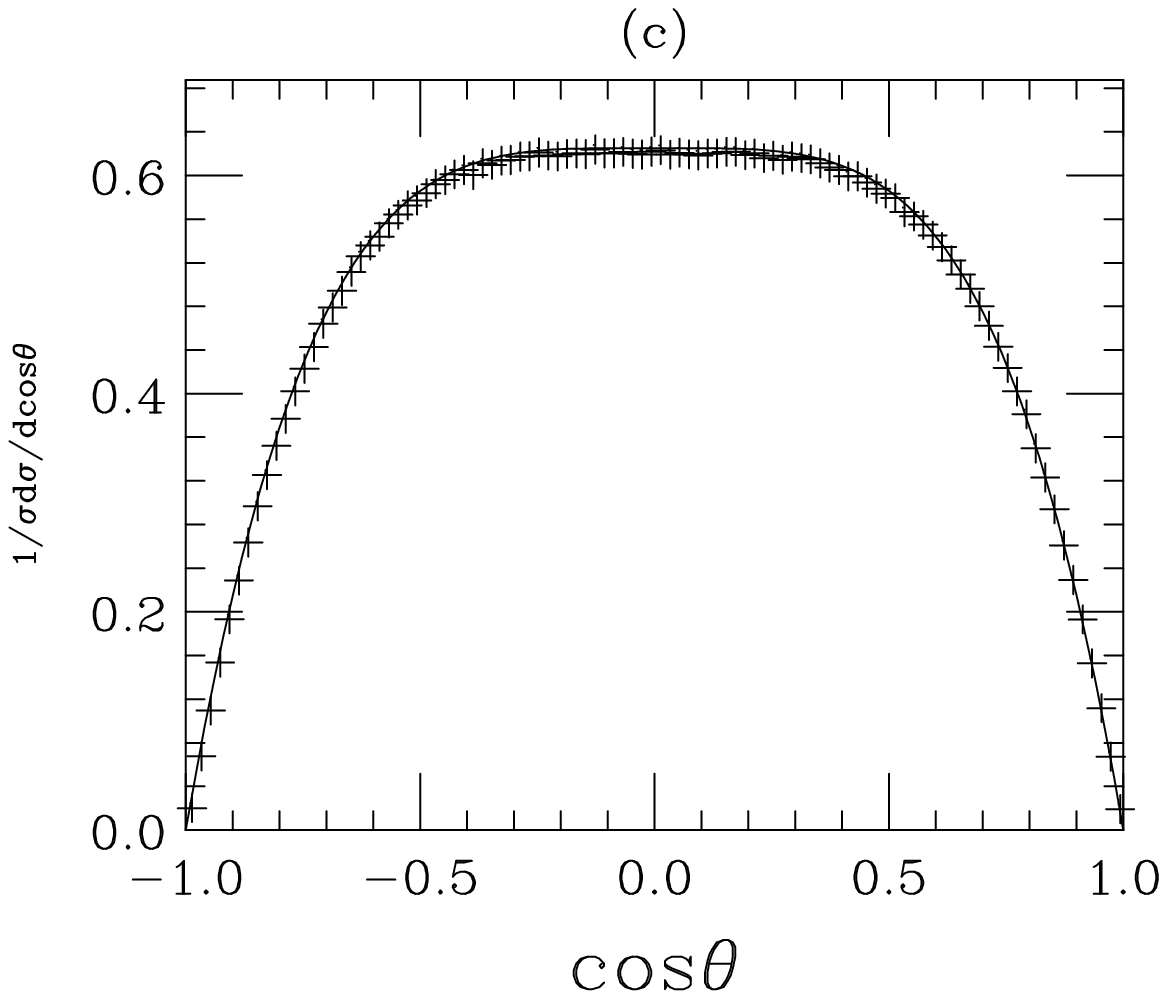}
  \caption{Angular distributions for fermion and boson production through
    a resonant graviton. The graviton has a mass of 1 TeV. 
    The black line denotes the analytical result and the crosses
    show the simulation data for (a) $gg\ra G\ra e^+e^-$
    (b) $u\bar{u}\ra G\ra e^+e^-$ and (c) $u\bar{u}\ra G \ra \gamma \gamma$.}
  \label{fig:GRes}
  \end{center}
  \end{figure}

  The LHC may give us the possibility of detecting narrow graviton resonances
  at the TeV scale through various hard subprocesses. To test our
  implementation of the RS Model we have picked three processes involving 
  graviton exchange, $gg\ra G\ra e^+e^-$, $u\bar{u}\ra G\ra e^+e^-$ and
  $u\bar{u}\ra G \ra \gamma \gamma$.
  The plots of the angular distribution of the outgoing fermion/boson with
  respect to the beam axis in the centre-of-mass frame are shown in 
  figure~\ref{fig:GRes}. There is good agreement here with the analytical 
  result from the matrix element and the numerical simulation indicating
  the correct implementation of the graviton Feynman rules and
  new matrix elements.

  These distributions show the characteristic behaviour of an exchanged 
  spin-2 particle. The angular dependence of an exchanged spin-1 boson 
  on the other hand is notably different and therefore this kind of 
  distribution is extremely useful in identifying the two cases and 
  eliminating possible background spin-1 exchange when searching for
  this new mode~\cite{Allanach:2002gn} in future experiments. This kind of
  behaviour may be important for the LHC since
  Randall-Sundrum type models predict the possibility of narrow graviton 
  resonances at the TeV-scale~\cite{Allanach:2000nr}. Discovery of the kind of behaviour shown in 
  figure~\ref{fig:GRes} 
  would certainly be a very strong indication of the
  existence of some type of extra dimensions model.

  \subsection{Squark Decay}\label{sec:sqDecay}
  The spin correlation algorithm discussed in section~\ref{sec:spin} was
  shown to work in the case of $t\bar{t}$ production and decay. One of the
  simplest cases to consider for a SUSY model is the different decay modes 
  of a left-handed squark. Considering the decay of the squark via the two
  modes (a) $\tilde{q}_L\ra \tilde{\chi}^0_2q\ra\tilde{l}^-_R l^+ u$  and 
  (b) $\tilde{q}_L\ra \tilde{\chi}^0_2q\ra\tilde{l}^+_R l^- u$ and plotting 
  the mass distribution of the produced quark and (anti-)lepton allows
  the effect of spin correlations to be shown.

\begin{figure*}
  \begin{center}
    \includegraphics[angle=90,clip,width=0.3\textwidth]{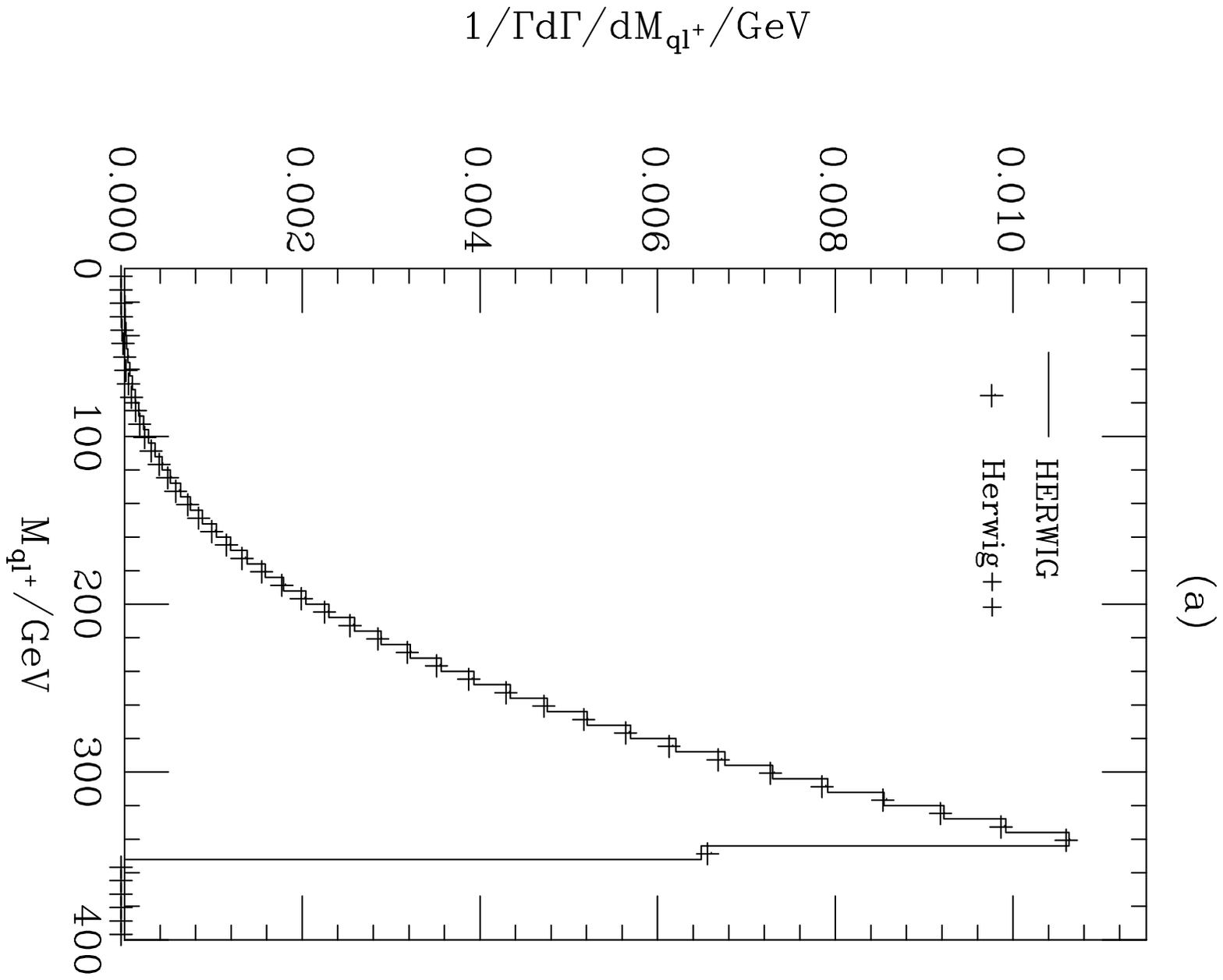}
    \includegraphics[angle=90,clip,width=0.3\textwidth]{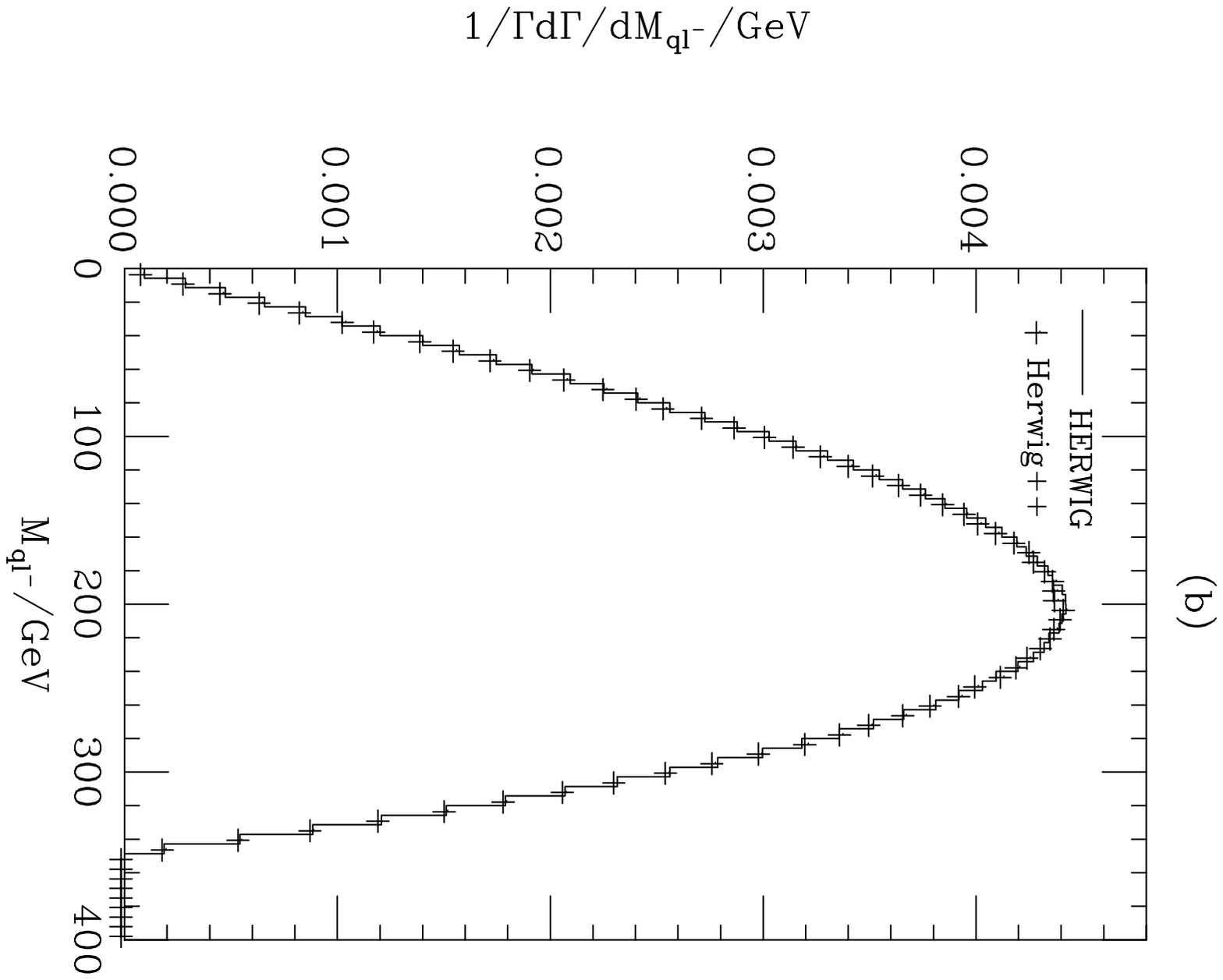}
    \caption{The invariant mass distribution of (a) the anti-lepton-quark and
    (b) lepton-quark produced in $\tilde{u}_L\ra\tilde{\chi}_2^0 u \ra
      e^{\pm}\tilde{e}_R^{\mp} u$.}
    \label{fig:qldecay}
  \end{center}
  \end{figure*}

  The plots in figure~\ref{fig:qldecay} were produced at Snowmass point 
  5~\cite{Allanach:2002nj}  where $\tan\beta=5$, $\text{sign}(\mu)=+$, $M_0=150\, 
  \textrm{GeV}$, 
  $M_{2}=300\,\textrm{GeV}$ and 
  $A_0=-1000\,\textrm{GeV}$.
  This parameter set gives the following particle 
  spectrum using \textsf{SOFTSUSY} 2.0.8~\cite{Allanach:2001kg} 
  $M_{\tilde{u}_L}=672.82\,\textrm{GeV}$, 
  $M_{\tilde{\chi}^0_2}=231.29\,\textrm{GeV}$ and 
  $M_{\tilde{l}_R}=192.87\,\textrm{GeV}$.

  There is a stark difference in the quark-lepton mass distribution for the 
  two decay modes considered above. The difference is due to the helicities of
  the external particles. At the mass scale of the squark the quark can be
  considered massless and left-handed while the produced lepton and anti-lepton
  will be right-handed. When back-to-back the lepton-quark system will have 
  net spin-1 and as such can not be produced in a scalar decay while the 
  anti-lepton-quark system will have spin-0 and is able to proceed.
  
  The end-point in both of the distributions is due to a kinematical cut-off
  where the invariant mass of the quark-lepton pair is at a maximum. The value
  of this end-point can be calculated by considering the mass-squared when
  the pair is back-to-back. The value is given in~\cite{Allanach:2000kt} as
  \beq
  (m)^2_{\text{max}} =\frac{ \left(m^2_{\tilde{u}_l} - m^2_{\tilde{\chi}^0_2}\right)
  \left(m^2_{\tilde{\chi}^0_2} - m^2_{\tilde{e}_R}\right)}{m^2_{\tilde{\chi}_2^0}}.
  \eeq
  Using the values for the sparticles above one finds a value for the cut-off 
  of 348.72\,GeV which is consistent with the plots in figure~\ref{fig:qldecay}.
  
  \subsection{Gaugino Production}
  Supersymmetry predicts the existence of Majorana fermions and it is
  necessary to ensure that their spin correlations are implemented correctly in 
  our new framework. We consider three production processes and  the angular 
  distributions of the leptons produced in the subsequent decays. 
  The SUSY spectrum for each was again generated using \textsf{SOFTSUSY} and the 
  masses for the points used in each process are given in the relevant section. 
  
  \subsubsection{$e^+e^-\ra \chi_2^0\chi_1^0$}
\begin{figure*}
  \begin{center}
    \includegraphics[angle=90,clip,width=0.25\textwidth]{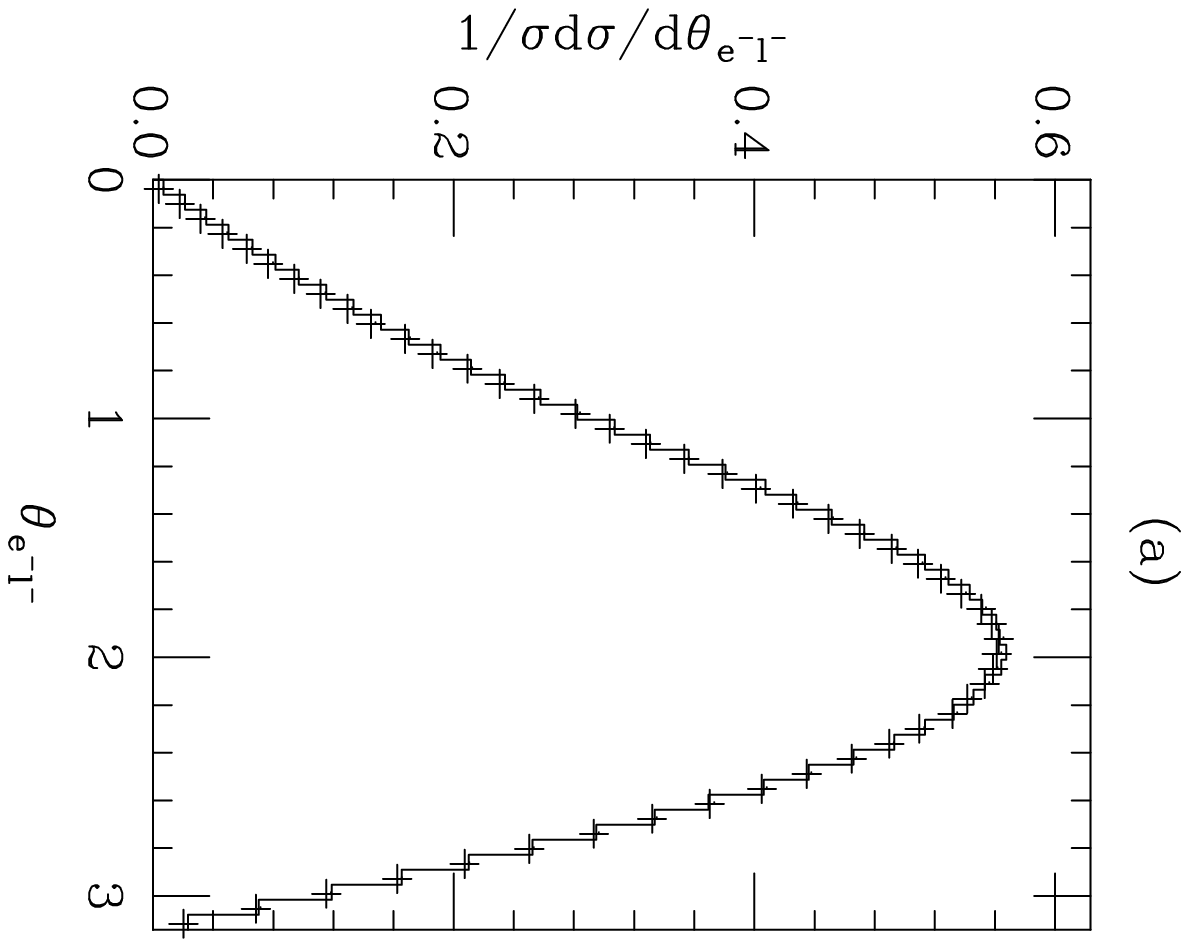}
    \includegraphics[angle=90,clip,width=0.25\textwidth]{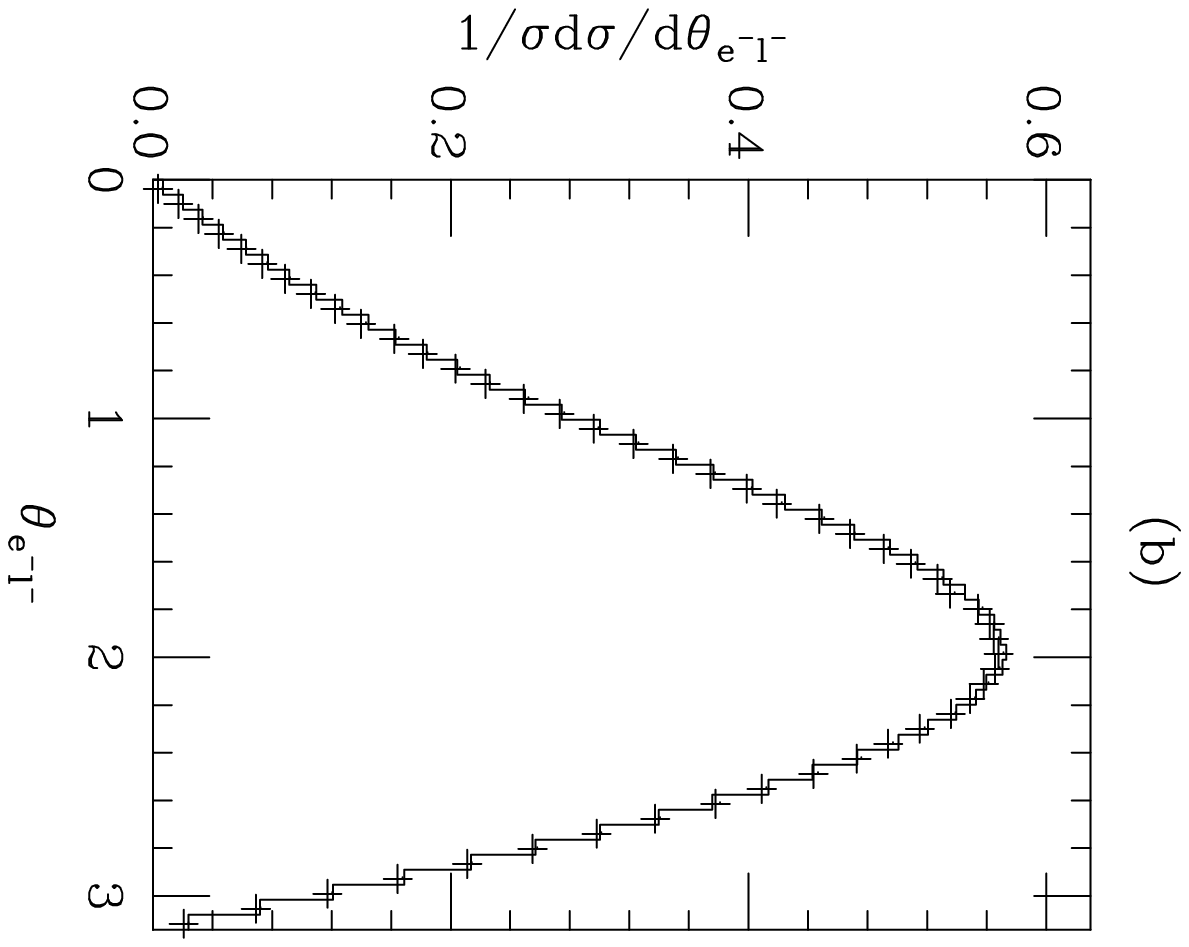}
    \includegraphics[angle=90,clip,width=0.25\textwidth]{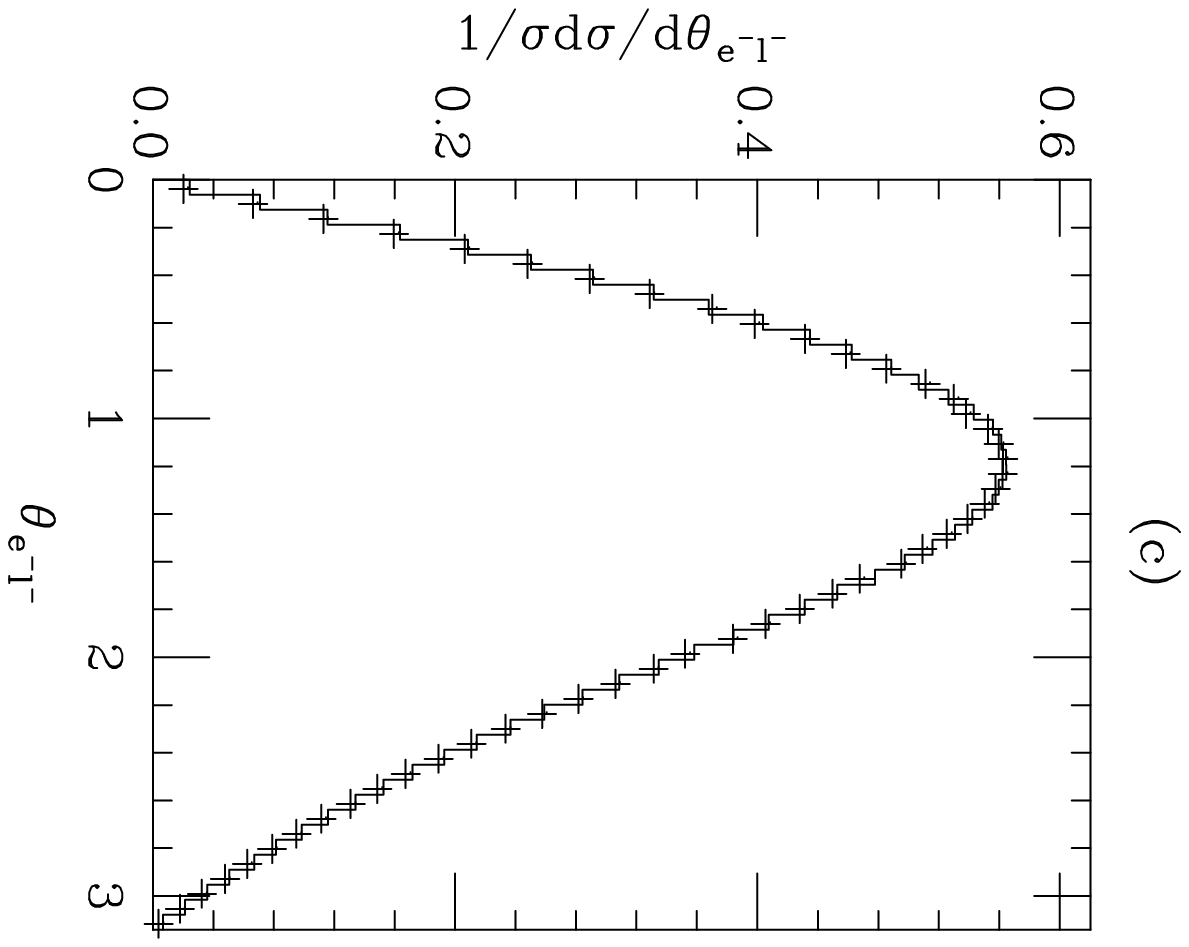}
    \caption{The angle between the lepton produced in 
      $e^+e^-\ra\chi_2^0\chi_1^0\ra\tilde{l}_R^+ l^-$ and the beam in the 
      lab frame for a centre-of-mass energy of 500 GeV and 
      (a) unpolarised incoming beams, (b) negatively polarised electrons and 
    positively polarised positrons and (c) positively polarised electrons and 
    negatively polarised positrons. The black histogram is from HERWIG and the
    crosses from \Hw\!\!.}
    \label{fig:n2n1lepton}
  \end{center}
  \end{figure*}

  Here we consider the production of the lightest and next-to-lightest 
  neutralinos with the $\chi_2^0$ decaying via the two modes
  (i) $\chi_2^0\ra\tilde{l}_R^+ l^-\ra l^+l^-\chi_1^0$, 
  (ii) $\chi_2^0\ra Z^0\chi_1^0\ra l^-l^+ \chi_1^0$ at SPS point 1b. The
  relevant sparticle masses are $M_{\chi_2^0}=306.55\,\text{GeV}$, 
  $M_{\chi_1^0}=161.78\,\text{GeV}$, $M_{\tilde{l}_R}=253.82\,\text{GeV}$.
  Figure~\ref{fig:n2n1lepton} shows how the polarisation of the beam affects
  the angular distribution of the lepton produced from the $\chi_2^0$ decay.

\begin{figure*}
  \begin{center}
    \includegraphics[angle=90,clip,width=0.25\textwidth]{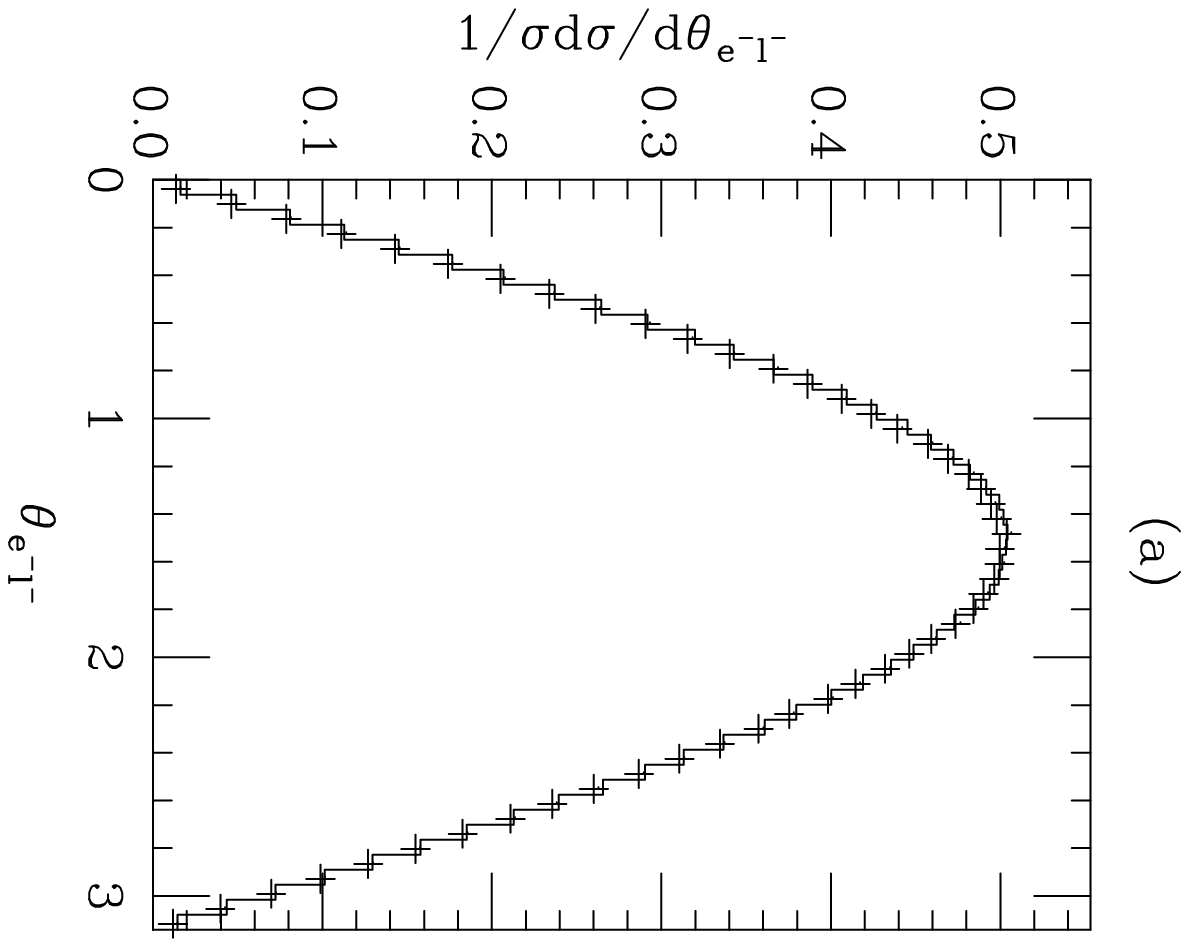}
    \includegraphics[angle=90,clip,width=0.25\textwidth]{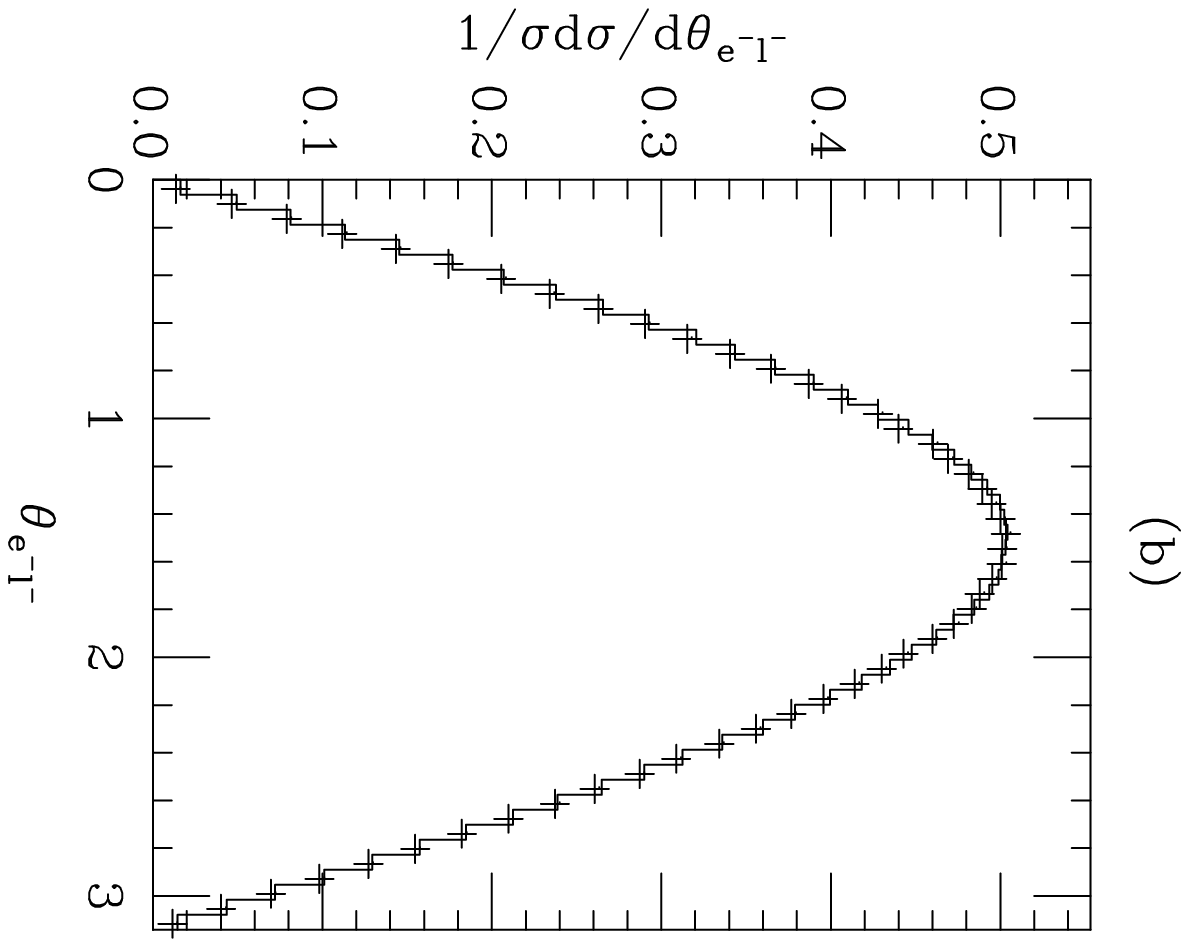}
    \includegraphics[angle=90,clip,width=0.25\textwidth]{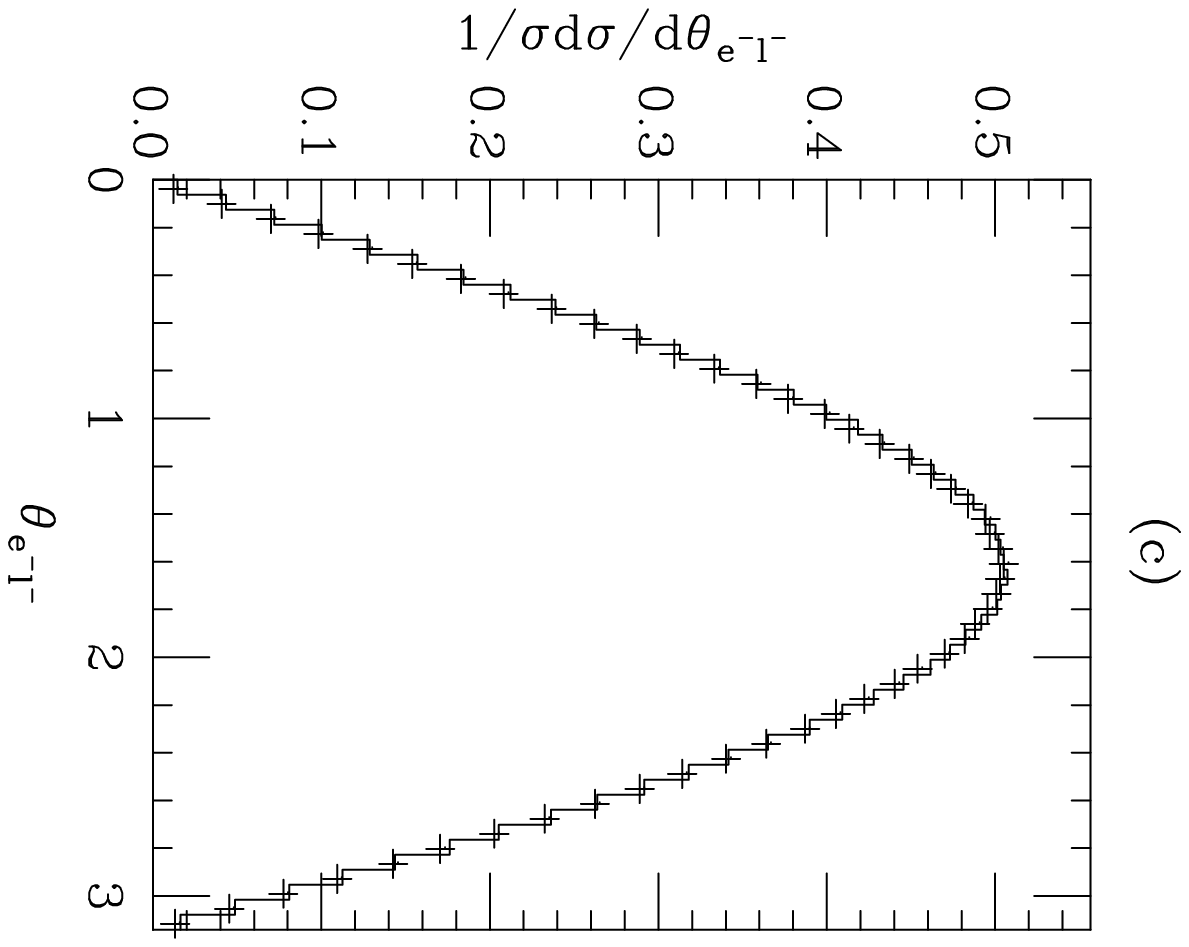}
    \caption{The angle between the lepton produced in 
      $e^+e^-\ra\chi_2^0\chi_1^0\ra Z^0 \chi_1^0\ra l^-l^+\chi_1^0$ and the 
      beam in the lab frame for a centre-of-mass energy of 500 GeV and 
      (a) unpolarised incoming beams, (b) negatively polarised electrons and 
    positively polarised positrons and (c) positively polarised electrons and 
    negatively polarised positrons. The black histogram is from HERWIG and the
    crosses from \Hw\!\!.}
    \label{fig:n2n1Zlept}
  \end{center}
  \end{figure*}

\begin{figure*}
  \begin{center}
    \includegraphics[angle=90,clip,width=0.25\textwidth]{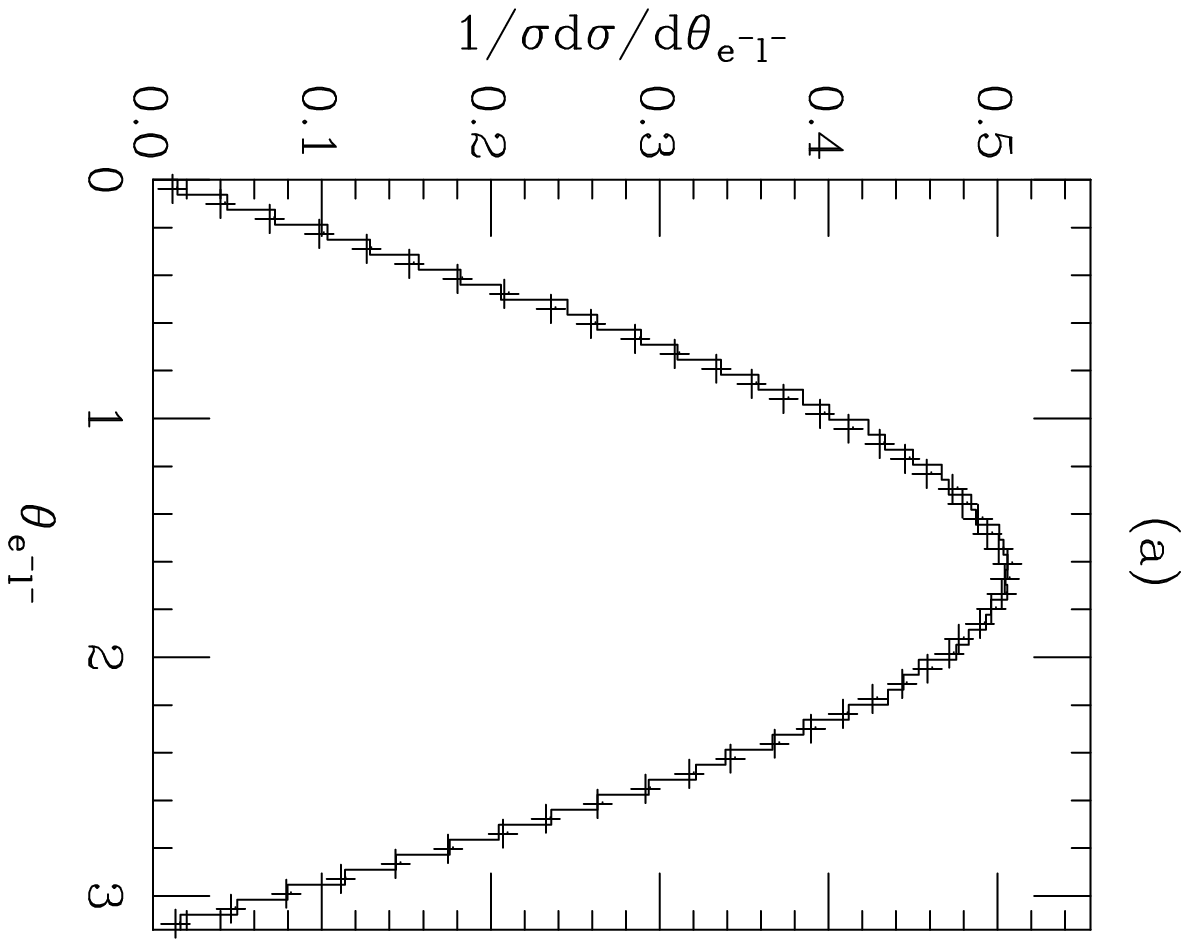}
    \includegraphics[angle=90,clip,width=0.25\textwidth]{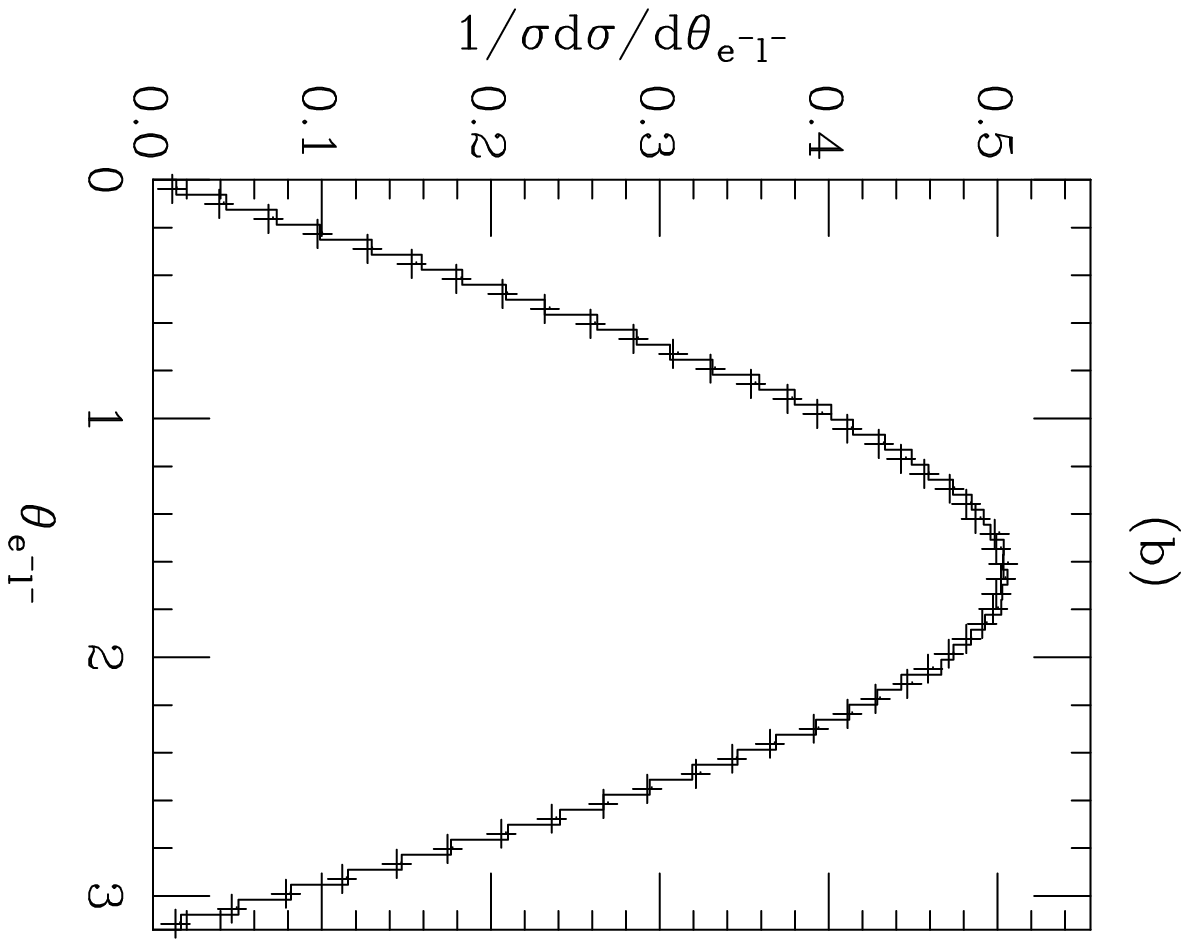}
    \includegraphics[angle=90,clip,width=0.25\textwidth]{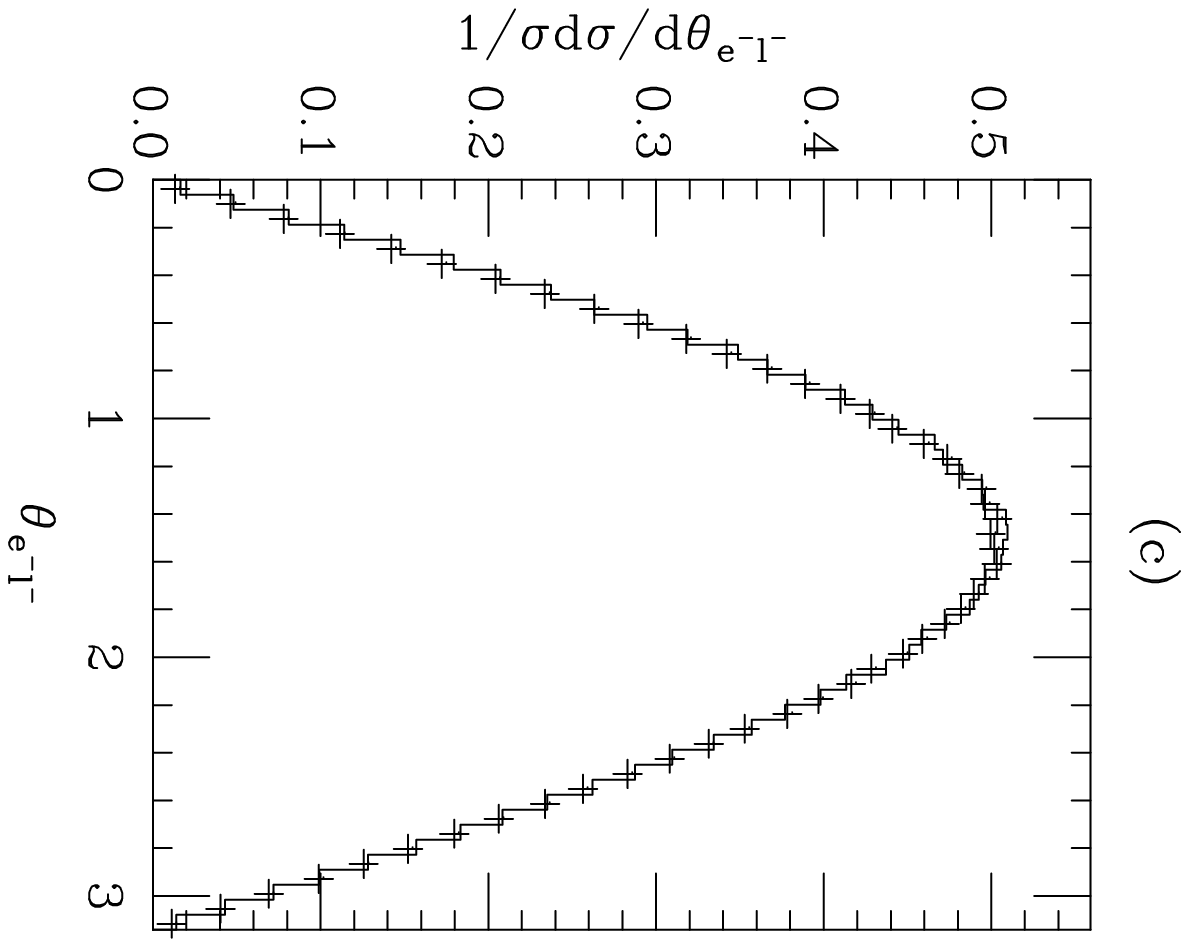}
    \caption{The angle between the lepton produced in 
      $e^+e^-\ra\chi_2^0\chi_1^0\ra\tilde{l}_R^- l^+\ra l^- l^+ \chi_1^0$ 
      and the beam in the lab frame for a centre-of-mass energy of 500 GeV and 
      (a) unpolarised incoming beams, (b) negatively polarised electrons and 
    positively polarised positrons and (c) positively polarised electrons and 
    negatively polarised positrons. The black histogram is from HERWIG and the
    crosses from \Hw\!\!.}
    \label{fig:n2n1sleptlept}
  \end{center}
  \end{figure*}

  The lepton shows a correlation with the beam polarisation due to the 
  neutralino being a fermion and preserving spin information when decaying.
  Figures~\ref{fig:n2n1Zlept} and~\ref{fig:n2n1sleptlept} show the angular
  dependence of the final-state lepton for the case of an intermediate
  $Z^0$ boson and $\tilde{l}_R$ respectively. As is to be expected for an
  intermediate slepton the incoming beam polarisation  has little effect on the 
  angular distribution of the final-state lepton due to its scalar nature\footnote{There is some residual effect due to the correlation of the $\tilde{l}_R$ direction with the beam in the $\chi_2^0$ decay.}.
  The plots are in good agreement with the HERWIG results.
 
  \subsubsection{$e^+e^-\ra \chi_i^+\chi_i^-$}
  We now consider the production of chargino pairs and their associated decays. 
  Two possible decay modes of the $\chi_i^\pm$ are (a) 
  $\chi_i^\pm\ra W^\pm \chi_1^0$ and (b) $\chi_i^\pm\ra \tilde{\nu}_\alpha l^\pm$.
  Here we use the $W$ decay mode for the lightest chargino and the
  sneutrino decay mode for the heaviest chargino in order to consider 
  final-states with differing spins. The mass spectrum was
  generated for SPS point 1a where $M_{\chi_2^+}=377.39\,\text{GeV}$,
  $M_{\chi_1^+}=181.53\,\text{GeV}$, $M_{\tilde{\nu}_L}=185.42\,\text{GeV}$ and 
  $M_{\chi_1^0}=97.00\,\text{GeV}$.
  
\begin{figure*}
  \begin{center}
    \includegraphics[angle=90,clip,width=0.25\textwidth]{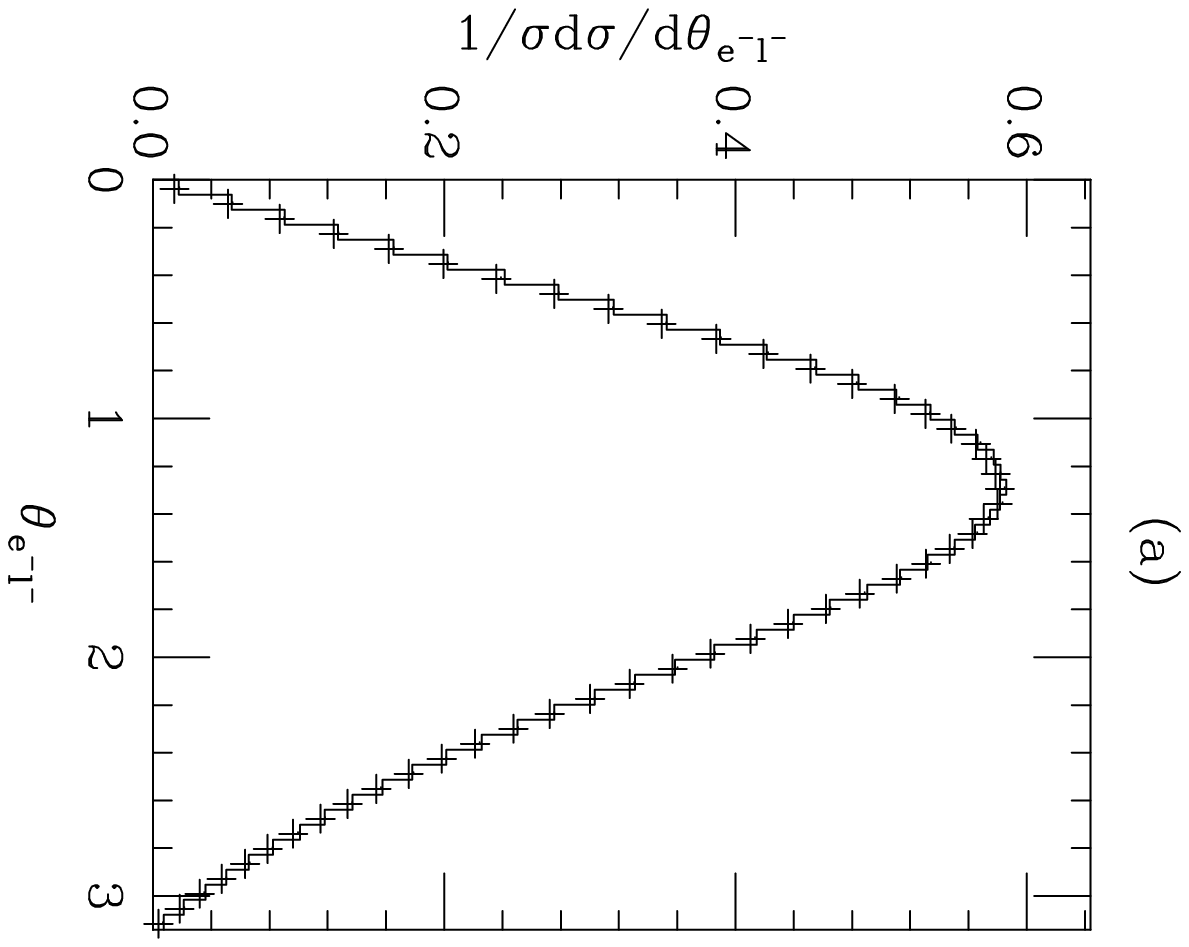}
    \includegraphics[angle=90,clip,width=0.25\textwidth]{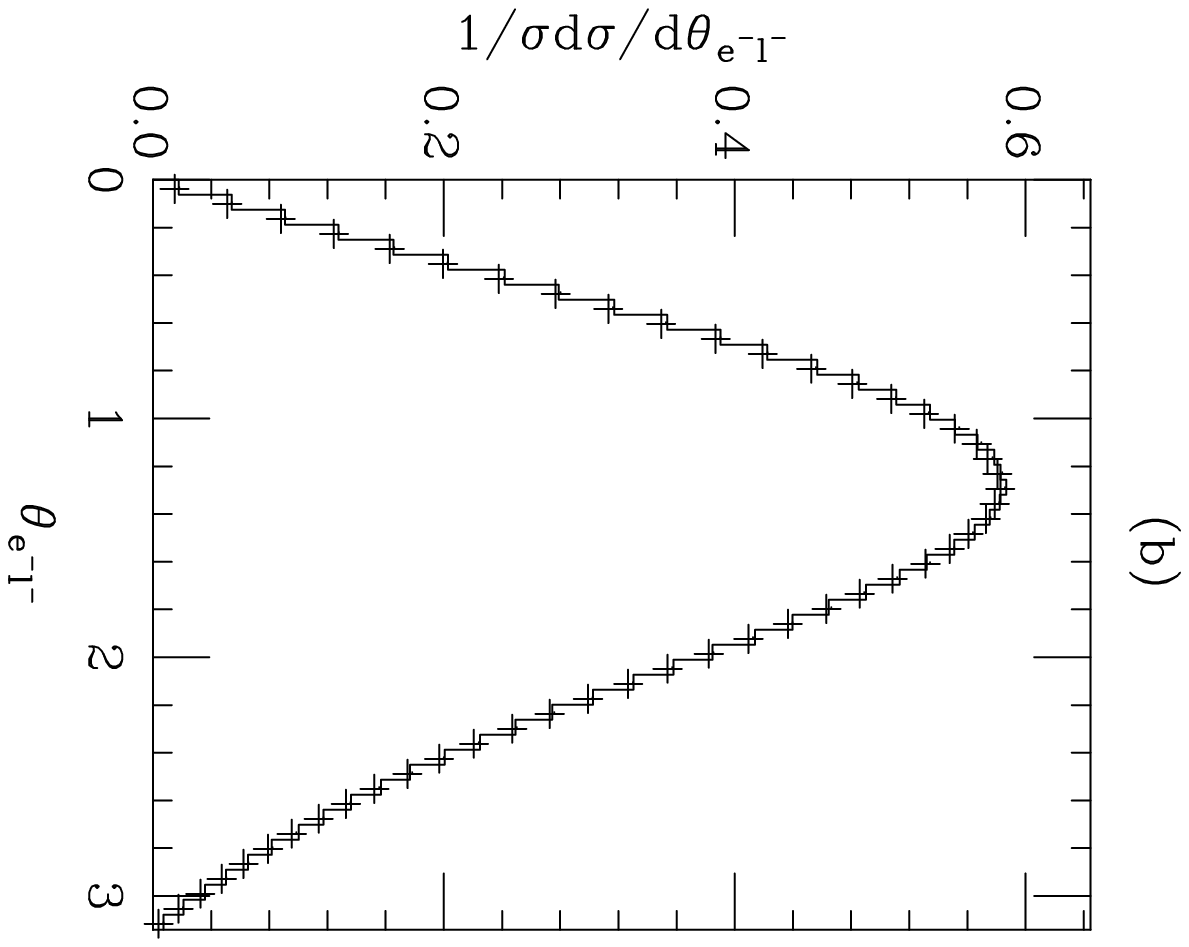}
    \includegraphics[angle=90,clip,width=0.25\textwidth]{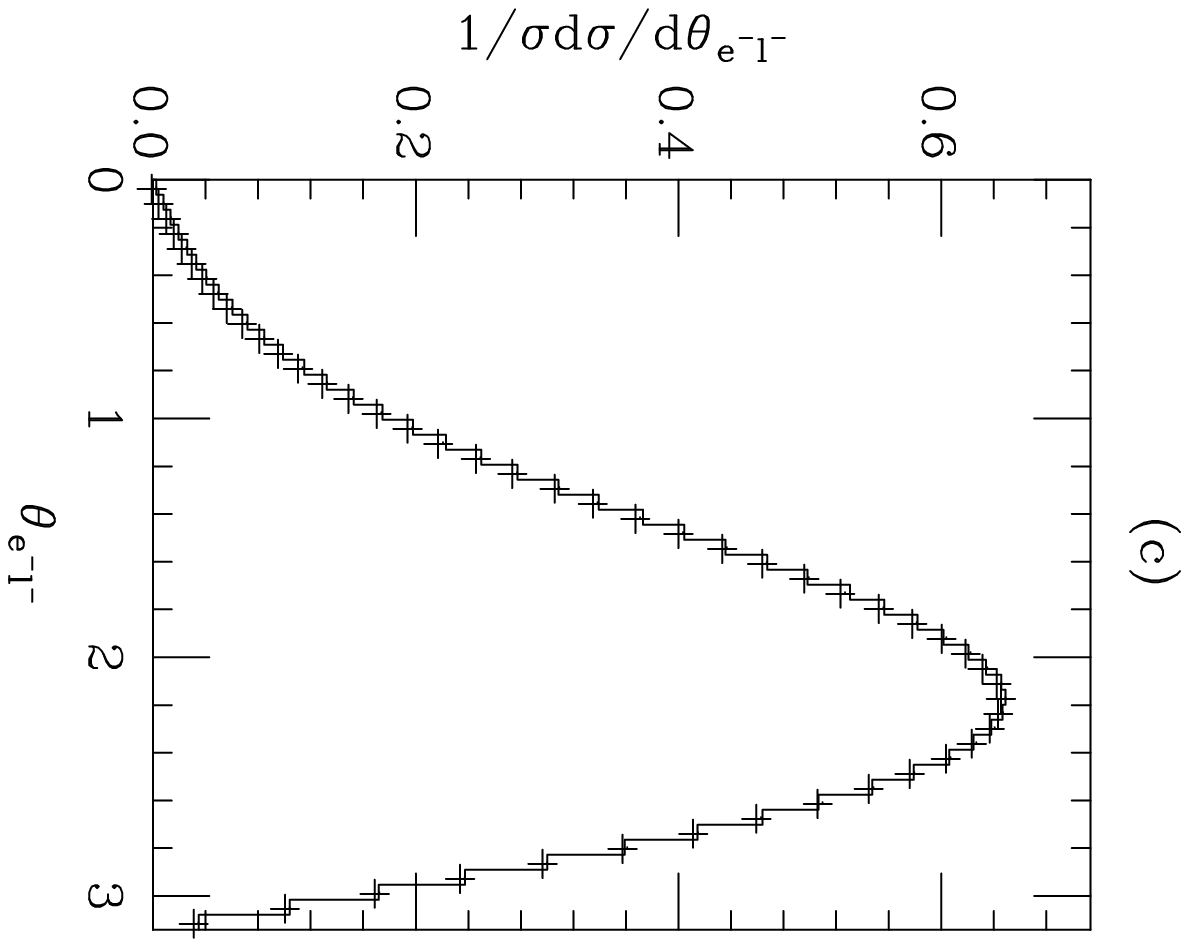}
    \caption{The angle between the lepton produced in 
      $e^+e^-\ra\chi_1^+\chi_1^-\ra W^+ W^-\chi_1^0 \chi_1^0 \ra l^- l^+ \nu_l 
      \bar{\nu}_l$ 
      and the beam in the lab frame for a centre-of-mass energy of 500 GeV and 
      (a) unpolarised incoming beams, (b) negatively polarised electrons and 
    positively polarised positrons and (c) positively polarised electrons and 
    negatively polarised positrons. The black histogram is from HERWIG and the
    crosses from \Hw\!\!.}
    \label{fig:c1c1Wlept}
  \end{center}
  \end{figure*}

\begin{figure*}
   \begin{center}
     \includegraphics[angle=90,clip,width=0.25\textwidth]{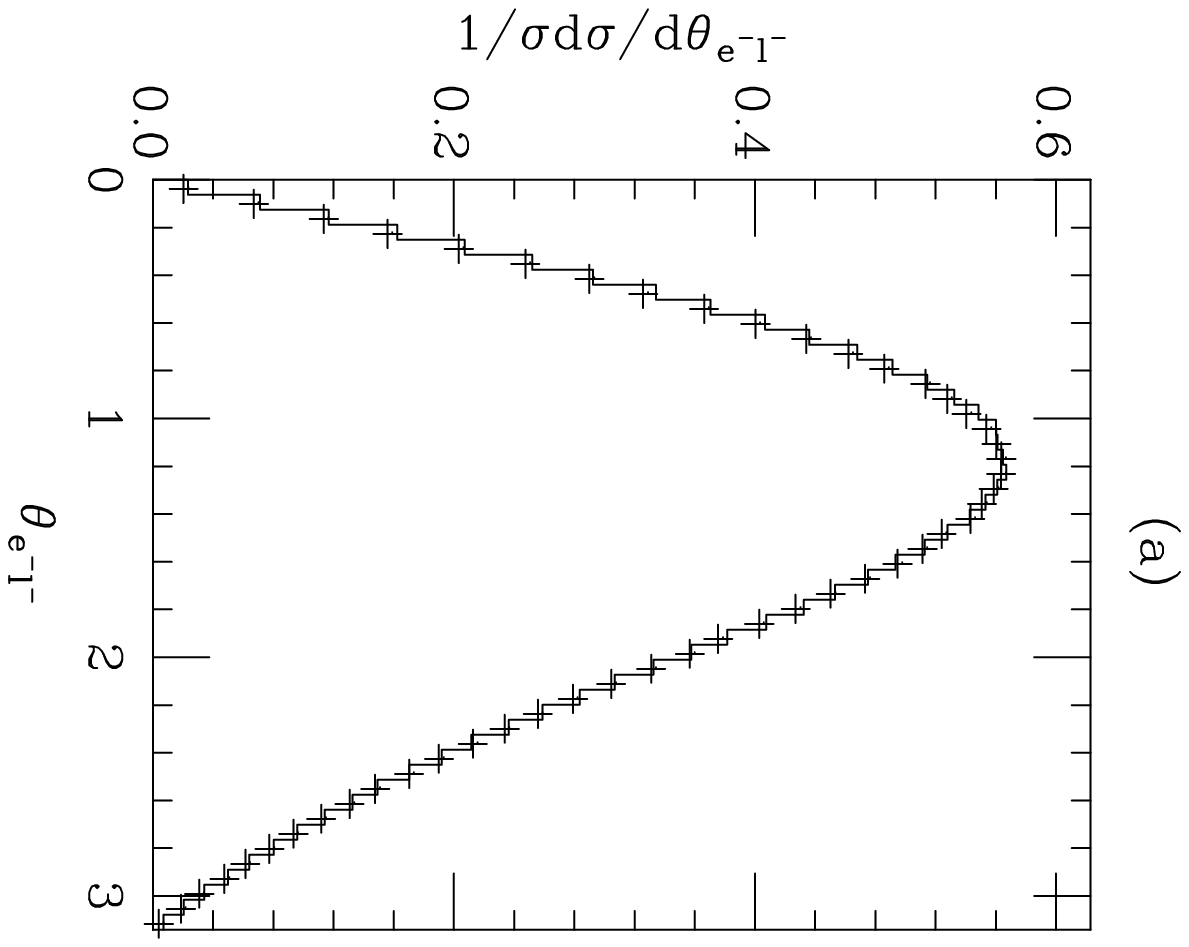}
     \includegraphics[angle=90,clip,width=0.25\textwidth]{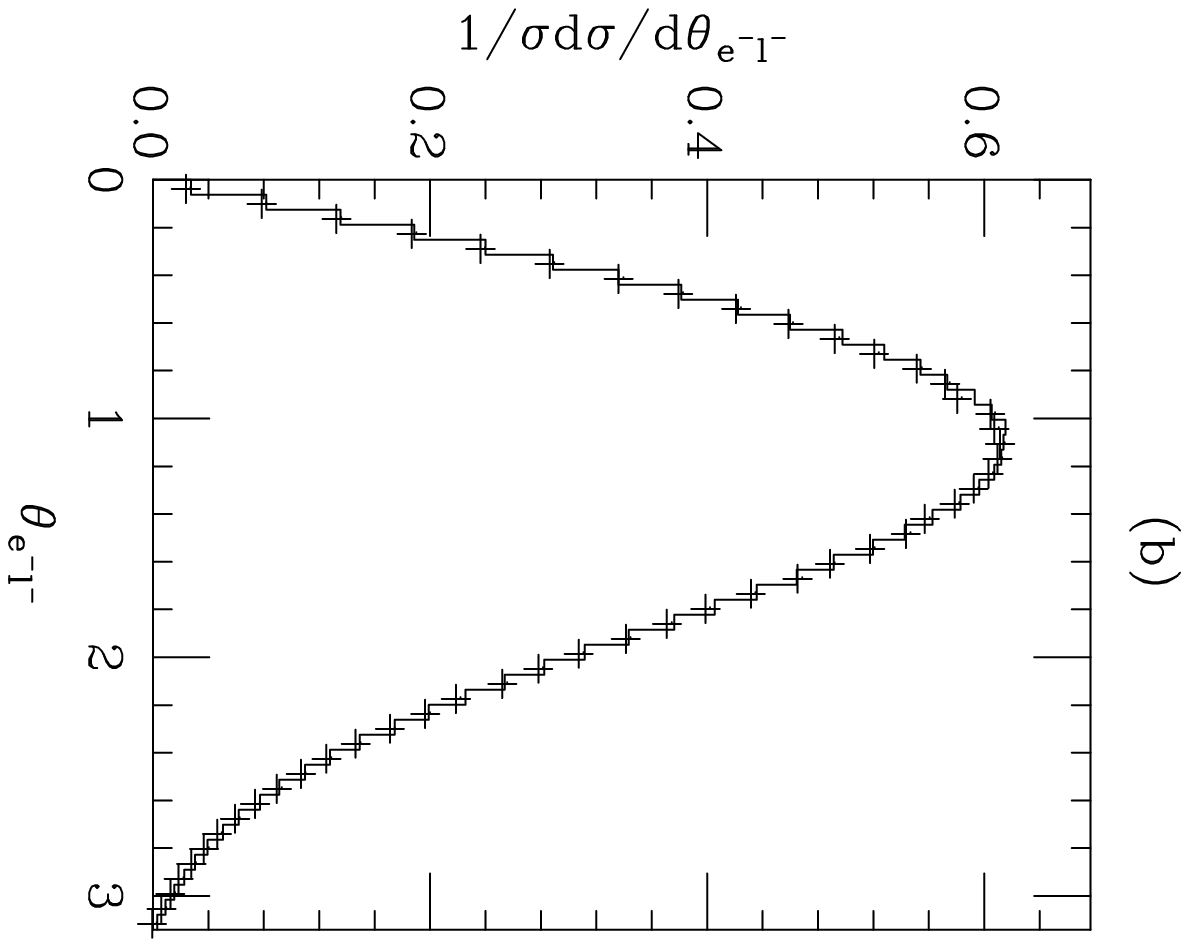}
     \includegraphics[angle=90,clip,width=0.25\textwidth]{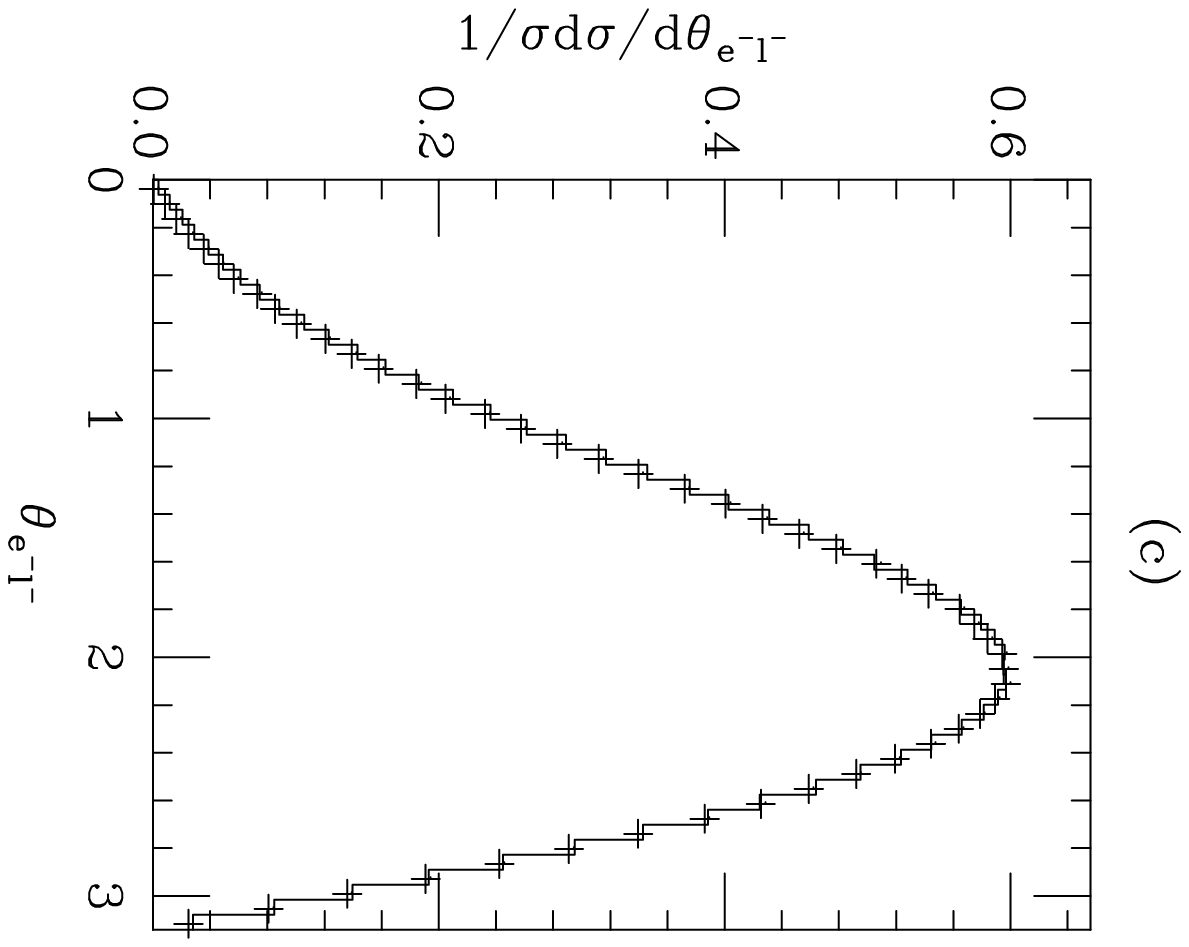}
    \caption{The angle between the lepton produced in 
      $e^+e^-\ra\chi_2^+\chi_2^-\ra \tilde{\nu}_L l^+ \tilde{\nu}_L l^- $
      and the beam in the lab frame for a centre-of-mass energy of 1 TeV and 
      (a) unpolarised incoming beams, (b) negatively polarised electrons and 
    positively polarised positrons and (c) positively polarised electrons and 
    negatively polarised positrons. The black histogram is from HERWIG and the
    crosses from \Hw\!\!.}
    \label{fig:c2c2sneu}
   \end{center}
   \end{figure*}

  Figure~\ref{fig:c1c1Wlept} shows the angle of the produced electron for the
  production of the lightest chargino. As is to be expected the beam polarisation
  affects the lepton distribution because of the intermediate $W$ boson
  carrying the spin correlations through to the final-state. The effects are 
  similar in the case of the sneutrino decay of the heaviest chargino shown in
  figure~\ref{fig:c2c2sneu}. The lepton
  accompanied with the $\tilde{\nu}_L$ still shows correlations with the beam
  on account of the chargino being a fermion. If the lepton had come from the 
  decay of a scalar then there would have been no such correlation.

  \subsection{Tau Decays}
  \subsubsection{One Prong Decays}
\begin{figure*}
  \begin{center}
    \includegraphics[angle=90,width=0.32\textwidth]{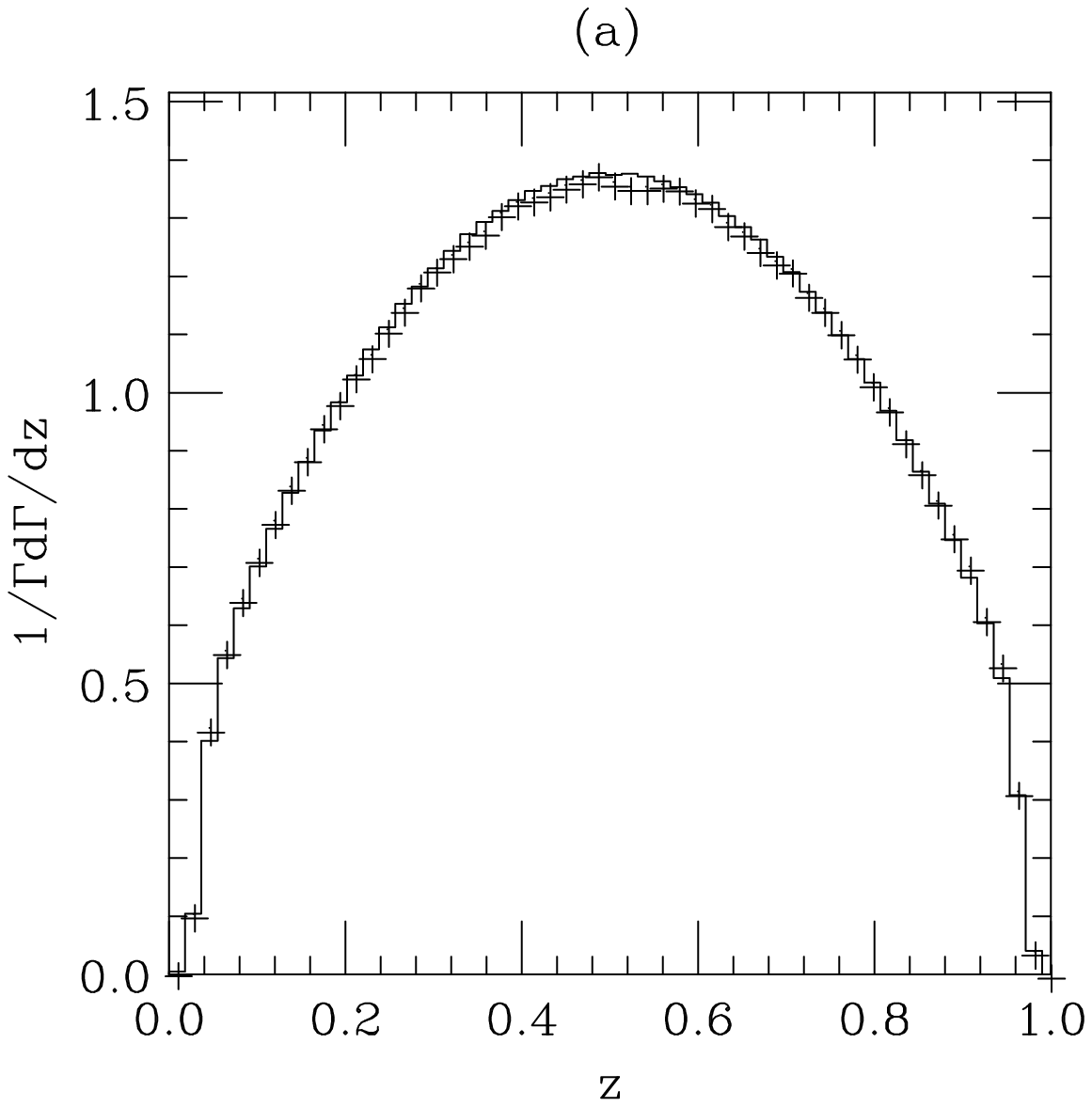}
    \includegraphics[angle=90,width=0.32\textwidth]{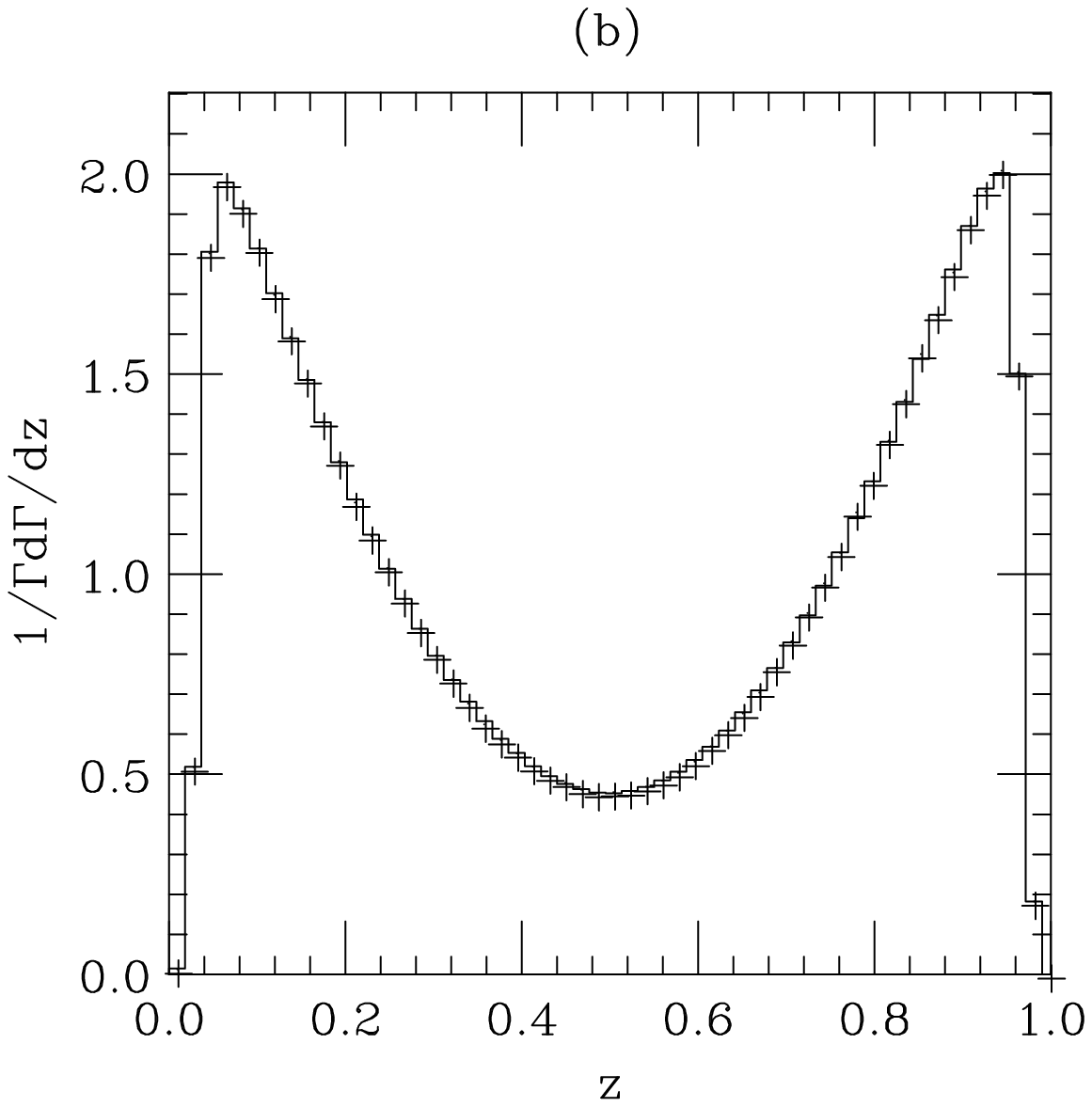}
    \caption{Energy fraction, $z$, carried away by the charged meson in the one prong
      $\tau$ decay $\rho^\pm\ra\pi^\pm\pi^0$ for (a) left-handed $\tilde{\tau}_1$
      and (b) right-handed $\tilde{\tau}_1$. The black histogram shows the
    results from HERWIG with the TAUOLA~\cite{Jadach:1993hs} decay package and 
    the crosses indicate the \Hw results.}
    \label{fig:Tau1Prong}
  \end{center}
  \end{figure*}

  The tau has a number of leptonic and hadronic
  decay modes. A more detailed analysis of these decays shows
  interesting features in the distribution of energy to the decay products.
  A typical tau decay involving several mesons has the form 
  $\tau^{\pm}\ra (nm^\pm)(qm^0) \nu_\tau$ where $nm^\pm$ denotes $n\geq 1$ 
  charged mesons, \ie the number of prongs, and $qm^0$ denotes $q \geq 0$ 
  neutral mesons. Here we will consider the one prong decay
  $\tau^\pm\ra \rho^\pm \nu_\tau\ra \pi^\pm \pi^0\nu_\tau$ where the $\tau$
  is produced from the decay of a $\tilde{\tau}_1$. Figure~\ref{fig:Tau1Prong}
  shows our results for the fraction of visible energy carried away by the 
  charged meson in the two cases where the $\tilde{\tau}_1$ is  
  (a) 100\% left-handed and (b) 100\% right-handed.

  There is a stark difference in the energy distribution for two possible 
  mixings of the $\tilde{\tau}_1$ in figure~\ref{fig:Tau1Prong} due to the 
  resulting helicity of the decaying $\rho$. For the case where the 
  $\tilde{\tau}_1$ is entirely $\tilde{\tau}_L$ the $\rho$ has a higher 
  probability of being transversely 
  polarised, from the results of~\cite{Bullock:1991fd}, which favours
  the equal splitting of energy between the two pions as confirmed by
  the first plot. A $\tilde{\tau}_1$ that is entirely $\tilde{\tau}_R$ however, 
  will give rise to mostly longitudinally polarised $\rho$ mesons that prefer to 
  distribute their energy unequally and favour a distribution where
  one meson receives most of the visible energy from the $\tau$ decay. This 
  is again confirmed in our second plot. The \Hw results are plotted together
  with those from HERWIG with the TAUOLA decay package~\cite{Jadach:1993hs}
  that is designed specifically for the decay of polarised $\tau$ leptons.

  \subsubsection{Squark Decay}
\begin{figure*}
  \begin{center}
    \includegraphics[angle=90,width=0.25\textwidth]{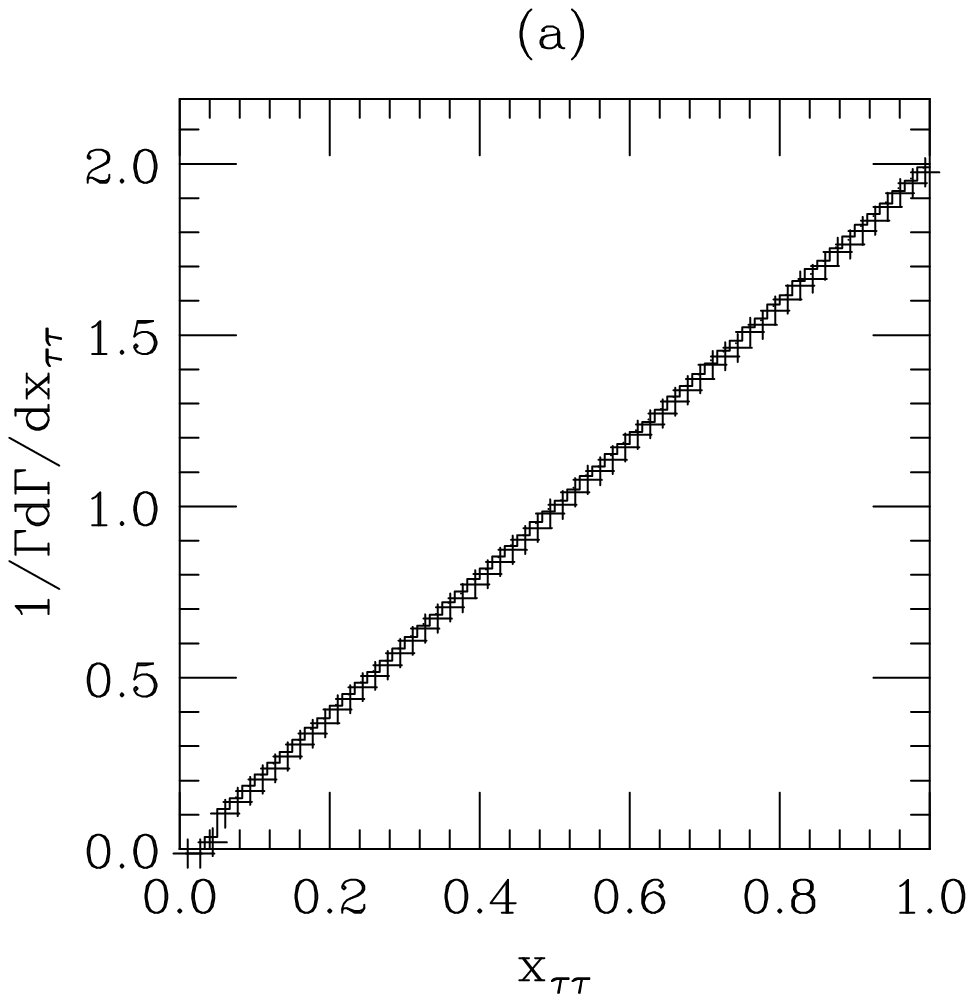}
    \includegraphics[angle=90,width=0.25\textwidth]{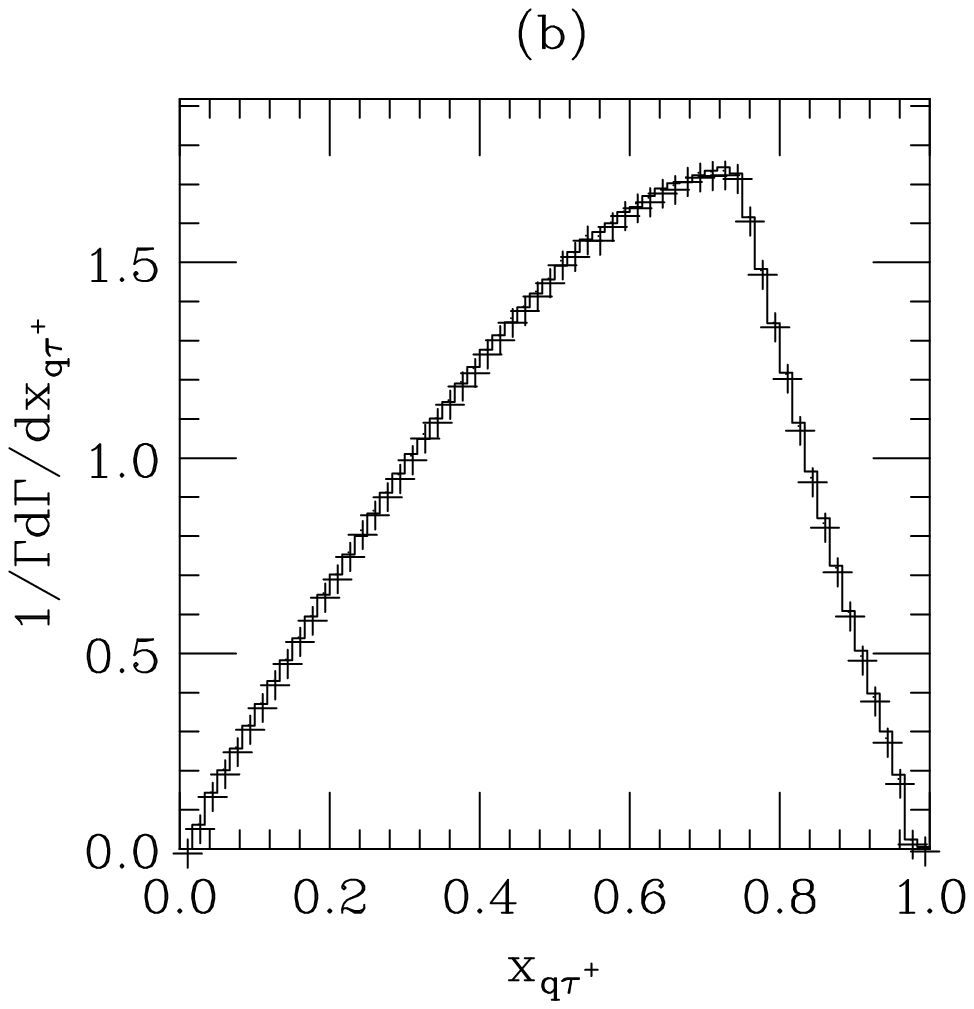}
    \includegraphics[angle=90,width=0.25\textwidth]{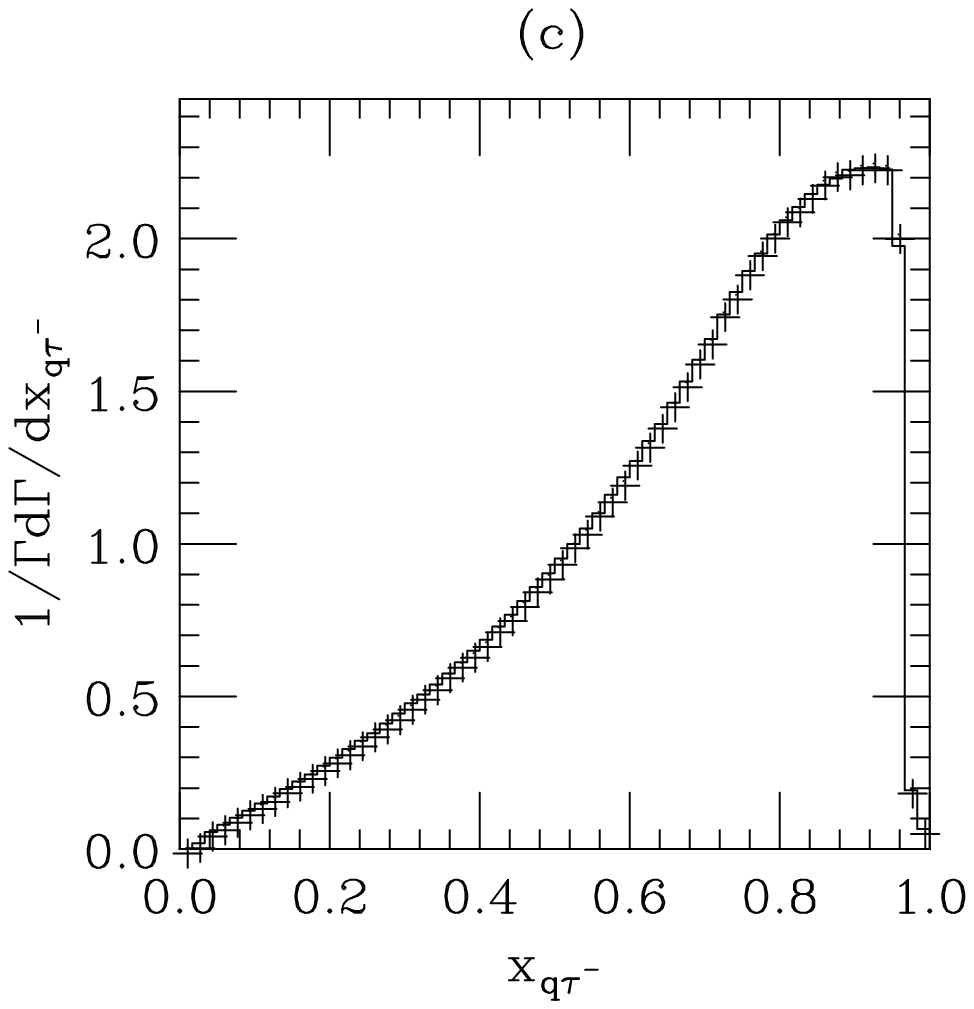}
  \end{center}
  \begin{center}
    \includegraphics[angle=90,width=0.25\textwidth]{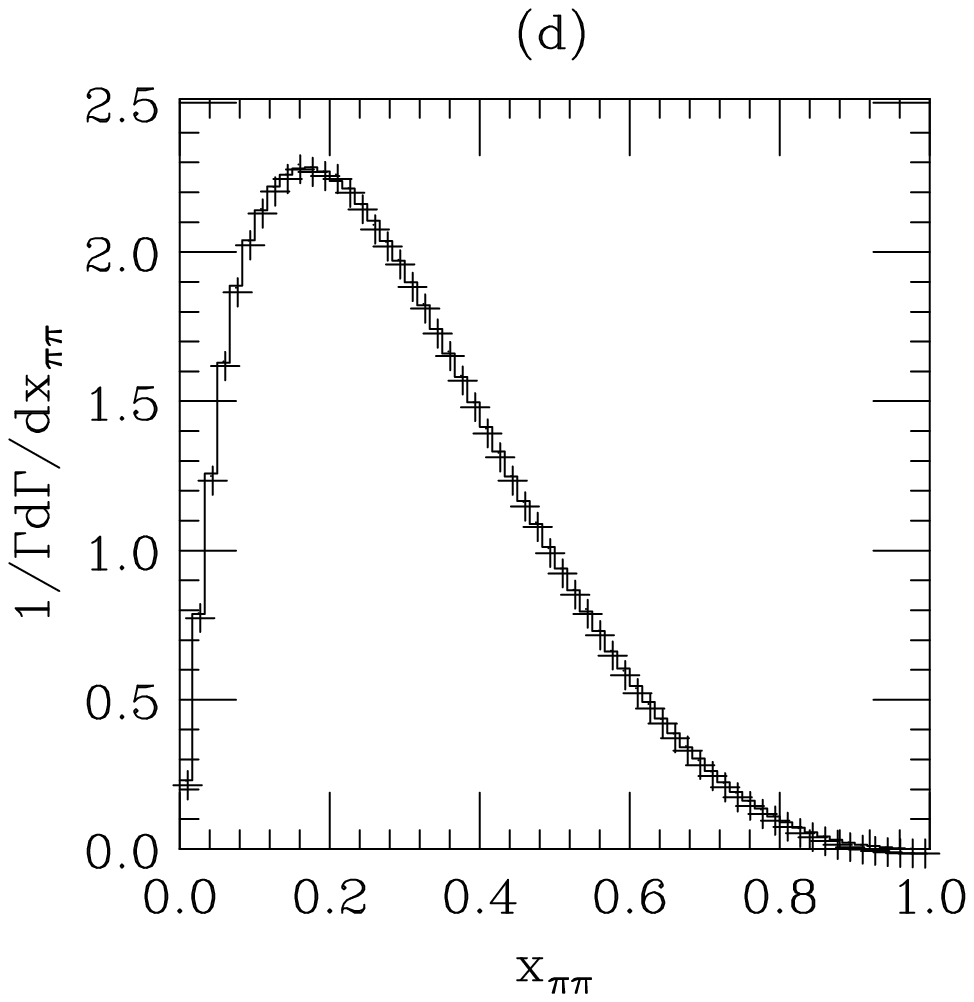}
    \includegraphics[angle=90,width=0.25\textwidth]{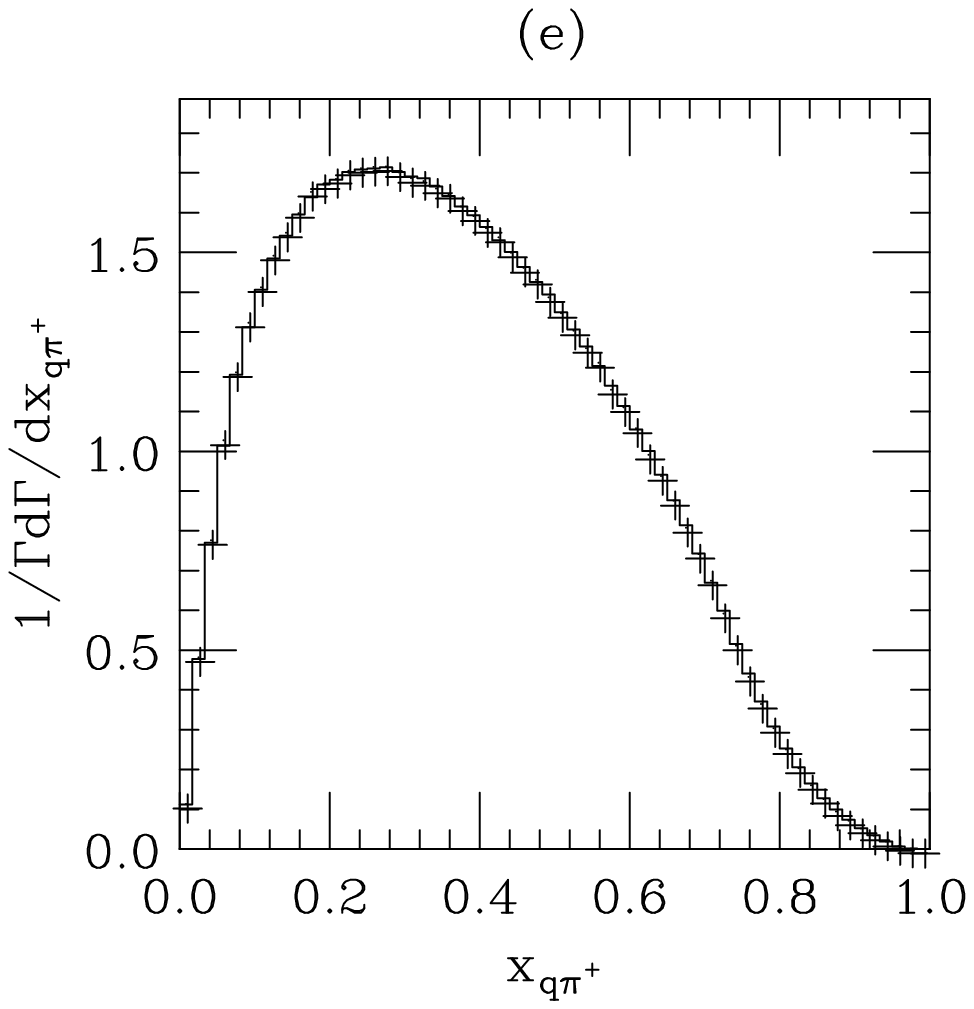}
    \includegraphics[angle=90,width=0.25\textwidth]{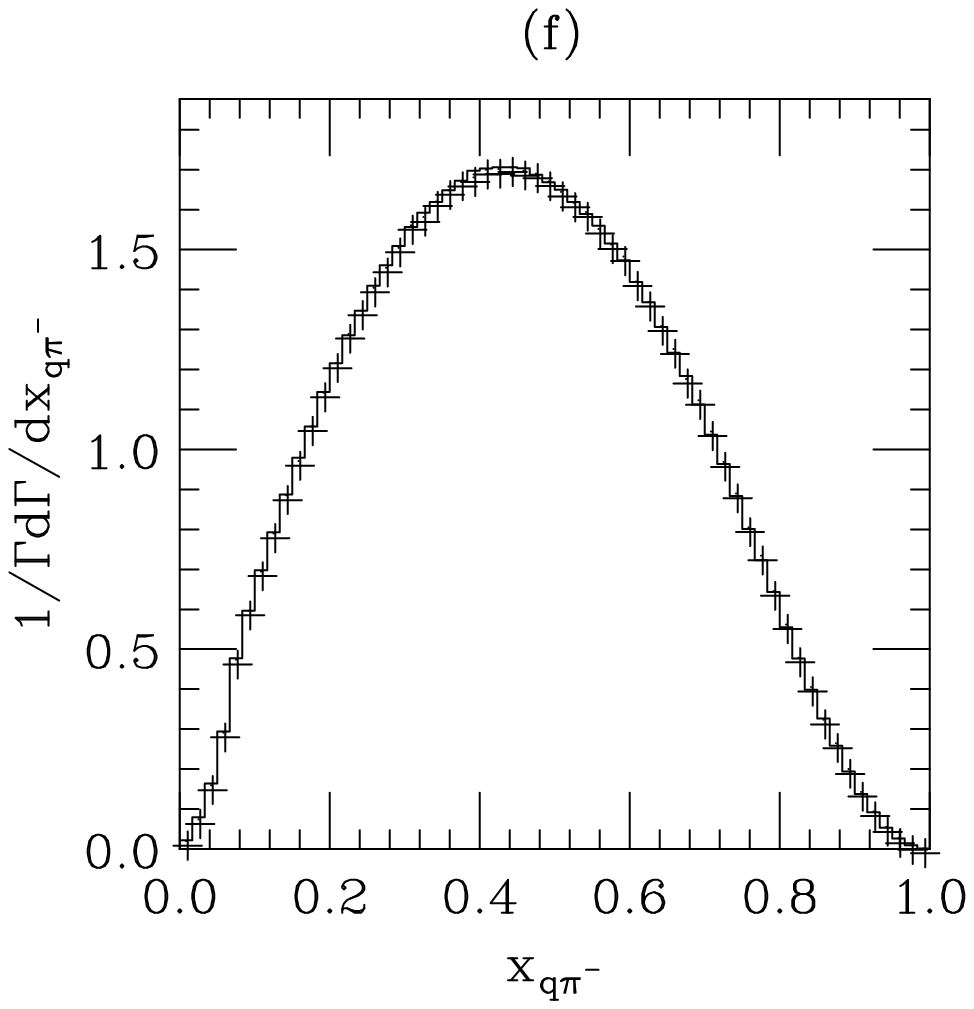}
    \caption{Normalised invariant mass distributions 
      $x_{ij}=m_{ij}/(m_{ij})_{max}$ for various pairs of decay products
      along the chain $\tilde{q}_\alpha\ra q \chi_2^0 \ra \tau^\pm_{near} 
      \tilde{\tau}^\mp_1\ra \tau_{far}^\mp\chi_1^0$ where the $\tau$ decays
      via $\tau \ra \pi\nu_{\tau}$ only.
      The black histogram denotes the results from HERWIG with the TAUOLA 
      package and the crosses are the results from \Hw\!\! for
      (a) $\tau\tau$, (b) $q\tau^+$\!, (c) $q\tau^-$\!, (d) $\pi\pi$\!, 
      (e) $q\pi^+$ and (f) $q\pi^-$.}
    \label{fig:tauInvMass}
  \end{center}
  \end{figure*}

  The use of the effects described above in the study of SUSY models has long
  been recognised. In~\cite{Choi:2006mt} a mechanism for the determining
  the spin properties of particles involved in SUSY cascade decays using $\tau$
  polarisation was suggested. The method involves analysing invariant mass 
  distributions of different particle pairs along the decay chain
  $\tilde{q}_\alpha\ra q \chi_2^0 \ra \tau^\pm_{\text{near}} \tilde{\tau}^\mp_1\ra 
  \tau_{\text{far}}^\mp\chi_1^0$ 
  with the $\tau$ decay restricted to $\tau^\pm\ra \pi^\pm \nu_\tau$. The 
  various normalised invariant mass distributions are shown in 
  figure~\ref{fig:tauInvMass} for $\tilde{q}_\alpha=\tilde{q}_L$ at SPS point 
  1a where $M_{q_L}=558.4\,\text{GeV}$, $M_{\chi_2^0}=180.96\,\text{GeV}$,
  $M_{\tilde{\tau}_1}=134.56\,\text{GeV}$ and $M_{\chi_1^0}=97.00\,\text{GeV}$.
  Since an experiment would be unable to distinguish a near or far $\tau$/$\pi$
  their distributions are combined. The normalisation is to the maximum of the
  invariant mass. For $m_{\tau\tau}$
  \begin{equation}
    \left(m_{\tau\tau}^2\right)_{\text{max}}=\left(m^2_{\tilde{\chi}_2^0} - m^2_{\tilde{\tau}_1}\right)
    \left(1 - m^2_{\tilde{\chi}_1^0}/m^2_{\tilde{\tau}_1}\right), \nonumber
  \end{equation}
  which is equal to $(m_{\pi\pi}^2)_{\text{max}}$. In the case of the $q\tau$ plots
  the maximum of 
  \vspace{-2mm}
  \begin{eqnarray}
    \left(m_{\tilde{q}_L}^2 - m^2_{\tilde{\chi}_2^0}\right)
    \left(1 - m^2_{\tilde{\tau}_1}/m^2_{\tilde{\chi}_2^0}\right),
    \nonumber \\
   \!\!\!\!\left(m_{\tilde{q}_L}^2 - m^2_{\tilde{\chi}_2^0}\right)
   \left(1 - m^2_{\tilde{\chi}_2^0}/m^2_{\tilde{\tau}_1}\right), \nonumber
  \end{eqnarray}
  is taken and again this equals $(m_{q\pi}^2)_{\text{max}}$.

  The differences in shape of the charge conjugate plots in 
  figure~\ref{fig:tauInvMass} for the $\tau$ and $\pi$ are 
  due to the different helicities of the $\tau$ and $\pi$ as explained in 
  section~\ref{sec:sqDecay}. The kinks in these distributions show the change 
  from near to far leptons or pions making up the main components of the event. 
  These kind of 
  invariant mass distributions serve as a good indication of the spin
  properties of the particles involved in cascade decays. This information
  is important when trying to confirm an exact model of new physics since 
  it is possible for two different BSM models to imitate each other in certain
  decays even though the new particles introduced into each model have 
  different spin assignments~\cite{Smillie:2005ar}.

  Again our results are plotted along with those from HERWIG using the 
  TAUOLA  package. There is excellent agreement 
  between the two sets of results and the distributions follow those of figure 3
  in~\cite{Choi:2006mt}.
    
  \section{Conclusions}\label{sec:concl}
  We have described a new method for including new physics models in the \Hw 
  event generator which is more general than the previous approach. It allows
  new models to be implemented with a minimal amount of work. For the models
  implemented the results are in good agreement with either analytical answers
  or those from the  HERWIG event generator. Any new model will automatically 
  have spin correlations included since the algorithm demonstrated in 
  section~\ref{sec:spin} is formulated independently of any specific model and
  has been shown to agree with expected results.

  In the future we plan to implement other BSM physics models in \Hw which will
  enable comparative studies of the phenomenological consequences of these
  models to be carried out under the framework of the same generator.
  The current release, 2.0, of \Hw\!\!~\cite{Gieseke:2006ga} 
  does not include any of the features discussed in this paper, it will 
  however appear in version 2.1 of the event generator.

\begin{acknowledgement}
    We would like to thank our callaborators on the \Hw project for many useful
    comments. This work was supported in part by the Science and Technology 
    Facilities Council and the European Union Marie Curie Research Training
    Network MCnet under contract MRTN-CT-2006-035606.
\end{acknowledgement}

  \appendix
  \section{Helicity Code}\label{app:HELAS}

  In the FORTRAN HERWIG program the helicity amplitudes were calculated using the
  formalism of Kleiss and Stirling~\cite{Kleiss:1985yh}. However this meant that:
\begin{itemize}
\item all the matrix elements had to be calculated in the same frame;
\item it was impossible to interface the correlations in the parton shower 
      and those in the decay of unstable fundamental particles;
\item each matrix element had to be separately calculated.
\end{itemize}
  In \textsf{Herwig++} we choose to use to use an approach based on the \textsf{HELAS} 
  formalism for the calculation of matrix elements. This approach has a number
  of advantages:
\begin{itemize}
\item we can use the spinors and polarisation vectors calculated when the particle
      is produced to calculate their decays in a different frame, after an appropriate
      Lorentz transformation, so that each step of the calculation can be done
      in the most relevant frame;
\item more complicated matrix elements can be calculated from the basic
      building blocks rather than coded from scratch;
\item the inclusion of particles other than scalars, spin-$\frac12$ fermions
      and massless spin-$1$ bosons, which is complicated in the Kleiss and Stirling
      formalism, is relatively simple.
\end{itemize}

  The implementation of the \textsf{HELAS} formalism in \textsf{Herwig++} is based on two
  fundamental types of objects, \textsf{WaveFunctionBase} and \textsf{VertexBase}.
  
  The \textsf{WaveFunctions} contain the momentum and a pointer to the
  properties of the particles together with the basis state for a given particle
  spin.\textsf{WaveFunctionBase} stores the momentum of the 
  particle and a pointer to the \textsf{ParticleData} object for the particle.
  The inheriting \textsf{ScalarWaveFunction}, \textsf{SpinorWaveFunction},
  \textsf{SpinorBarWaveFunction}, \textsf{VectorWaveFunction}, 
  \textsf{RSSpinorWaveFunction}, \textsf{RSSpinorBarWaveFunction} and 
  \textsf{TensorWaveFunction} classes then contain storage of the 
  wavefunctions for spin-0, -$\frac12$, -1, -$\frac32$ and -2 particles together
  with methods to calculate the wavefunctions for a given helicity state.

  The \textsf{Vertices} contain methods to combine the wavefunctions
  for a specific vertex to give either off-shell wavefunctions, which can be used
  in further calculations, or the matrix element. The \textsf{VertexBase} class
  contains storage of the particles which interact at a given vertex. A number
  of classes then inherit from this class and implement the calculation of the 
  matrix element and off-shell wavefunctions for a given Lorentz structure of the 
  vertex in terms of arbitrary couplings which are calculated by 
  virtual member functions.

  This strategy is essentially identical to that adopted in the original 
  \textsf{HELAS} 
  approach. However, the new structure has the benefit that the user need only
  decide which vertex to use with the structure then supplying the relevant 
  couplings, whereas these had to be specified by the hand when using 
  the \textsf{HELAS} library.

  \subsection{Conventions}
  
  To numerically evaluate the matrix elements using the 
  \textsf{HELAS} formalism we need a specific choice of the Dirac matrices,  
  we currently support two options. The conventional low-energy choice, used
  in for example~\cite{Haber:1994pe}
\begin{eqnarray}
&\gamma^{\rm HABER}_i=\left(\begin{array}{cc}
                          0 & \sigma_i \\
                          -\sigma_i & 0
\end{array}\right),\hspace{3mm}
\gamma^{\rm HABER}_0=\left(\begin{array}{cc}
                  1 & 0 \\
                  0 & -1
\end{array}\right), \nonumber \\
&\gamma^{\rm HABER}_5=\left(\begin{array}{cc}
  0 & 1 \\
  1 & 0
\end{array}\right),
\end{eqnarray}
and the original choice of \textsf{HELAS} which is more appropriate at high energies
\begin{eqnarray}
& \gamma^{\rm HELAS}_i=\left(\begin{array}{cc}
                          0 & \sigma_i \\
                          -\sigma_i & 0
                        \end{array}\right),
 \gamma^{\rm HELAS}_0=\left(\begin{array}{cc}
                  0 & 1 \\
                  1 & 0
                \end{array}\right), \nonumber \\
& \gamma^{\rm HELAS}_5=\left(\begin{array}{cc}
                  -1 & 0 \\
                  0 & 1
                \end{array}\right).
\end{eqnarray}
These two representations are related by the transformation
\begin{equation}
 \psi_{\rm HELAS} = S \psi_{\rm Haber}\quad\mbox{where}\quad
 S=\frac{1}{\sqrt{2}}\left(\begin{array}{cc}
                             1 & -1 \\
                             1 & 1
                     \end{array}\right).
\end{equation}
  A number of container classes are implemented in the \textsf{ThePEG} 
  framework~\cite{Lonnblad:2006pt}, on which \textsf{Herwig++} is built, to store 
  the basis states for the different spins:
\begin{itemize}
\item \textsf{LorentzSpinor}:
  storage of the spinor, $u$ or $v$, for a spin-$\frac12$ fermion;
\item \textsf{LorentzSpinorBar}:
  storage of the barred spinor, $\bar{u}$ or $\bar{v}$, for a spin-$\frac12$ fermion;
\item \textsf{LorentzPolarizationVector}:
  storage of the polarization vector, $\epsilon^\mu$, for a spin-$1$ boson;
\item \textsf{LorentzRSSpinor}:
  storage of the spinor, $u^\mu$ or $v^\mu$, for a spin-$\frac32$ fermion;
\item \textsf{LorentzRSSpinorBar}:
  storage of the barred spinor, $\bar{u}^\mu$ or $\bar{v}^\mu$, for a spin-$\frac32$
  fermion;
\item \textsf{LorentzTensor}:
  storage of the polarization tensor, $\epsilon^{\mu\nu}$, for a spin-$2$ boson.
\end{itemize}
  In addition to providing storage of the basis state these classes 
  implement the Lorentz transformations, both boosts and rotations, for the objects.
  In the case of fermions the Dirac basis used to calculate the spinor, together
  with whether the spinor is $u$ or $v$ type is also stored. Methods to convert
  between the two supported Dirac matrix definitions are implemented together with
  the transformation between $u$ and $v$ spinors for Majorana particles.

\subsection{Lorentz Transformations}
\label{app:lorentz}
  In addition to the storage of the basis states we need to be able to 
  transform them between different Lorentz frames.
  
  The Lorentz transformation for a spinor is given by
\begin{equation}
 \psi'(x') = \psi'(ax) = S(a)\psi(x),
\end{equation}
  where $a^\nu_\mu=\frac{\partial x^{\prime\nu}}{\partial x^\mu}$.
  For a Lorentz boost along the direction specified
  by the unit vector $\hat{\bf{n}}$ with a magnitude $\beta$ the transformation is
  given by
\begin{equation}
   S_{\rm{boost}} = \cosh\left(\frac\chi2\right)+\sinh\left(\frac\chi2\right)\hat{n}_i\gamma^0\gamma^i,
\end{equation}
   where $\tanh\chi=\beta$.
  For a rotation by an angle $\phi$ about the unit vector $\hat{\bf{n}}$ the Lorentz
  transformation is given by 
\begin{equation}
  S_{\rm{rotation}} = \cos\left(\frac\phi2\right)+\sin\left(\frac\phi2\right)\epsilon^{ijk}\hat{n}_k\gamma^i\gamma^j.\label{eqn:spinorrot}
\end{equation}
  The Lorentz transformations for a four vector is given by
\begin{equation}
\epsilon^\mu(x') = L(a)^\mu_\nu\epsilon^\nu(x).
\end{equation}
  If we wish to boost by a factor $\beta$ along
  a unit vector $\hat{\bf n}$ the transformation is
\begin{eqnarray}
L^\mu_\nu = \left(
\begin{array}{cccc}
   \gamma & -\gamma\beta\hat{n}_1 & -\gamma\beta\hat{n}_2 & -\gamma\beta\hat{n}_3\\
-\gamma\beta\hat{n}_1 & 1-\hat{n}_1\hat{n}_1\omega & \phantom{1}-\hat{n}_1\hat{n}_2\omega & \phantom{1}-\hat{n}_1\hat{n}_3\omega \\
-\gamma\beta\hat{n}_2 & \phantom{1}-\hat{n}_2\hat{n}_1\omega & 1-\hat{n}_2\hat{n}_2\omega & \phantom{1}-\hat{n}_2\hat{n}_3\omega \\
-\gamma\beta\hat{n}_3 & \phantom{1}-\hat{n}_3\hat{n}_1\omega & \phantom{1} -\hat{n}_3\hat{n}_2\omega &1-\hat{n}_3\hat{n}_3\omega \\ 
\end{array}
\right),\hspace{5mm}
\end{eqnarray}
  where $\omega=1-\gamma$ and $\gamma=\frac1{\sqrt{1-\beta^2}}$.
  The Lorentz transformation for a rotation by an angle $\delta$ about 
a unit vector $\hat{\bf n}$ is
\begin{eqnarray}
L^\mu_\nu = &  \nonumber \\
&\hspace{-8mm} \left( 
\begin{array}{cccc} 
1 & 0& 0& 0\\
0 & \lambda\hat{n}_1\hat{n}_1 +  c_\delta
  & \lambda\hat{n}_1\hat{n}_2 -  s_\delta\hat{n}_3
  & \lambda\hat{n}_1\hat{n}_3 +  s_\delta\hat{n}_2\\
0 & \lambda\hat{n}_2\hat{n}_1 +  s_\delta\hat{n}_3
  & \lambda\hat{n}_2\hat{n}_2 +  c_\delta
  & \lambda\hat{n}_2\hat{n}_3 -  s_\delta\hat{n}_1\\
0 & \lambda\hat{n}_3\hat{n}_1 -  s_\delta\hat{n}_2
  & \lambda\hat{n}_3\hat{n}_2 +  s_\delta\hat{n}_1
  & \lambda\hat{n}_3\hat{n}_3 +  c_\delta
\end{array}
\right),\hspace{2mm}
\end{eqnarray}
 where $c_\delta=\cos\delta$, $s_\delta=\sin\delta$ and $\lambda=1-\cos\delta$.
  The transformations for the higher-spin particles can then be constructed as products
  of the spin-$\frac12$ and spin-$1$ transformations, \ie
\begin{subequations}
\begin{eqnarray}
\psi^\mu(x') &=& L(a)^\mu_\nu S(a)\psi^\nu(x),\\
\epsilon^{\mu\nu}(x') &=& L(a)^\mu_\alpha L(a)^\nu_\beta \epsilon^{\al\be}.
\end{eqnarray}
\end{subequations}

\subsection{WaveFunctions}

\subsubsection{ScalarWaveFunction}

  The \textsf{ScalarWaveFunction} class inherits from the \textsf{WaveFunctionBase}
  class and implements the storage of the wavefunction of a scalar particle
  as a complex number. For external particles this is just 1, however it
  can assume different values when the \textsf{WaveFunction} is the result
  is an off-shell internal line from a \textsf{Vertex} class.

\subsubsection{SpinorWaveFunction and SpinorBarWaveFunction}

  As with the \textsf{ScalarWaveFunction} the \textsf{SpinorWaveFunction}
  and \textsf{SpinorBarWaveFunction} classes inherit from the \textsf{WaveFunctionBase}
  class. The spinor is stored as either a \textsf{LorentzSpinor} or 
  \textsf{LorentzSpinorBar}. In addition the calculation of the 
  spinors for external particles is implemented. 
  The spinors are calculated in terms of two-component spinors, as 
  in~\cite{Murayama:1992gi}
\begin{subequations}
\begin{eqnarray}
\chi_+(p) &=& \frac1{\sqrt{2|{\bf p}|\left(|{\bf p}|+p_z\right)}}
\left(\begin{array}{c}|{\bf p}|+p_z\\ p_x+ip_y\end{array}\right),\\
\chi_-(p) &=& \frac1{\sqrt{2|{\bf p}|\left(|{\bf p}|+p_z\right)}}
\left(\begin{array}{c}-p_x+ip_y\\|{\bf p}|+p_z\end{array}\right),
\end{eqnarray}
 where $p_{x,y,z}$ are the $x$, $y$ and $z$ components of the momentum respectively,
 $E$ is the energy of the particle and $|{\bf p}|$ is the magnitude of the three
 momentum.
\end{subequations}
  For the \textsf{HELAS} choice of the Dirac representation the spinors are given by
\begin{subequations}
\begin{eqnarray}
u(p) = \left(\begin{array}{c}\phantom{-\lambda}\omega_{-\lambda}(p)\chi_{\lambda}(p)\\
			     \phantom{-\lambda}\omega_{\lambda}(p)\chi_{\lambda}(p)
\end{array}\right),\\
v(p) = \left(\begin{array}{c}-\lambda\omega_{\lambda}(p)\chi_{-\lambda}(p)\\
			     \phantom{-}\lambda\omega_{-\lambda}(p)\chi_{-\lambda}(p)
\end{array}\right),
\end{eqnarray}
\end{subequations}
where $\omega_\pm(p)=\sqrt{E\pm|{\bf p}|}$ and the helicity $\lambda=\pm1$.
Similarly for the low energy definition
\begin{subequations}
\begin{eqnarray}
u(p) = \left(\begin{array}{c}\phantom{-\lambda}\omega_{+}(p)\chi_{\lambda}(p)\\
			     \phantom{-}\lambda\omega_{-}(p)\chi_{\lambda}(p)
\end{array}\right),\\
v(p) = \left(\begin{array}{c}\phantom{-\lambda}\omega_{-}(p)\chi_{-\lambda}(p)\\
			     -\lambda\omega_{+}(p)\chi_{-\lambda}(p)
\end{array}\right).
\end{eqnarray}
\end{subequations}

\subsubsection{VectorWaveFunction}

  The \textsf{VectorWaveFunction} class inherits from the \textsf{WaveFunctionBase}
  class and implements the storage of the polarization vector using the 
  \textsf{LorentzPolarizationVector} class.
  The polarization vectors of a spin-1 particle can be calculated using,
  as in~\cite{Murayama:1992gi}
  \begin{subequations} 
\begin{eqnarray}
\epsilon^\mu_1(p) &=& \frac1{|{\bf p}|p_T}\left(0,p_xp_z,p_yp_z,-p_T^2\right),\\
\epsilon^\mu_2(p) &=& \frac1{p_T}         \left(0,-p_y,p_x,0\right),\\
\epsilon^\mu_3(p) &=& \frac{E}{m|{\bf p}|}\left(\frac{|{\bf p}|^2}{E},p_x,p_y,p_z\right),
\end{eqnarray}
\end{subequations}
  where $m$ is the mass and $p_T=\sqrt{p_x^2+p_y^2}$.
  We include two choices of the polarization vectors
\begin{subequations}
\begin{eqnarray}
\epsilon^\mu(p,\lambda=         \pm 1) &=& \frac1{\sqrt{2}}\left(\mp\epsilon^\mu_1(p)
                                                               -i\epsilon^\mu_2(p)
                                                         \right),\\
\epsilon^\mu(p,\lambda=\phantom{\pm}0) &=& \epsilon^\mu_3(p),
\end{eqnarray}
which is the choice used in \textsf{HELAS}. However, while this option is
available in the \textsf{Herwig++} by default we include the additional phase
factor $\exp(i\lambda\phi)$ as in \cite{Haber:1994pe} in order to make the inclusion of
spin correlations in the parton shower easier.
\end{subequations}

\subsubsection{RSSpinorWaveFunction and RSSpinorBarWaveFunction}

  Although there are currently no fundamental spin-$\frac32$ particles included in
  \textsf{Herwig++} the Rarita-Schwinger spinors for spin-$\frac32$ 
  particles are included
  both to allow the simulation of spin-$\frac32$ hadronic resonances and for the 
  possible future inclusion of the gravitino. The \textsf{RSSpinorWaveFunction}
  and \textsf{RSSpinorBarWaveFunction} inherit from the \textsf{WaveFunctionBase}
  class and implement the storage of the Rarita-Schwinger spinors
  using the \textsf{LorentzRSSpinor} and \textsf{LorentzRSSpinorBar} classes
  respectively.

  The spinors are calculated using the Clebsch-Gordon decomposition:
\begin{subequations}
\begin{eqnarray}
\psi^\mu(p,\lambda=-2) &=& \epsilon^\mu(p,-1)\psi(p,-1); \\
\psi^\mu(p,\lambda=-1) &=& \sqrt{\frac13}\epsilon^\mu(p,-1)\psi(p,1) \\
                       & &  + \sqrt{\frac23}\epsilon^\mu(p,0)\psi(p,-1);\\
\psi^\mu(p,\lambda=1)  &=& \sqrt{\frac13}\epsilon^\mu(p,1)\psi(p,-1)\\
                       & &  + \sqrt{\frac23}\epsilon^\mu(p,0);\\
\psi^\mu(p,\lambda=2)  &=& \epsilon^\mu(p,1)\psi(p,1).
\end{eqnarray}
\end{subequations}
  For massive particles the spinors are calculated in the rest frame of the 
  the particle and then boosted to the required frame in order that
  the Clebsch-Gordon decomposition can be easily applied. For massless spin-$\frac32$
  particles, which only have the $\pm2$ helicity states, the spinors are calculated
  in the same frame as the momentum.

\subsubsection{TensorWaveFunction}

  The \textsf{TensorWaveFunction} class inherits from the \textsf{WaveFunctionBase} 
  class and implements the storage of the 
  polarization tensor for spin-$2$ particles using the \textsf{LorentzTensor} class.

  The wavefunction is calculated using the Clebsch-Gordon decomposition:
\begin{subequations}
\begin{eqnarray}
\epsilon^{\mu\nu}(p,\lambda=-2) &=& \epsilon^\mu(p,-1)\epsilon^\nu(p,-1);\\
\epsilon^{\mu\nu}(p,\lambda=-1) &=& \sqrt{\frac12}\left[
\epsilon^\mu(p,-1)\epsilon^\nu(p,0)\right. \\
& & \hspace{10mm} +\left.\epsilon^\mu(p,0)\epsilon^\nu(p,-1)\right];\\
\epsilon^{\mu\nu}(p,\lambda=0) &=& \sqrt{\frac12}[
\epsilon^\mu(p,1)\epsilon^\nu(p,-1)
+\epsilon^\mu(p,-1)\epsilon^\nu(p,1) \nonumber\\
& &\hspace{10mm}+2\epsilon^\mu(p,0)\epsilon^\nu(p,0)];\\
\epsilon^{\mu\nu}(p,\lambda=1) &=& \sqrt{\frac12}\left[
\epsilon^\mu(p,1)\epsilon^\nu(p,0)\right.\\
& & \hspace{10mm}\left.+\epsilon^\mu(p,0)\epsilon^\nu(p,1)\right];\\
\epsilon^{\mu\nu}(p,\lambda=2) &=& \epsilon^\mu(p,1)\epsilon^\nu(p,1).
\end{eqnarray}
\end{subequations}
  Here this is applied in the frame in which the momentum is specified.

\subsection{Vertices}
  
  The \textsf{Vertices} all inherit from the \textsf{VertexBase} class.
  In general for all the vertices
  all the particles and momenta are defined to be incoming.

\subsubsection{Scalar Vertices}

  There are a number of vertices involving scalar bosons.

\begin{description}
\item[FFSVertex] The vertex for the coupling of a fermion and antifermion to
                 a scalar boson is defined to have the perturbative form
\begin{equation}
ic\bar{f_2}a^\lambda P_\lambda f_1\phi_3,
\end{equation}
 where $c$ is the overall normalisation, $a^\lambda$ are the left/right couplings,
 $P_\lambda$ are the helicity projection operators,
 $f_1$ is the wavefunction for the fermion, $\bar{f_2}$ is the wavefunction 
 for the antifermion and $\phi_3$ is the wavefunction for the scalar boson.

\item[GeneralSVVVertex] In addition to the perturbative form for the vertex
coupling a scalar and two vector bosons described below we include a general
form for this interaction so that effective vertices, for example $h^0\to gg$, can be
implemented. The form of the vertex is 
\begin{eqnarray}
&ic\left[a_{00}g^{\mu\nu}+a_{22}p_2^\mu p_2^\nu+a_{23}p_2^\mu p_3^\nu
+a_{32}p_3^\mu p_2^\nu+a_{33}p_3^\mu p_3^\nu \right.  \nonumber \\
&\hspace{3mm}\left. +a_{\epsilon}\epsilon^{\mu\nu\alpha\beta}p_{2\alpha} p_{3\beta}\right]
\epsilon_{2\mu}\epsilon_{3\nu}\phi_1,
\end{eqnarray}
  where $p_{2,3}$ are the momenta of the vector bosons,
  $\epsilon_{2,3}$ are the wavefunctions of the vector bosons,
  $\phi_1$ is the wavefunction of the scalar boson,
  $c$ is the overall coupling and
  $a_{ij}$ are the couplings of the different terms. 

\item[SSSVertex] The vertex for the coupling of three scalar bosons is defined
                 to have the perturbative form
\begin{equation}
ic\phi_1\phi_2\phi_3,
\end{equation}
  where $\phi_{1,2,3}$ are the wavefunctions for the scalar bosons and
  $c$ is the coupling.

\item[SSSSVertex] The vertex for the coupling of four scalar bosons is defined
                 to have the perturbative form
\begin{equation}
ic\phi_1\phi_2\phi_3\phi_4,
\end{equation}
  where $\phi_{1,2,3,4}$ are the wavefunctions for the scalar bosons and
  $c$ is the coupling.

\item[VSSVertex] The vertex for the coupling of a vector boson and two scalar
bosons is defined to have the perturbative form
\begin{equation}
-ic\left(p_2-p_3\right)\cdot\epsilon_1\phi_2\phi_3,
\end{equation}
  where $\epsilon_1$ is the wavefunction of the vector boson,
  $\phi_{2,3}$ are the wavefunctions for the scalar bosons and
  $p_{2,3}$ are the momenta of the scalar bosons and
  $c$ is the coupling.
 
\item[VVSSVertex] The vertex for the interaction of two vector and two
scalar bosons is defined to have the perturbative form
\begin{equation}
icg^{\mu\nu}\epsilon_{1\mu}\epsilon_{2\nu}\phi_3\phi_4,
\end{equation}
 where $\epsilon_{1,2}$ are the wavefunctions of the vector bosons and
  $\phi_{3,4}$ are the wavefunctions for the scalar bosons and
  $c$ is the coupling.

\item[VVSVertex]The vertex for the interaction of two vector and
a scalar boson is defined to have the perturbative form
\begin{equation}
icg^{\mu\nu}\epsilon_{1\mu}\epsilon_{2\nu}\phi_3,
\end{equation}
where $\epsilon_{1,2}$ are the wavefunctions of the vector bosons and
  $\phi_3$ is the wavefunction for the scalar boson and
  $c$ is the coupling.
\end{description}

\subsubsection{Vector Vertices}

  There are a number of vertices involving vector bosons.
\begin{description}
\item[FFVVertex] The interaction of a fermion, antifermion and a vector boson
is taken to have the perturbative form
\begin{equation}
ic\bar{f_2}\gamma^\mu a^\lambda P_\lambda f_1\epsilon_{3\mu},
\end{equation}
 where $c$ is the overall normalisation, $a^\lambda$ are the left/right couplings,
 $f_1$ is the wavefunction for the fermion, $\bar{f_2}$ is the wavefunction 
 for the antifermion and $\epsilon_3$ is the wavefunction for the vector boson.

\item[VVVVertex] The interaction of three vector bosons is
taken to have the perturbative form
\begin{eqnarray}
&ig\left[  (p_1-p_2)^\gamma g^{\alpha\beta }
               +(p_2-p_3)^\alpha g^{\beta \gamma}\right. \nonumber \\
&\hspace{5mm}\left. +(p_3-p_1)^\beta  g^{\alpha\gamma}
    \right]\epsilon_{1\alpha}\epsilon_{2\beta}\epsilon_{3\gamma},
\end{eqnarray}
where $\epsilon_{1,2,3}$ are the wavefunctions of the vector bosons
and $p_{1,2,3}$ are the momenta of the vector bosons.

\item[VVVVVertex] The interaction of four vector bosons is taken to have the form
\begin{equation}
ic^2\left[
    2\epsilon_1\cdot\epsilon_2\epsilon_3\cdot\epsilon_4-
    \epsilon_1\cdot\epsilon_3\epsilon_2\cdot\epsilon_4-
    \epsilon_1\cdot\epsilon_4\epsilon_2\cdot\epsilon_3
  \right],
\end{equation}
where $\epsilon_{1,2,3,4}$ are the wavefunctions of the vector bosons
and $p_{1,2,3,4}$ are the momenta of the vector bosons. For the quartic gluon
vertex this is the contribution of one colour structure. The others can
be obtained by an appropriate reordering of the input wavefunctions.

\end{description}

\subsubsection{Tensor Vertices}\label{app:TensorVert}
There are a number of vertices involving spin-2 particles. The form
of the Feynman rules follows that of~\cite{Han:1998sg}.
\begin{description}
\item[FFTVertex] The interaction of a pair of fermions with a tensor is taken to
  have the perturbative form
  \begin{eqnarray}
&    \hspace{-4mm}-\frac{i\kappa}8\bar{f_2}\left[
      \gamma_\mu(p_1-p_2)_\nu+\gamma_\nu(p_1-p_2)_\mu \right. \nonumber \\
& \hspace{4mm}\left. -2g_{\mu\nu}(p\!\!\!\!\!\not\,\,_1-p\!\!\!\!\!\not\,\,_2)
      +4g_{\mu\nu}m_{f}
      \right]f_1\epsilon^{\mu\nu}_3
  \end{eqnarray}
  where $\kappa$ is the defined as $2/\Lambda_{cut-off}$, $p_{1,2}$ are the 
  momenta of the fermions, $f_1$ is the fermion wavefunction, $\bar{f}_2$ is
  the antifermion wavefunction and $\epsilon_3^{\mu\nu}$ is the polarisation tensor
  for the spin-2 particle.
\item[VVVTVertex] The interaction of three vector bosons with a tensor is taken to
  have the perturbative form
  \begin{eqnarray}
&\hspace{-4mm}  g\frac\kappa2\left[
    C_{\mu\nu,\rho\sigma}(p_1-p_2)_\lambda
    +C_{\mu\nu,\rho\lambda}(p_3-p_1)_\sigma \right. \nonumber \\ 
    & \left.  +C_{\mu\nu,\sigma\lambda}(p_2-p_3)_\rho
    +F_{\mu\nu,\rho\sigma\lambda}
    \right]\epsilon_1^\rho\epsilon^\sigma_2\epsilon^\lambda_3
  \epsilon^{\mu\nu}_4
  \end{eqnarray}
  where  $\kappa$ is
  $2/\Lambda_{cut-off}$, $p_{1,2,3}$ are the momenta of the vector bosons,
  $\epsilon_{1,2,3}^\mu$ are the polarisation vectors and $\epsilon_4^{\mu\nu}$
  is the polarisation tensor. The $C$ and $F$ symbols are defined as
\begin{eqnarray}
  C_{\mu\nu,\rho\sigma} & = &g_{\mu\rho}g_{\nu\sigma}+g_{\mu\sigma}g_{\nu\rho}
  -g_{\mu\nu}g_{\rho\sigma}, \\
  F_{\mu\nu,\rho\sigma\lambda} & = & 
    g_{\mu\rho}g_{\sigma\lambda}(p_2-p_3)_\nu
   +g_{\mu\sigma}g_{\rho\lambda}(p_3-p_1)_\nu \nonumber \\
   &&   \hspace{5mm}+g_{\mu\lambda}g_{\rho\sigma}(p_1-p_2)_\nu 
   +(\mu\leftrightarrow\nu).
  \end{eqnarray}
\item[VVTVertex] The interaction of two vector bosons with a tensor is taken to
  have the perturbative form
  \beq
  -\frac{i\kappa}2\left[(m^2_v+p_1\cdot p_2)C_{\mu\nu,\rho\sigma}+D_{\mu\nu,\rho\sigma}\right]
  \epsilon_1^\rho\epsilon_2^\sigma \epsilon_3^{\mu\nu},
  \eeq
  where $\kappa$ is the defined as $2/\Lambda_{cut-off}$, $m_v$ is the mass of 
  the gauge boson $p_{1,2}$ are the 
  momenta of the vector bosons, $\epsilon_{1,2,3}^\mu$ are the polarisation 
  vectors and $\epsilon_3^{\mu\nu}$ is the polarisation tensor. 
  The $C$ symbol is defined as above and $D$ is defined as
  \begin{eqnarray}
    D_{\mu\nu,\rho\sigma} & = &
    g_{\mu\nu}k_{1\sigma}k_{2\rho} 
    -\left[g_{\mu\sigma}k_{1\nu}k_{2\rho}+
      g_{\mu\rho}k_{1\sigma}k_{2\nu}\nonumber \right. \\
  &&  \left. -g_{\rho\sigma}k_{1\mu}k_{2\nu} +(\mu\leftrightarrow\nu)\right].
  \end{eqnarray}
\item[FFVTVertex] The interaction of a pair of fermions with a vector boson
  and a tensor is taken to have the perturbative form
  \beq
  \frac{ig\kappa}4\bar{f_2}(C_{\mu\nu,\rho\sigma}-g_{\mu\nu}g_{\rho\sigma})
  \gamma^\sigma f_1\epsilon_3^\rho\epsilon_4^{\mu\nu}
  \eeq
  where $\kappa$ is the defined as $2/\Lambda_{cut-off}$, $\epsilon_{3}^\rho$ is
  the polarisation vector for the boson, $f_1$ is the fermion wavefunction, 
  $\bar{f}_2$ is the antifermion wavefunction and 
  $\epsilon_4^{\mu\nu}$ the polarisation
  tensor for the spin-2 particle. The $C$ symbol is defined above.
\item[SSTVertex] The interaction of a pair of scalars with a tensor is taken to
  have the perturbative form
  \beq
  -\frac{i\kappa}2\left[m^2_Sg_{\mu\nu}-p_{1\mu}p_{2\nu}-p_{1\nu}p_{2\mu}
    +g_{\mu\nu}p_1\cdot p_2\right]\epsilon^{\mu\nu}_3\phi_1\phi_2
  \eeq
  where $\kappa$ is the defined as $2/\Lambda_{cut-off}$, $\Lambda_{cut-off}$ is
  the ultraviolet cutoff scale,  $m_S$ is the mass of 
  the scalar, $p_{1,2}$ are the momenta of the scalars, $\epsilon_3^{\mu\nu}$ 
  the polarisation tensor for the spin-2 particle and $\phi_{1,2}$ are the
  scalar wavefunctions.
\end{description}
 
\section{\textsf{MatrixElement} Classes}\label{app:matrix}
  Table~\ref{tab:matrix} gives a list of the implemented $2\ra2$ 
  \textsf{MatrixElement} classes with a description of the types 
  of hard subprocess they can handle. $\Psi$ stands for a fermion, $V$ a 
  vector boson and $\phi$ a scalar. Charge conjugate modes are handled by the class 
  responsible for standard mode.
  \begin{table}
    \centering
    \begin{tabular}{|c|c|}
      \hline
      Class Name & Process Type \\
      \hline
      MEff2ff & $\Psi\mbox{ }\Psi^{'}\ra \Psi^{''}\mbox{ }\Psi^{'''}$ \\
      \hline
      MEff2ss & $\Psi\mbox{ }\Psi^{'}\ra \phi\mbox{ }\phi^{'}$ \\
      \hline
      MEff2vv & $\Psi\mbox{ }\Psi^{'}\ra V\mbox{ }V^{'}$ \\
      \hline
      MEfv2fs & $\Psi\mbox{ }V\ra \Psi\mbox{ }\phi $ \\
      \hline
      MEvv2ss & $V\mbox{ }V^{'}\ra \phi\mbox{ }\phi^{'}$ \\
      \hline
      MEvv2ff & $V\mbox{ }V^{'}\ra \Psi\mbox{ }\Psi^{'}$ \\
      \hline
      MEfv2vf & $\Psi\mbox{ }V^{}\ra V^{'}\mbox{ }\Psi^{'}$ \\
      \hline
      MEvv2vv & $V\mbox{ }V^{'}\ra V^{''}\mbox{ }V^{'''}$ \\
      \hline
    \end{tabular}
    \caption{The \textsf{MatrixElement} classes.}
    \label{tab:matrix}	
  \end{table}
  
\section{\textsf{Decayer} Classes}\label{app:decayers}
Table~\ref{tab:decclasses} is a list of the implemented two-body \textsf{Decayer} 
  classes with a description of the types of decay that they are designed to
  handle. $\Psi$ stands for a fermion, $V$ a vector boson, $\phi$ a scalar and
  $T$ a tensor. Charge conjugate modes are handled by the class responsible for 
 standard decay mode.

 \begin{table}
   \begin{center}
     \begin{tabular}{|c|c|}
	\hline
	Class Name & Decay Type \\
	\hline
	FFSDecayer & $\Psi\ra \Psi^{'}\mbox{ }\phi$ \\
	\hline
	FFVDecayer & $\Psi\ra \Psi^{'}\mbox{ }V$ \\ 
	\hline
	SFFDecayer & $\phi\ra \Psi\mbox{ }\Psi^{'}$ \\
	\hline
	SVVDecayer & $\phi\ra V\mbox{ }V^{'}$\\
	\hline
	SSSDecayer & $\phi\ra \phi^{'}\mbox{ }\phi^{''}$ \\
	\hline
	SSVDecayer & $\phi\ra \phi\mbox{ }V$ \\
	\hline
	TFFDecayer & $T\ra \Psi\mbox{ } \Psi^{'}$ \\
	\hline
	TSSDecayer & $T\ra \phi\mbox{ } \phi^{'}$ \\
	\hline
	TVVDecayer & $T\ra V\mbox{ }V^{'}$ \\
	\hline
	VFFDecayer & $V\ra \Psi\mbox{ } \Psi^{'}$ \\
	\hline
	VSSDecayer & $V\ra \phi\mbox{ } \phi^{'}$ \\
	\hline
	VVVDecayer & $V\ra V^{'}\mbox{ }V^{''}$ \\
	\hline
     \end{tabular}
     \end{center}
    \caption{The two-body \textsf{Decayer} classes.}
    \label{tab:decclasses}
 \end{table} 

 \vspace{10mm}
 
  \section{Implemented Vertices}\label{app:vert}
  Below is a list of vertices and associated Feynman rules, for BSM physics, as 
  they are currently implemented. The rules involving colour are written with
  the colour dependence explicitly pulled to the front since this is not 
  included in the code for a specific vertex as explained in
  section~\ref{sec:hardproc}. If a structure  constant $f^{abc}$ is involved 
  an additional factor of $i$ is pulled out along with it due to the 
  commutator relation $\left[ t^a,t^b \right]=if^{abc} t^c$.

  \subsection{RSModel}
  The Feynman rules as implemented in \Hw are given in section~\ref{app:TensorVert}.
  
  \subsection{MSSM}\label{app:mssm}
  We give the Feynman rules for the MSSM as implemented in \Hw\!\!. The sfermion
  mixing matrices are denoted by $Q^{k}_{\alpha\beta}$  and $L^k_{\alpha\beta}$
  for the squarks and leptons respectively where $k$ is the generation number,
  $\alpha$ the left/right eigenstate and $\beta$ the mass eigenstate. 
  $N_{ij}$, $U_{ij}$ and $V_{ij}$ are the 
  neutralino and chargino mixing matrices respectively. 
  The primed matrices in the neutralino rules are related to the unprimed ones 
  via
  \begin{subequations}
  \begin{eqnarray}
    N'_{i1} & = & N_{i1}\cw + N_{i2}\sw, \\
    N'_{i2} & = & N_{i2}\cw - N_{i1}\sw, \\
    N'_{i3} & = & N_{i3}, \\
    N'_{i4} & = & N_{i4}.
  \end{eqnarray}
  \end{subequations}  
  \begin{figure}
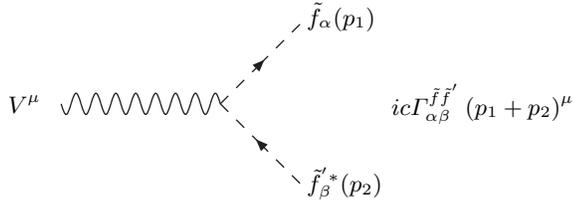

  \begin{center}
    \bpc(100,80)(50,10)
    \Photon(15,50)(75,50){-4}{8}
    \DashArrowLine(75,50)(105,80){5} \Text(105,83)[l]{{ $\tilde{f}_{\alpha}(p_1)$}}
    \DashArrowLine(105,20)(75,50){5} \Text(105,20)[l]{{ $\tilde{f}_{\beta}^{'*}(p_2)$}}
    \Text(140,50)[l]{$\displaystyle{ic}
      \Gamma^{\tilde{f}\tilde{f}^{'}}_{\alpha\beta} 
      \left(p_1 + p_2 \right)^{\mu}$}
    \Text(-5,50)[l]{$V^\mu$}
    \epc
    \end{center}
  \caption{Feynman rule for the interaction of a gauge boson with a pair
  of sfermions. The definition of $\Gamma$ for the various types of 
  gauge boson and sfermion is given in table ~\ref{tab:gaugeSFcp}. 
  The momenta are to be taken as in the direction of the arrows.}
  \end{figure}
  \begin{table}
    \begin{center}
      \renewcommand{\arraystretch}{1.5}
      \begin{tabular}{|c|c||c|c|}
	\hline
	$V^\mu$ &  ${\tilde{f}\tilde{f}^{'}}$ & {\large $c$}  & 
	  {\large$\Gamma^{\tilde{f}\tilde{f}^{'}}_{\alpha\beta}$}  \\
      \hline
      \hline
      $\gamma$ & $\tilde{q}\tilde{q}^{'} $  & $\displaystyle{-ee_q}$ & 
      $\delta^{\tilde{q}\tilde{q}^{'}} 
      _{\alpha\beta}$ \\
      \hline
      $\gamma$ & $\tilde{l}\mbox{ \!}\tilde{l}^{'} $  & $\displaystyle{e}$ & 
      $\delta^{\tilde{l}\mbox{ \!}\tilde{l}^{'}} 
      _{\alpha\beta}$ \\
      \hline
      $g$ & $\tilde{q}\tilde{q}^{'} $  & $\displaystyle{-gt^{a}}$ & 
      $\delta^{\tilde{q}\tilde{q}^{'}} 
      _{\alpha\beta}$ \\
      \hline
      $Z^0$ & $\tilde{u}_\alpha\tilde{u}_\beta$ & $\frac{g}{\cos\theta_W}$ & 
      $\frac{1}{2}\left(-Q^{2i}_{1\alpha}Q^{2i}_{1\beta}
       + 2e_u \ssw \delta_{\alpha\beta}\right)$ \\
       \hline
       $Z^0$ & $\tilde{d}_\alpha\tilde{d}_\beta$ & $\frac{g}{\cos\theta_W}$ & 
      $\frac{1}{2}\left(Q^{2i-1}_{1\alpha}Q^{2i-1}_{1\beta}
       + 2e_d \ssw \delta_{\alpha\beta}\right)$ \\
       \hline
       $Z^0$ & $\tilde{l}_\alpha\tilde{l}_\beta$ & $\frac{g}{\cos\theta_W}$ & 
      $\frac{1}{2}\left(L^{2i-1}_{1\alpha}L^{2i-1}_{1\beta}
       - 2\ssw \delta_{\alpha\beta}\right)$ \\
       \hline
       $Z^0$ & $\tilde{\nu}_\alpha\tilde{\nu}_\beta$ & $\frac{g}{\cos\theta_W}$ & 
       $-\frac{1}{2}\delta_{11}$ \\
       \hline
       $W^-$ & $\tilde{q}_\alpha\tilde{q}^{'}_\beta$ & $\frac{-g}{\sqrt{2}}$ &
	 $Q_{1\alpha}^{2i}Q_{1\beta}^{2i-1} $\\
       \hline
       $W^-$ & $\tilde{\nu}\tilde{l}_\beta$ & $\frac{-g}{\sqrt{2}}$ &
       $L^{2i-1}_{1\beta} $\\
       \hline	 
    \end{tabular}
  \end{center}
    \caption{Couplings for the gauge bosons and sfermions.}
    \label{tab:gaugeSFcp}
  \end{table}

 \begin{figure}
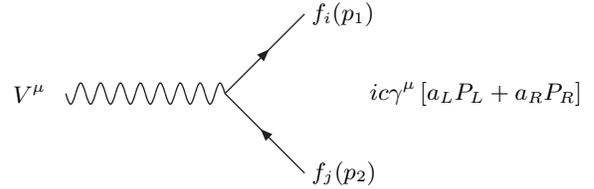

  \begin{center}
    \bpc(100,80)(50,10)
    \Photon(15,50)(75,50){-4}{8}
    \ArrowLine(75,50)(105,80) \Text(105,80)[l]{{ $f_{i}(p_1)$}}
    \ArrowLine(105,20)(75,50) \Text(105,20)[l]{{ $f_{j}(p_2)$}}
    \Text(-5,50)[l]{$V^\mu$}
    \Text(130,50)[l]{$\displaystyle{ic}\gamma^{\mu}\left[a_L P_L + 
	a_R P_R\right]$}
    \epc
    \caption{Feynman rule for the interaction of a gauge boson and a pair of
    gauginos. The couplings are defined in table~\ref{tab:gaugeMFcp}.}
    \end{center}
  \end{figure}

  \begin{table*}
    \renewcommand{\arraystretch}{1.5}
    \begin{center}
      \begin{tabular}{|c|c||c|c|c|}
        \hline
      $V^\mu$ &  ${\tilde{f}\tilde{f}^{'}}$ & {\large $c$}  & 
      {\large $a_{\scriptscriptstyle{L}}$ } & {\large $a_{\scriptscriptstyle{R}}$ } \\
      \hline
      \hline
      $\gamma$ & $\tilde{\chi}^{+}_i \tilde{\chi}^{-}_{j} $ & $-e$ & $1$ 
      & $1$ \\
      \hline
      $W^+$ & $\tilde{\chi}^{0}_i \tilde{\chi}^{+}_j$ & $g$ 
      & $-\frac{1}{\sqrt{2}}N_{i4}V^*_{j2} + N_{i2}V^*_{j1}$ &
      $\frac{1}{\sqrt{2}}N^*_{i3}U_{j2} + N^*_{i2}U_{j1} $ \\
      \hline
      $Z^{0} $ & $\tilde{\chi}^0_i \tilde{\chi}^0_j $ 
      & $\frac{g}{\cos\theta_W}$ 
      & $-\frac{1}{2}N_{i3}N^*_{j3} + \frac{1}{2}N_{i4}N^*_{j4}$ &
      $-a^*_L$\\
      \hline
      $Z^{0} $ & $\tilde{\chi}^-_i \tilde{\chi}^+_j $ 
      & $\frac{g}{\cos\theta_W}$ 
      & $-V_{i1}V_{j1}^* - \frac{1}{2}V_{i2}V_{j2}^* $ 
        & $-U_{i1}^*U_{j1} - \frac{1}{2}U_{i2}^*U_{j2} $
      \\
      & & & $ + \delta_{ij}\sin^2 \theta_W$ & $ +\delta_{ij}\sin^2 \theta_W$ \\
      \hline
      $ g_a $ & $\tilde{g}_b\mbox{ }\! \tilde{g}_c $ 
      & $\left\{if_{abc}\right\}g_s $ & $1$ & $1$ \\
      \hline
    \end{tabular}
    \caption{Feynman rules for the coupling of a gauge boson to a 
      pair of electroweak gauginos. All momenta are to be taken as outgoing.}
    \label{tab:gaugeMFcp}
    \end{center}
  \end{table*}
   \begin{figure}
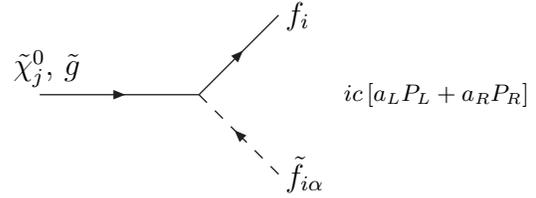

    \begin{center}
      \bpc(100,80)(50,10)
      \ArrowLine(15,50)(75,50)  \Text(5,60)[l]{{\large $ \tilde{\chi}^0_j$,
	  $\tilde{g}$}}
      \ArrowLine(75,50)(105,80) \Text(108,80)[l]{{\large $f_i$}}
      \DashArrowLine(105,20)(75,50){5} 
      \Text(108,20)[l]{{\large $\tilde{f}_{i\alpha}$}}
      \Text(130,50)[l]{$\displaystyle{ic}\left[a_L P_L + 
	  a_R P_R\right]$}
      \epc
  
    \caption{Feynman rule of the coupling of the neutralinos and gluino with 
      a Standard Model
    fermion and an sfermion. Table~\ref{tab:neutcp} gives the definitions of
    $a_L$ and $a_R$.}
      \end{center}
  \end{figure}
    \begin{table*}
      \renewcommand{\arraystretch}{1.5}
      \begin{center}
      \begin{tabular}{|c||c|l|}
	\hline
	$f_i\mbox{ }\!\tilde{f}_{i\alpha}$ & {\large $c$} 
	& \\ 
	\hline
	\hline
        $u_i\mbox{ }\!\tilde{u}_{i\alpha}$ & -$\sqrt{2}$
	& $a_{\scriptscriptstyle{L}}\!:$ 
	$\frac{gm_{u_i}N^*_{j4}}{2M_W\sin\beta}Q^{2i}_{1\alpha} - 
	Q^{2i}_{2\alpha}\left(ee_u N'^{*}_{j1} -
	\frac{ge_u \sin^2 \theta_W N'^*_{j2}}{\cos\theta_W}\right) $ \\
	& & $a_{\scriptscriptstyle{R}}\!:$ 
	$\frac{gm_{u_i}N_{j4}}{2M_W\sin\beta}Q^{2i}_{2\alpha} + 
	Q^{2i}_{1\alpha}\left(ee_u N'_{j1} +
	\frac{g(\frac{1}{2} - e_u\ssw)N'_{j2}}{\cos\theta_W}\right)$ \\
	\hline
        $d_i\mbox{ }\!\tilde{d}_{i\alpha}$ & -$\sqrt{2}$
	& $a_{\scriptscriptstyle{L}}\!:$ 
	 $\frac{gm_{d_i}N^*_{j3}}{2M_W\cos\beta}Q^{2i-1}_{1\alpha} - 
	Q^{2i-1}_{2\alpha}\left(ee_d N'^{*}_{j1} -
	\frac{ge_d \sin^2 \theta_W N'^*_{j2}}{\cos\theta_W}\right) $ \\
	& & $a_{\scriptscriptstyle{R}}\!:$ 
	 $\frac{gm_{d_i}N_{j3}}{2M_W\cos\beta}Q^{2i-1}_{2\alpha} + 
	Q^{2i-1}_{1\alpha}\left(ee_d N'_{j1} -
	\frac{g(\frac{1}{2} + e_d\ssw)N'_{j2}}{\cos\theta_W}\right)$ \\
	\hline
	$l_i \mbox{ }\!\tilde{l}_{i\alpha}$  
	& -$\sqrt{2} $ & $a_{\scriptscriptstyle{L}}\!:$ 
	 $\frac{gm_{l_i}N^{*}_{j3}}{2M_W\cos\beta} 
 	L^{2i-1}_{1\alpha}  + L^{2i-1}_{2\alpha}\left(eN'^*_{j1} - 
	\frac{g\ssw}{\cos\theta_W}N'^*_{j2}   \right) $ \\
	& & $a_{\scriptscriptstyle{R}}\!:$ 
	 $\frac{gm_{l_i}N_{j3}}{2M_W\cos\beta} 
 	L^{2i-1}_{2\alpha}  - L^{2i-1}_{1\alpha}\left(eN'_{j1} + 
	\frac{g(\frac{1}{2} - \ssw)}{\cos\theta_W}N'_{j2}   \right) $ \\
	\hline
	$\nu_i \mbox{ }\!\tilde{\nu}_i$
	& -$\sqrt{2} $ & $a_{\scriptscriptstyle{L}}\!:$ 
	0  \hspace{33pt}$a_{\scriptscriptstyle{R}}\!:$
	 $\frac{gN'_{j2}}{2\cos\theta_W}$ \\
 	\hline
	$q^b_i\mbox{ }\!\tilde{q}^c_{i\alpha}$ &  
	$\left\{ t^a_{bc}\right\}(-g_s\sqrt{2})$ 
	& $a_{\scriptscriptstyle{L}}\!:$ $-Q^k_{2\alpha}$
	\hspace{10pt}$a_{\scriptscriptstyle{R}}\!:$ $Q^k_{1\alpha}$ \\
	\hline
      \end{tabular}
      \caption{Neutralino and gluino couplings. In the case of the 
	gluino $k=2i$ for up-type quarks and $k=2i-1$ for
      down-type quarks.}
      \label{tab:neutcp}
    \end{center}
    \end{table*}

   \begin{figure}
    \begin{center}
      \bpc(300,200)(0,0)
      \ArrowLine(75,50)(15,50)  \Text(0,50)[l]{{\large $ \tilde{\chi}^+_j $}}
      \ArrowLine(75,50)(105,80) \Text(108,80)[l]{{\large $d_i$, $l^-_i$}}
      \DashArrowLine(105,20)(75,50){5} 
      \Text(108,20)[l]{{\large $\tilde{u}_{i\alpha}$, $\tilde{\nu}_i$}}
      \Text(160,50)[l]{$\displaystyle{ic}\left[a_L P_L + 
	  a_R P_R\right]C$}
      \SetOffset(0,105)
      \ArrowLine(15,50)(75,50)  \Text(0,50)[l]{{\large $\tilde{\chi}^+_j $}}
      \ArrowLine(75,50)(105,80) \Text(108,80)[l]{{\large $u_i$, $\nu_i$}}
      \DashArrowLine(105,20)(75,50){5} 
      \Text(108,20)[l]{{\large $\tilde{d}_{i\alpha}$, $\tilde{l}_{i\alpha}$}}
      \Text(160,50)[l]{$\displaystyle{ic}\left[a_L P_L + 
	  a_R P_R\right]$}
      \epc
    \end{center}
    \caption{Feynman rule of the coupling of a chargino with a Standard Model
    fermion and an sfermion. $C$ is the charge conjugation matrix.
    Table~\ref{tab:chargcp} gives the definitions of
    $a_L$ and $a_R$.}
    \end{figure}
    \begin{table}
      \renewcommand{\arraystretch}{1.5}
      \begin{center}
      \begin{tabular}{|c|c||c|c|}
	\hline
	$f_i\mbox{ }\!\tilde{f}_{i\alpha}^{'}$ & {\large $c$} 
	& {\large $a_{\scriptscriptstyle{L}}$ } 
	& {\large $a_{\scriptscriptstyle{R}}$ } \\
	\hline
	\hline
        $u_i\mbox{ }\!\tilde{d}_{i\alpha}$ & $-g$
	& $-\frac{m_{u_i}V^*_{j2}}{\sqrt{2}M_W\sin\beta}Q^{2i-1}_{1\alpha}$
	& $U_{j1}Q^{2i-1}_{1\alpha} - 
	\frac{m_{d_i}U_{j2}}{\sqrt{2}M_W\cos\beta}Q^{2i-1}_{2\alpha}$ \\
	\hline
	$d_i\mbox{ }\!\tilde{u}_{i\alpha}$ & $-g$ 
	& $-\frac{m_{d_i}U^*_{j2}}{\sqrt{2}M_W\cos\beta}Q^{2i}_{1\alpha}$
	& $V_{j1}Q^{2i}_{1\alpha} - 
	\frac{m_{u_i}V_{j2}}{\sqrt{2}M_W\sin\beta}Q^{2i}_{2\alpha}$ \\
	\hline
	$\nu_i\mbox{ }\!\tilde{l}_{i\alpha} $ & $-g$ 
	& 0 & $U_{j1}L_{1\alpha}^{2i-1} - 
	\frac{m_{l_i}U_{j2}}{\sqrt{2}M_W\cos\beta}L^{2i-1}_{2\alpha} $ \\
	\hline
	$l_i\mbox{ }\!\tilde{\nu}_{i\alpha}$ & $-g$
	& $-\frac{m_{l_i}U^*_{j2}}{\sqrt{2}M_W\cos\beta}$ & $V_{j1}$ \\
	\hline
      \end{tabular}
      \caption{Chargino couplings for the $i^{th}$ fermion generation.}
      \label{tab:chargcp}
    \end{center}
    \end{table}
    \begin{figure}
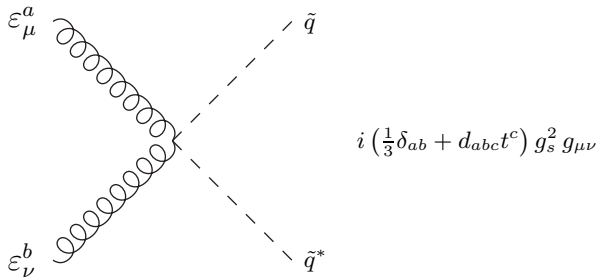

    \begin{center}
      \bpc(300,130)(-10,20)
      \Gluon(15,130)(60,85){4}{8}
      \Text(-2,130)[l]{{\large$\varepsilon^a_{\mu}$}}
      \Gluon(15,40)(60,85){-4}{8} 
      \Text(-2,40)[l]{{\large$\varepsilon^b_{\nu}$}}
      \DashLine(60,85)(105,130){5} \Text(110,130)[l]{$\tilde{q}$}
      \DashLine(105,40)(60,85){5} \Text(110,40)[l]{$\tilde{q}^*$}
      \Text(130,85)[l]{$i\left(\frac{1}{3}\delta_{ab} + 
	d_{abc}t^c\right)g_s^2\mbox{ }\!g_{\mu\nu}$}
      \epc
      \caption{Feynman rule for the coupling of two gluons to a pair of
      squarks.}
    \end{center}
    \label{fig:ggsqsqRule}
    \end{figure}
    \begin{figure}
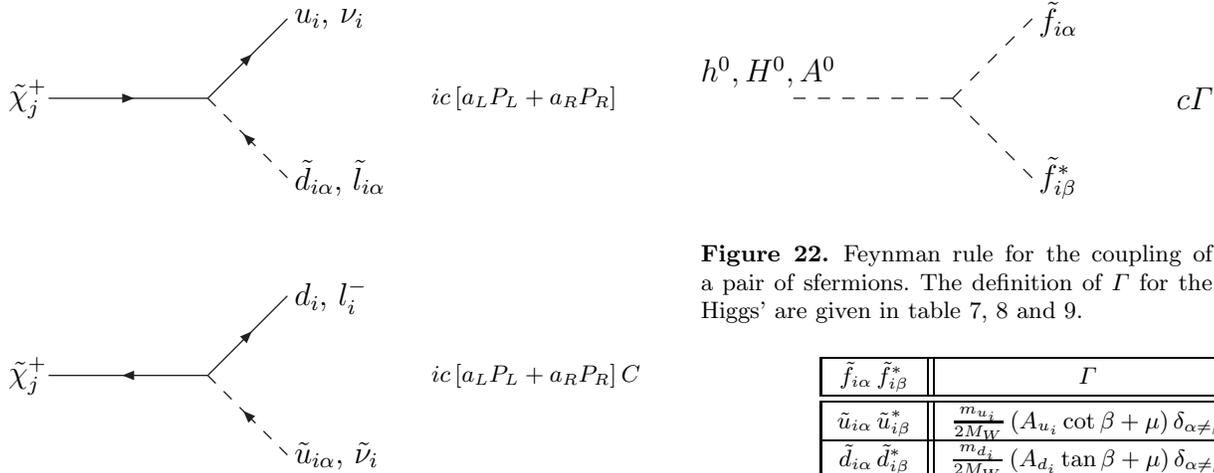

    \begin{center}
      \bpc(300,85)(-20,10)
      \DashLine(75,50)(15,50){5}  \Text(-20,60)[l]{{\large $h^0, H^0, A^0$}}
      \DashLine(75,50)(105,80){5} \Text(108,80)[l]{{\large $\tilde{f}_{i\alpha}$}}
      \DashLine(105,20)(75,50){5} 
      \Text(108,20)[l]{{\large $\tilde{f}^*_{i\beta}$}}
      \Text(160,50)[l]{{\large $c\Gamma$ }}
      \epc
    \end{center}
    \caption{Feynman rule for the coupling of a higgs with a pair of 
      sfermions. The definition of $\Gamma$ for the three neutral Higgs' are given
      in table~\ref{tab:smhiggscp},~\ref{tab:higgs2cp} and~\ref{tab:higgs3cp}.}
    \end{figure}
    \begin{table*}
       \begin{center}
      \renewcommand{\arraystretch}{1.5}
	\begin{tabular}{|c||l|}
	  \hline
	   $\tilde{f}_{i\alpha}\mbox{ }\!\tilde{f}^*_{i\beta}$ 
	  & \hspace{140pt} $\Gamma$ \\
	  \hline
	  \hline
	  $\tilde{u}_{i\alpha}\mbox{ }\!\tilde{u}^*_{i\beta}$ 
	  & $\frac{M_Z\sin(\alpha + \beta)}{\cos\theta_W}
	  \left[ Q^{2i}_{1\alpha}Q^{2i}_{1\beta}(\frac{1}{2} - e_u\ssw)
	    + e_u\ssw Q^{2i}_{2\alpha}Q^{2i}_{2\beta}\right]$ \\
	  & $-\frac{m_{u_i}^2\cos\alpha}{M_W\sin\beta}
	  \left[Q^{2i}_{1\alpha}Q^{2i}_{1\beta} + 
	    Q^{2i}_{2\alpha}Q^{2i}_{2\beta}\right]$ \\
	  & $- \frac{m_{u_i}}{2M_W\sin\beta}(A_{u_i}\cos\alpha + \mu\sin\alpha)$ 
	  $\left[Q^{2i}_{2\alpha}Q^{2i}_{1\beta} + 
	    Q^{2i}_{1\alpha}Q^{2i}_{2\beta}  \right]$ \\
	  \hline
	  $\tilde{d}_{i\alpha}\mbox{ }\!\tilde{d}^*_{i\beta}$ 
	  & $-\frac{M_Z\sin(\alpha + \beta)}{\cos\theta_W}
	  \left[ Q^{2i-1}_{1\alpha}Q^{2i-1}_{1\beta}(\frac{1}{2} + e_d\ssw)
	    - e_d\ssw Q^{2i-1}_{2\alpha}Q^{2i-1}_{2\beta}\right]$ \\
	  & $+\frac{m_{d_i}^2\sin\alpha}{M_W\cos\beta}
	  \left[Q^{2i-1}_{1\alpha}Q^{2i-1}_{1\beta} + 
	    Q^{2i-1}_{2\alpha}Q^{2i-1}_{2\beta}\right]$ \\
	  & $+ \frac{m_{d_i}}{2M_W\cos\beta}(A_{d_i}\sin\alpha + \mu\cos\alpha)$ 
	  $\left[Q^{2i-1}_{2\alpha}Q^{2i-1}_{1\beta} + 
	    Q^{2i-1}_{1\alpha}Q^{2i-1}_{2\beta}\right]$ \\
	  \hline
	  $\tilde{l}_{i\alpha}\mbox{ }\!\tilde{l}^*_{i\beta}$ 
	  & $-\frac{M_Z\sin(\alpha + \beta)}{\cos\theta_W}
	  \left[ L^{2i-1}_{1\alpha}L^{2i-1}_{1\beta}(\frac{1}{2} - \ssw)
	    + \ssw L^{2i-1}_{2\alpha}L^{2i-1}_{2\beta}\right]$ \\
	  & $+\frac{m_{l_i}^2\sin\alpha}{M_W\cos\beta}
	  \left[L^{2i-1}_{1\alpha}L^{2i-1}_{1\beta} + 
	    L^{2i-1}_{2\alpha}L^{2i-1}_{2\beta}\right]$ \\
	  & $+\frac{m_{l_i}}{2M_W\cos\beta}(A_{e_i}\sin\alpha + \mu\cos\alpha)$ 
	  $\left[L^{2i-1}_{2\alpha}L^{2i-1}_{1\beta} + 
	    L^{2i-1}_{1\alpha}L^{2i-1}_{2\beta}\right]$ \\
	  \hline
	    $\tilde{\nu}_{i\alpha}\mbox{ }\!\tilde{\nu}^*_{i\beta}$ 
	  & $\frac{M_Z\sin(\alpha + \beta)}{2\cos\theta_W} $ \\
	  \hline
	\end{tabular}
	\caption{$h^0$ couplings to sfermion pairs with  $c=ig$.}
	\label{tab:smhiggscp}
      \end{center}
    \end{table*}

    \begin{table*}
      \renewcommand{\arraystretch}{1.5}
      \centering
       \begin{center}
	\begin{tabular}{|c||l|}
	  \hline
	   $\tilde{f}_{i\alpha}\mbox{ }\!\tilde{f}^*_{i\beta}$ 
	  & \hspace{140pt} $\Gamma$ \\
	  \hline
	  \hline
	  $\tilde{u}_{i\alpha}\mbox{ }\!\tilde{u}^*_{i\beta}$ 
	  & $-\frac{M_Z\cos(\alpha + \beta)}{\cos\theta_W}
	  \left[ Q^{2i}_{1\alpha}Q^{2i}_{1\beta}(\frac{1}{2} - e_u\ssw)
	    + e_u\ssw Q^{2i}_{2\alpha}Q^{2i}_{2\beta}\right]$ \\
	  & $-\frac{m_{u_i}^2\sin\alpha}{M_W\sin\beta}
	  \left[Q^{2i}_{1\alpha}Q^{2i}_{1\beta} + 
	    Q^{2i}_{2\alpha}Q^{2i}_{2\beta}\right]$ \\
	  & $-\frac{m_{u_i}}{2M_W\sin\beta}(A_{u_i}\sin\alpha - \mu\cos\alpha)$ 
	  $\left[Q^{2i}_{2\alpha}Q^{2i}_{1\beta} + 
	    Q^{2i}_{1\alpha}Q^{2i}_{2\beta}  \right]$ \\
	  \hline
	  $\tilde{d}_{i\alpha}\mbox{ }\!\tilde{d}^*_{i\beta}$ 
	  & $\frac{M_Z\cos(\alpha + \beta)}{\cos\theta_W}
	  \left[ Q^{2i-1}_{1\alpha}Q^{2i-1}_{1\beta}(\frac{1}{2} + e_d\ssw)
	    - e_d\ssw Q^{2i-1}_{2\alpha}Q^{2i-1}_{2\beta}\right]$ \\
	  & $-\frac{m_{d_i}^2\cos\alpha}{M_W\cos\beta}
	  \left[Q^{2i-1}_{1\alpha}Q^{2i-1}_{1\beta} + 
	    Q^{2i-1}_{2\alpha}Q^{2i-1}_{2\beta}\right]$ \\
	  & $+ \frac{m_{d_i}}{2M_W\cos\beta}(\mu\sin\alpha - A_{d_i}\cos\alpha)$ 
	  $\left[Q^{2i-1}_{2\alpha}Q^{2i-1}_{1\beta} + 
	    Q^{2i-1}_{1\alpha}Q^{2i-1}_{2\beta}\right]$ \\
	  \hline
	  $\tilde{l}_{i\alpha}\mbox{ }\!\tilde{l}^*_{i\beta}$ 
	  & $\frac{M_Z\cos(\alpha + \beta)}{\cos\theta_W}
	  \left[ L^{2i-1}_{1\alpha}L^{2i-1}_{1\beta}(\frac{1}{2} - \ssw)
	    + \ssw L^{2i-1}_{2\alpha}L^{2i-1}_{2\beta}\right]$ \\
	  & $-\frac{m_{l_i}^2\cos\alpha}{M_W\cos\beta}
	  \left[L^{2i-1}_{1\alpha}L^{2i-1}_{1\beta} + 
	    L^{2i-1}_{2\alpha}L^{2i-1}_{2\beta}\right]$ \\
	  & $+\frac{m_{l_i}}{2M_W\cos\beta}(\mu\sin\alpha - A_{e_i}\cos\alpha )$ 
	  $\left[L^{2i-1}_{2\alpha}L^{2i-1}_{1\beta} + 
	    L^{2i-1}_{1\alpha}L^{2i-1}_{2\beta}\right]$ \\
	  \hline
	    $\tilde{\nu}_{i\alpha}\mbox{ }\!\tilde{\nu}^*_{i\beta}$ 
	  & $-\frac{M_Z\cos(\alpha + \beta)}{2\cos\theta_W} $ \\
	  \hline
	\end{tabular}
	\caption{$H^0$ couplings to sfermion pairs with  $c=ig$.}
	\label{tab:higgs2cp}
      \end{center}
    \end{table*}

    \begin{table}
      \renewcommand{\arraystretch}{1.3}
	\begin{center}
      \begin{tabular}{|c||c|}
	\hline $\tilde{f}_{i\alpha}\mbox{ }\!\tilde{f}^*_{i\beta}$ &  $\Gamma$ \\
	\hline
	\hline
	$\tilde{u}_{i\alpha}\mbox{ }\!\tilde{u}^*_{i\beta}$ 
	& $\frac{m_{u_i}}{2M_W}\left(A_{u_i}\cot\beta + \mu\right)\delta_{\alpha\neq\beta}$ \\
	\hline
	$\tilde{d}_{i\alpha}\mbox{ }\!\tilde{d}^*_{i\beta}$ 
	& $\frac{m_{d_i}}{2M_W}\left(A_{d_i}\tan\beta+\mu\right)\delta_{\alpha\neq\beta}$ \\
	\hline
	$\tilde{l}_{i\alpha}\mbox{ }\!\tilde{l}^*_{i\beta}$ 
	& $\frac{m_{l_i}}{2M_W}\left(A_{e_i}\tan\beta + \mu\right)\delta_{\alpha\neq\beta}$ \\
	\hline
      \end{tabular}
      \caption{$A^0$ couplings to sfermion pairs with  $c=g$.}
      \label{tab:higgs3cp}
      \end{center}
    \end{table}


\providecommand{\href}[2]{#2}\begingroup\raggedright\endgroup

\end{document}